\documentclass[twocolumn]{aastex631}

\usepackage{multirow}

\begin{document}

\title{Early Planet Formation in Embedded Disks (eDisk). XIX. Structures of molecular outflows 
}

\author[0000-0002-3003-7977]{Anton Feeney-Johansson}
\affiliation{National Astronomical Observatory of Japan, Osawa 2-21-1, Mitaka, Tokyo 181-8588, Japan}
\affiliation{Department of Astronomy, Graduate School of Science, The University of Tokyo, 113-0033 Tokyo, Japan}

\author{Yuri Aikawa}
\affiliation{Department of Astronomy, Graduate School of Science, The University of Tokyo, 113-0033 Tokyo, Japan}

\author{Shigehisa Takakuwa}
\affiliation{Department of Physics and Astronomy, Graduate School of Science and Engineering, Kagoshima University, 1-21-35 Korimoto, Kagoshima, Kagoshima 890-0065,
Japan}
\affiliation{Academia Sinica Institute of Astronomy \& Astrophysics, 11F of Astronomy-Mathematics Building, AS/NTU, No. 1, Sec. 4, Roosevelt Rd., Taipei 106319, Taiwan, R.O.C.}

\author[0000-0003-0998-5064]{Nagayoshi Ohashi}
\affiliation{Academia Sinica Institute of Astronomy \& Astrophysics, 11F of Astronomy-Mathematics Building, AS/NTU, No. 1, Sec. 4, Roosevelt Rd., Taipei 106319, Taiwan, R.O.C.}

\author{Adele Plunkett}
\affiliation{National Radio Astronomy Observatory, 520 Edgemont Road, Charlottesville, VA 22903, USA;}

\author{Jes K. J\o rgensen}
\affiliation{Niels Bohr Institute, University of Copenhagen, Jagtvej 155A, DK-2200, Copenhagen N, Denmark}

\author{Hsien Shang}
\affiliation{Institute of Astronomy and Astrophysics, Academia Sinica, Taipei 106216, Taiwan}

\author{Zhi-Yun Li}
\affiliation{University of Virginia, 530 McCormick Rd., Charlottesville, Virginia 22903, USA}

\author{Rajeeb Sharma}
\affiliation{Niels Bohr Institute, University of Copenhagen, Jagtvej 155A, DK-2200, Copenhagen N, Denmark}

\author{Woojin Kwon}
\affiliation{Department of Earth Science Education, Seoul National University, 1 Gwanak-ro, Gwanak-gu, Seoul 08826, Republic of Korea}
\affiliation{SNU Astronomy Research Center, Seoul National University, 1 Gwanak-ro, Gwanak-gu, Seoul 08826, Republic of Korea}

\author{Jeong-Eun Lee}
\affiliation{Department of Physics and Astronomy, Seoul National University, 1 Gwanak-ro, Gwanak-gu, Seoul 08826, Korea}

\author{Leslie W. Looney}
\affiliation{Department of Astronomy, University of Illinois, 1002 West Green St, Urbana, IL 61801, USA}

\author{Yao-Lun Yang}
\affiliation{RIKEN Cluster for Pioneering Research, Wako-shi, Saitama, 351-0198, Japan}

\author{Mayank Narang}
\affiliation{Institute of Astronomy and Astrophysics, 
Academia Sinica, Taipei 106216, Taiwan}

\author{Itziar de Gregorio-Monsalvo}
\affiliation{European Southern Observatory, Alonso de Cordova 3107, Casilla 19, Vitacura, Santiago, Chile}

\author{eDisk team}


\begin{abstract}

As part of the ALMA Large Program ``Early Planet Formation in Embedded Disks'' (eDisk), $^{12}$CO (2-1) was observed towards 19 nearby low-mass protostars. Of these objects, 15 sources are found to show molecular outflow emission.
Based on their morphological and kinematical structures, the CO outflows are classified into three types: a wind-driven shell, where ambient material is swept up by a wide-angle wind from the star, a bow shock,
and a slow disk wind,
which is a conical or parabolic flow with onion-like velocity structure.
We categorize 11 outflows as a slow disk wind, 7 as a wind-driven shell, and 1 as a bow shock. Four of these outflows were found to show signs of both slow disk wind and wind-driven shell characteristics.
Five objects show misalignment between the red- and
blue-shifted outflows. Seven objects show significant misalignment between the outflow axis (either or both of the red- and blue-shifted outflows) and the minor axis of the dust continuum emission around the protostar.
For the objects showing wind-driven shell emission, we compare simple parametrized models with the observations to derive physical properties of the observed shells, such as their dynamical ages. This shows evidence of a time variability in the outflows, such as changes in their direction. In some objects, large differences are seen between the properties of the red- and blue-shifted outflows, possibly indicating differences in the properties of the ambient medium with which the outflow interacts. 

\end{abstract}



\begin{table*}
\begin{small}
\begin{center}
\caption{Properties of the eDisk sources}\label{tab:sum}
\begin{tabular}{lccccccccc}\toprule
Source Name & Class & $T_{\mathrm{bol}}$ & $i_{\rm cont}$\tablenotemark{c} & Emission & Emission & Previous Jet/Outflow Detections & ref.\tablenotemark{g, h}\\
 & & K & $^{o}$ & Blue & Red & & \\
\hline
BHR71 IRS2 &0 & 39 &31 & SDW\tablenotemark{d} & SDW & $^{12}$CO & [2]\textbf{[38]}\\
B335 &0 &41 &37 & SDW & SDW & $^{12}$CO, IR ([Fe II], HI, H$_2$) &[1][2][3][4][5][37][39][40] \\
L1527 IRS &0 & 41 & $\sim90$ & SDW & SDW & $^{12}$CO, Radio & [2][3][6][24]\textbf{[25]}\\
IRAS 16253-2429 &0 & 42 &68 & SDW & SDW & $^{12}$CO, IR ([Fe II], [Ne II], [Ni II], [H I]) & [7][8]\textbf{[26]}\\
IRAS 16544-1604 &0 &52 &73 & WS\tablenotemark{e} & WS & $^{12}$CO & [4][9]\textbf{[27]}\\
GSS30 IRS3 &0 & 50 & 64 & SDW/WS & WS & $^{12}$CO, Radio & [10][11]\textbf{[41]} \\
IRAS 15398-3359 &0 & 50 &51 & BS\tablenotemark{f} & BS & $^{12}$CO, IR ([Fe II], [Ne II], [S I], H$_2$) & [1][4][12][13]\textbf{[28]}\\
R CrA IRS5N &0 & 59 &65 & -- & -- & Radio?/ H$_2$? \tablenotemark{i} & [14][15]\textbf{[29]}\\
IRAS 04166+2706 &0 & 61 &47 & SDW & SDW/WS  & $^{12}$CO & [3][4][16]\textbf{[45]}\\
R CrA IRAS 32\tablenotemark{a} &0 & 64 &69 & SDW/WS & SDW/WS & $^{12}$CO & [4][12]\textbf{[30]}\\
BHR71 IRS1 &0 & 66 &39 & SDW & SDW & $^{12}$CO & [1][2]\textbf{[38]}\\
Ced 110 IRS4\tablenotemark{b}  & 0 & 68 & 75 & -- & -- & $^{12}$CO, Radio, IR(H$_2$) & [2][12][17][18][19]\textbf{[31]}[44]\\
R CrA IRS 7B\tablenotemark{b}  &I & 88 & 68 & -- & -- &  $^{12}$CO, Radio / H$_2$? & [12][14][15]\textbf{[32]} \\
IRAS 04302+2247 & I & 88 & 84 & WS? & -- & $^{12}$CO, IR & [3][20]\textbf{[33]}\\
IRAS 04169+2702 & I & 163 &44 & SDW & SDW? & $^{12}$CO, Optical ([S II]) & [3][21]\textbf{[43]}\\
TMC-1A & I & 183 &52  & SDW/WS & SDW & $^{12}$CO, Radio?, IR ([Fe II], H$_2$) & [1][2][3][23][36]\\
Oph IRS43\tablenotemark{b}  & I & 193 & 78 & -- & -- & $^{12}$CO & [3]\textbf{[34]}\\
L1489 IRS & I & 213 &71 & SDW & SDW  & $^{12}$CO, Radio & [2][22]\textbf{[35]}\\
Oph IRS63 & I & 348 &47 & WS & WS & $^{12}$CO & [2][3][4][9]\textbf{[42]}\\

\hline
\end{tabular}

\end{center}
\end{small}
\tablenotetext{a}{R CrA IRAS 32 is a binary. We assume that the outflow is launched from IRAS 32A to quantify the properties of the outflow, e.g. PA.}
\tablenotetext{b}{Binary sources. The inclination is for the primary source.}
\tablenotetext{c}{Inclination angle derived from the dust continuum emission, assuming that it traces a disk; zero inclination means face-on. The inclination angles are estimated by 2D Gaussian fitting of the dust continuum \citep{Ohashi2023}, except for L1527 IRS, which is known from continuum and molecular line emission to be nearly edge-on \citep[see Section 4.3,][]{VantHoff2023}. Since the disk can be geometrically thick and since the envelope emission can also contribute, these values are considered a lower limit.}
\tablenotetext{d}{``SDW'' stands for slow disk wind.}
\tablenotetext{e}{``WS'' stands for wind-driven shell}
\tablenotetext{f}{``BS'' stands for bow shock.}
\tablenotetext{g}{
[1]\citet{Yang2018}, [2]\citet{Yildiz2015}, [3]\citet{Bontemps1996}, [4]\citet{Dunham2014}, [5]\citet{Yen2010}, [6]\citet{Reipurth2004},
[7]\citet{Hsieh2016}, [8]\citet{Narang2024}, [9]\citet{Fukui1989}, [10]\citet{Friesen2018}, [11]\citet{Coutens2019}, [12]\citet{VanKempen2009},
[13]\citet{Yang2022b}, [14]\citet{Miettinen2008}, [15]\citet{Kumar2011},
[16]\citet{Wang2014}, [17]\citet{Belloche2006}, [18]\citet{Lehtinen2003},
[19]\citet{Bally2006}, [20]\citet{Lucas_1997}, [21]\citet{Gomez1997},
[22]\citet{Girart2002}, [23]\citet{Harsono2023}, [24]\citet{Aso2017}, [25]\citet{VantHoff2023}
[26]\citet{Aso2023}, [27]\citet{Kido2023}, [28]\citet{Thieme2023}, [29]\citet{Sharma2023}, [30]\citet{Encalada2024}, [31]\citet{Sai2023}, [32]\citet{Ohashi2023}, [33]\citet{Lin2023}, [34]\citet{Narayanan2023}, [35]\citet{Yamato2023}, [36]\citet{Bjerkeli2016}, [37]\citet{Bjerkeli2019}, [38]\citet{Gavino2024}, [39]\citet{Federman2024}, [40]\citet{Hodapp2024}
[41]\citet{Santamaria-Miranda2024}
[42]\citet{Flores2023}
[43]\citet{Han2025}
[44]\citet{Narang2025}
[45]\citet{Phuong2025}
}
\tablenotetext{h}{eDisk first-look papers are in bold font}
\tablenotetext{i}{Radio emission at the position of IRS 5N was detected by \citet{Miettinen2008}, however it is possibly gyrosynchrotron emission due to the stellar magnetosphere rather than thermal emission from the ionized jet. IRS 5N is listed as a possible driving source for several molecular hydrogen emission-line objects (MHOs) by \citet{Kumar2011} but it is highly uncertain whether it is the driving source for these.}
\end{table*}

\section{Introduction} \label{sec:intro}
Protostellar outflows play an important role in the process of star formation \citep{Shu1993,Arce2002,Frank2014, Bally2016,Pascucci2023}. They enable the circumstellar material to accrete onto the star by extracting angular momentum \citep[e.g.][]{Shu2000,Bacciotti2002}. They also play a role in the evolution of molecular clouds and cores by injecting energy and momentum into the region surrounding the star \citep[e.g.][]{McKee2007,nakamura2007}. 
Protostellar outflows are made up of several components that can be identified using various types of emission. Jets, which are highly collimated and have high-velocity
($\sim$ a few 100 km s$^{-1}$), 
are traced in atomic and ionized emission lines at optical and near-infrared (NIR) wavelengths \citep[e.g.][]{Burrows1996, Reipurth2001}. For more embedded sources, they are also traced by molecular lines at millimeter (mm) and sub-millimeter (sub-mm) wavelengths \citep[e.g.][]{Bachiller1996,Tafalla2010} and at IR wavelengths \citep[e.g.][]{McCaughrean1994,Bally2016,Ray2023}. Emission from the ionized material at the base of the jet is seen at radio wavelengths \citep[e.g.][]{Anglada1996,Anglada1998, Scaife2012,Feeney-Johansson2023}. On the other hand, the less collimated component of the outflow is typically slow ($\lesssim$ a few 10 km s$^{-1}$) compared to the jet and is usually traced by molecular lines such as CO \citep{Snell1980, Bachiller1999,Reipurth2001, Bally2016}.

Several explanations have been proposed for the origin of the mm/sub-mm CO emission seen in protostellar outflows. The classic and most frequently discussed origin is a bipolar molecular flow of swept-up ambient gas. Such outflows have been seen on scales as large as several parsecs in young stellar objects (YSOs) \citep[e.g][]{Arce2007,Frank2014, Bally2016}. The properties of the swept-up gas depend on the driving flow.
In the wind-driven shell model \citep{Shu1991,Lee2000}, a wide-angle wind from the protostar blows into the ambient medium, creating a shock that sweeps the ambient material forming a thin wide-angle shell. \citet{Zhang2019}, for example, found multiple shells, which could be formed by a series of outbursts from an intermittent wide-angle wind in the HH 46/47 molecular outflow.
In the jet-driven bow shock model \citep{Lee2000}, a collimated high-velocity jet propagates into the ambient medium. When the jet interacts with the ambient medium, bow shocks can arise both as the tips of the terminal shocks or as internal working surfaces,
forming a dense shell of shocked molecular gas around the jet. 
Bow shocks can be produced by variation of velocities or other physical properties; e.g.
variations in the mass-loss rate in the jet can  produce a series of shocks along the jet axis.

In recent years, sub-mm observations with ALMA have also revealed flows known as slow molecular winds.
These are slow ($\lesssim$ a few 10 km s$^{-1}$) molecular flows observed at scales of $\leq$ 2000 au towards several Class 0 and I objects \citep[][and references therein]{Pascucci2023}. These flows typically show conical or parabolic shapes with wide opening angles ($10 - 40 \degr$) and an `onion-like' velocity structure, where the opening angle of the cone is smaller for higher velocities, implying that material further away from the outflow axis is moving at slower velocities. In addition, for some sources \citep[e.g. DG Tau B;][]{DeValon2020}, it has been shown that the flow rotates around the outflow axis in the same direction as the disk. These properties suggest that these flows are tracing material ejected directly from the surface of the Keplerian disk and appear to originate from the disk at radii $\sim 10 - 100$ au. 
\citet{Bjerkeli2016}, for example, observed TMC1A with a spatial resolution of 6 au to find that gas is ejected from a region extending up to a radial distance of 25 au from the central protostar, and that angular momentum is removed from an extended region of the disk. 


While ALMA has revealed many interesting details about the molecular outflows close to the protostar by high spatial resolution imaging as described above, 
so far most of them have been on individual objects. Systematic studies of high-resolution molecular outflow emission from a large sample of objects will allow us to measure the fraction of objects that show molecular outflow emission, as well as the frequency with which different types of outflow emission are seen \citep{Vazzano2021, Hsieh2023}. It will also enable us to measure the frequency of shell structure and detect evidence of variability in mass-loss rate of outflows.

The ``Early Planet Formation in Embedded Disks'' (eDisk) project \citep{Ohashi2023} is an ALMA Large Program with the goal of observing substructures, such as rings and gaps, in the disks around young (Class 0/I) protostars in nearby star-forming regions. This involves high-resolution ($\sim 7\ \mathrm{au}$) dust continuum observations of 19 nearby Class 0 and Class I sources. The eDisk spectral setting also includes a number of intriguing molecular lines, such as the $^{12}$CO ($J = 2 - 1$) line, a well-known tracer of molecular outflows. The target protostars are listed in Table \ref{tab:sum} in the order of increasing bolometric temperature. Of the 19 sources in the survey, 18 have previously been found to be associated with molecular outflows, with an exception of R CrA IRS 5N (IRS 5N hereinafter) (see also \S 5.2). 
These outflows should be detectable if they have structures on the scales that eDisk observations are sensitive to (see \S \ref{sec:obs}), while large-scale structures may be filtered out. Therefore, eDisk offers an excellent opportunity for a systematic survey of the molecular outflow emission from embedded sources at a high spatial resolution.

\begin{table*}
    \centering
    \caption{Properties of the $^{12}$CO images used}
    \begin{tabular}{lcccc}
    \toprule
    Source & Briggs weighting & $uv$ taper & Synthesized Beam & Noise level \\
     & & $\mathrm{k\lambda}$ & $(\theta_{\mathrm{maj}} \times \theta_{\mathrm{min}}, \mathrm{PA})$ & mJy beam$^{-1}$ \\
    \hline
    BHR71 IRS 2 & 2.0 & 2000 & $(0\farcs13 \times 0\farcs12, -1.31\degr)$ & 0.8 \\
    B335 & 0.5 & 1500 & $(0\farcs09 \times 0\farcs08, 55.27\degr)$ & 1.4 \\
    L1527IRS & 2.0 & 2000 & $(0\farcs17 \times 0\farcs13, -19.45\degr)$ & 0.8 \\
    IRAS 16253-2429 & 2.0 & 2000 & $(0\farcs35 \times 0\farcs25, 76.37\degr)$ & 1.2 \\
    IRAS 16544-1604 & 2.0 & 2000 & $(0\farcs24 \times 0\farcs18, 78.76\degr)$ & 1.0 \\
    GSS30 IRS3 & 2.0 & 2000 & $(0\farcs34 \times 0\farcs24, 74.96\degr)$ & 1.0 \\
    IRAS 15398-3359 & 2.0 & 2000 & $(0\farcs17 \times 0\farcs15, -79.76\degr)$ & 1.1 \\
    R Cr A IRS 5N & 2.0 & 2000 & $(0\farcs15 \times 0\farcs12, -85.10\degr)$ & 1.0 \\
    IRAS 04166+2706 & 2.0 & 2000 & $(0\farcs17 \times 0\farcs14, 7.33\degr)$ & 1.0 \\
    R CrA IRAS 32 & 2.0 & 2000 & $(0\farcs16 \times 0\farcs12, -85.08\degr)$ & 1.0 \\
    BHR71 IRS 1 & 2.0 & 2000 & $(0\farcs15 \times 0\farcs14, -11.07\degr)$ & 0.9 \\
    Ced 110 IRS 4 & 2.0 & 2000 & $(0\farcs18 \times 0\farcs13, -21.33\degr)$ & 0.9 \\
    R CrA IRS7B & 2.0 & -- & $(0\farcs12" \times 0\farcs10, 85.20\degr)$ & 0.8\\
    IRAS 04302+2247 & 2.0 & 2000 & $(0\farcs14 \times 0\farcs11, -18.19\degr)$ & 1.0 \\
    IRAS 04169+2702 & 2.0 & 2000 & $(0\farcs18 \times 0\farcs15, 6.46\degr)$ & 1.0 \\
    TMC-1A & 0.5 & 2000 & $(0\farcs10 \times 0\farcs06, 36.00\degr)$ & 1.7 \\
    Oph IRS43 & 2.0 & 2000 & $(0\farcs23 \times 0\farcs18, -63.95\degr)$ & 0.8 \\
    L1489 IRS & 2.0 & 2000 & $(0\farcs16 \times 0\farcs12, 12.21\degr)$ & 1.0 \\
    Oph IRS63 & 0.5 & -- & $(0\farcs40 \times 0\farcs28, 78.87\degr)$ & 1.5 \\
    \hline
    \end{tabular}
    \label{tab:image_properties}
\end{table*}

The paper is structured as follows. The observations and data reduction process for the eDisk data is described in Section \ref{sec:obs}. Specific features of CO outflow emission used to categorize the outflows are described in Section \ref{sec:models}. We categorize the observed CO emission features and present the observational data of representative objects in Section \ref{sec:results}. In Section \ref{sec:discussion}, we discuss time-variability, objects without clear outflows, and the relationship between outflow features and the evolutionary stage of the central protostars. Section 6 summarizes our main results.

\section{Observations and Data Reduction}
\label{sec:obs}
The ALMA Large Program eDisk (2019.1.00261.L; PI N. Ohashi) observed 17 nearby Class 0/I disks between 2021 April and 2022 July. Additional observations were carried out through the ALMA Director's Discretionary Time (DDT) program (2019.A.00034.S; PI: J. Tobin). Two more sources, B335 and TMC-1A, were added from the ALMA archive. These datasets are taken mainly from the programs targeting the outflow launching region \citep{Bjerkeli2016, Bjerkeli2019}, providing a similar baseline coverage and spectral setup to those of eDisk.
Then the total sample size is 19 sources, i.e. 12 Class 0 sources and 7 Class I sources with $L_{\rm bol}\sim 0.1-17 ~L_{\odot}$ and $T_{\rm bol}=39-348$ K (Table \ref{tab:sum}). A detailed description of the observation and data reduction process is given in the eDisk overview paper by \citet{Ohashi2023}. Here, we summarize the details relevant to this paper.

The eDisk observations were optimized to observe nearby Class 0/I disks, aiming to spatially resolve the disks around the sample protostars. All of the sources in the sample are nearby ($d < 200\ \mathrm{pc}$) and relatively bright ($L_{\mathrm{bol}} > 0.1 ~ L_{\odot}$). In order to achieve high angular resolution, long baseline observations were carried out in the extended antenna configuration C43-8, while short baseline observations were carried out in the more compact configuration C43-5, in order to observe more extended structure. This gives a total range of baselines between $\sim$15 m and $\sim$ 12.6 km, resulting in an angular resolution of $\sim 0\farcs04$ and a maximum recoverable scale of $2 \arcsec - 3 \arcsec$. The correlator was set up to observe the 1.3 mm (225 GHz) continuum as well as several molecular lines, including $^{12}$CO ($J = 2 -1$) at 230.54 GHz, which will be the focus of the present study as it is a tracer of molecular outflows. The velocity resolution of the $^{12}$CO line data is $0.63\ \mathrm{km\ s^{-1}}$.

The data were first calibrated with the standard ALMA calibration pipeline version 2021.2.0.128 using the Common Astronomy Software Application \citep[CASA;][]{CASA2022} version 6.2.1. Next, self-calibration was performed using a set of scripts designed for the eDisk project, as described in \citet{Ohashi2023}. First, a continuum image was made for each execution block separately, and positional offsets of the continuum emission were corrected. The azimuthally averaged visibilities were then compared between different execution blocks and a rescaling factor was then applied to each block to correct for possible flux calibration errors. This results in data with an expected flux density scale accuracy of $\sim 5 - 10 \%$. After the flux rescaling, self-calibration was performed on the continuum, first only on the short baseline data, and then on the short-baseline and long-baseline data combined. For both steps, several rounds of phase only self-calibration and amplitude + phase calibration were performed until the signal-to-noise ratio in the image saturated. The resulting self-calibration solutions were then applied to the continuum data and the line data, resulting in the final data sets. Specific aspects of the calibrations depend on the sources and are detailed in each of the first look papers (see the references in bold font in Table \ref{tab:sum}).

The $^{12}$CO line data was imaged using Briggs weightings of 0.5 and 2 with a $uv$-taper of 2000 k$\lambda$.The resulting spatial resolutions and rms noise levels are summarized in Table \ref{tab:image_properties}. For this paper, unless otherwise noted, the images with a Briggs weighting of 2 were used in order to maximise the S/N and detect fainter outflow emission. For B335 and TMC-1A, the images with weightings of 0.5 were used since the images with weightings of 2 were found to have large sidelobes, which made it difficult to accurately analyze the structure in the emission. For Oph IRS63, the long baseline observations were not carried out due to scheduling restraints. For the continuum images of this source, archival data were used to obtain the long baselines (Program ID: 2015.1.01512.S), while for the molecular lines, including $^{12}$CO, only the short baselines were available. As a result, the $^{12}$CO image with a Briggs weighting of 0.5 was used in order to optimize the angular resolution and S/N.

\begin{figure*}
    \centering
    \includegraphics[width=\linewidth]{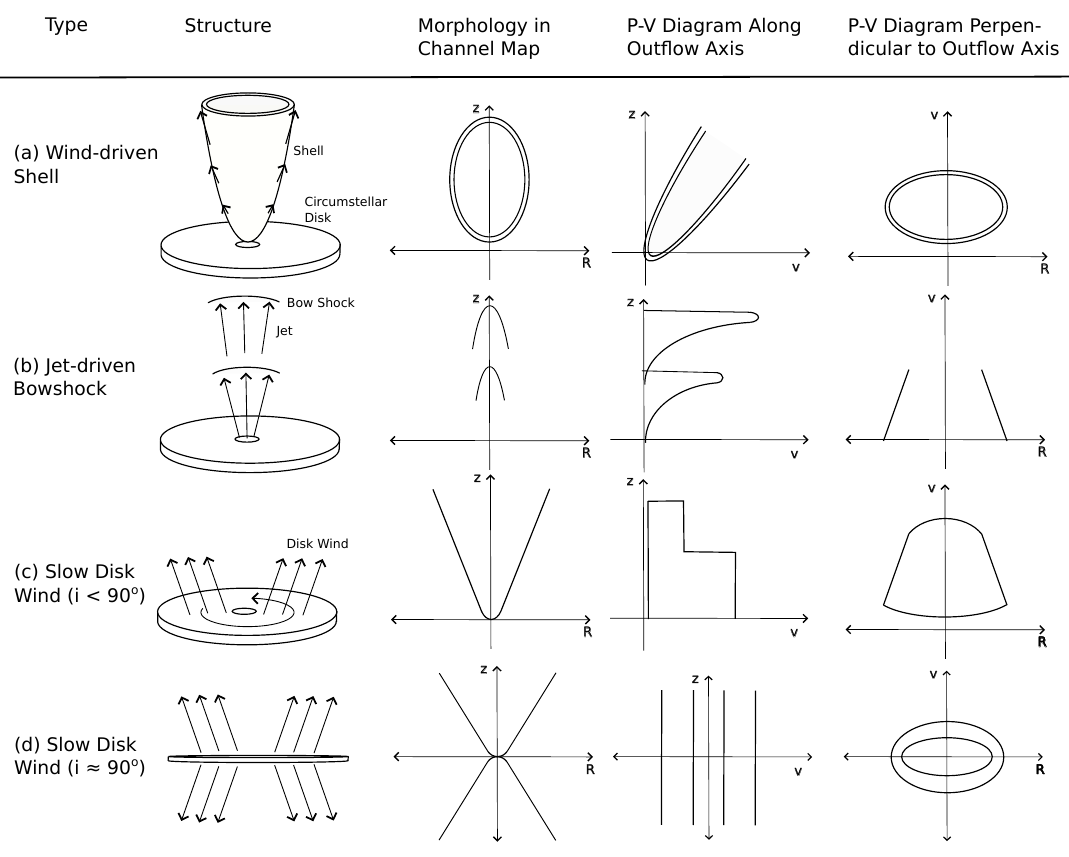}
    \caption{Categorization of molecular outflows based on the channel map and the P-V diagrams. References: (a) \citet{Shu1991,Li1996,Lee2000,Lee2001,Arce2002, Shang2006}
    (b) \citet{Lee2000,Lee2001,Cliffe1996,Arce2002}
    (c) \citet{Blandford1982,Tomisaka1998,Machida2008, DeValon2022}
    (d) \citet{Hirota2017,Tabone2017}}
    \label{tab:schematic}
\end{figure*}

\section{Characteristic features of outflows}
\label{sec:models}

\citet{Arce2002} schematically described molecular outflow properties predicted by different entrainment models. Inspired by their work, we analyzed channel maps and position-velocity (P-V) diagrams of eDisk targets to see if we can extract some similar features among them. We found that we can categorize our targets roughly into the three types which are described in the following subsections. We interpret them as (i) wind-driven shell, (ii) bow shock, (iii) slow disk wind (Figure \ref{tab:schematic}). It should, however, be noted that our categorization is basically qualitative and that we do not aim to validate or exclude specific theoretical models. 

\subsection{Wind-driven Shell}
\label{sec:wind_driven_model}
In the velocity channel maps, several of our targets show ellipses or shell-like structures that become larger and shift further away from the protostar at higher velocity channels. We interpret them as wind-driven shells.

When a wide-angle wind from the central protostar blows into ambient material, it sweeps up material into a momentum-conserving shell that is expanding with a Hubble law velocity structure \citep[e.g.][]{{Shu1991,Li1996,Lee2000,Lee2001,Arce2002, Shang2006}}.
A cut along the outflow axis results in a parabola on the P-V diagram, whose tilt and opening angle depend on the opening angle and inclination $i$ of the shell. Meanwhile, a cut perpendicular to the outflow axis results in an ellipse on the P-V diagram with the ellipse increasing in size further along the outflow axis.

\subsection{Bow shocks}
\label{sec:bow_shock_model}
When a shell-like structure appears in almost the same position in multiple velocity channels, it means that the shell contains gas of relatively large velocity range. We interpret them as bow shocks.

While bow shocks are ubiquitous in various astrophysical objects, bow shocks caused by the interaction between a fast collimated jet and ambient material are often investigated/modeled in the context of protostellar outflows \citep[e.g.][]{Cliffe1996,Lee2000, Lee2001, Arce2002}.
A shell structure produced by a jet-driven bow shock is often very elongated along the jet axis \citep[e.g.][]{Lee2001}, with the largest velocity pointed along the $z$-direction (Figure \ref{tab:schematic}). 
In velocity channel maps, the emission appears as an inverted V-shaped curve, with its tip at the position of the bow shock.
The P-V diagram along the jet axis shows a spur-like feature, with the velocities increasing rapidly up to the position of the bow shock, where a broad range of velocities is present. The steepness of the spur-like feature close to the bow shock depends on the inclination angle of the jet.

\subsection{Slow disk wind}
\label{sec:disk_wind_model}
Many of our targets show conical or parabolic flow in the velocity channel maps of $|v| \lesssim 20$ km s$^{-1}$ with respect to the systemic velocity ($v_{\rm sys}$).
They also show an onion-like velocity structure; i.e. the opening angle of the conical feature is narrower at higher velocity channels. These features could be interpreted as centrifugal winds whose velocity peaks decrease with distance from the central star.

Theoretical models and numerical simulations predict centrifugal winds launching from a rotating disk \citep[e.g.][]{Blandford1982,Shu1994,Tomisaka1998,Machida2008, DeValon2022}, with varying launch radii ranging from the inner disk to outer regions.  The overall shape of the wind is conical or parabolic. Since the disk rotation is faster at the inner radii, the wind velocity is higher in the flow originating at the inner disk radii. The actual theoretical models and simulations vary in their respective predictions and features.
To draw the schematic morphology of channel maps and P-V diagrams in Figure \ref{tab:schematic}, we referred to 
the simple disk wind emission model of \citet{DeValon2022}. They calculated synthetic images by assuming an axisymmetric flow having reached its terminal velocity at the observed spatial scale and a constant poloidal velocity along its trajectory. The code used to create these synthesized images is available online 
\footnote{https://github.com/Alois-deValon/Axoproj}.

The emission predicted based on the slow disk wind models highly depends on the inclination. When the disk inclination is $i \approx 90\degr$, i.e. the axis of the disk wind is on the plane of the sky, only low-velocity emission is seen, as we only observe the radial and rotation component of the outflow velocity. In P-V diagrams perpendicular to the outflow axis, the emission traces an ellipse symmetric around the origin ($v=0, \mathrm{offset}=0$), if the radial velocity dominates over rotation. If rotation is significant, the ellipse is tilted in the P-V diagram \citep[e.g.][]{Hirota2017,Tabone2017}.

For outflows inclined to the plane of the sky ($i < 90\degr$), the velocity maps show a conical or parabolic shape with much higher velocities than the $i=90\degr$ case, as we are also seeing the velocity component along the outflow axis.  In P-V diagrams perpendicular to the outflow axis, the emission appears as an upside-down cone that is roughly symmetrical around offset$=0$. This indicates that the velocity of the emission decreases with distance from the outflow axis, reflecting the onion-like velocity structure with the higher velocity gas located closer to the outflow axis. 

Note the emission patterns on the channel maps and the P-V diagrams perpendicular to the outflow axis could be similar for the wind-driven shell and slow disk wind cases at an inclination of 90$\degr$, especially for the base cavity. It is thus important to refer the P-V diagrams in both directions, along and perpendicular to the outflow axis, as well as the channel map when categorizing the outflow.

Having said this, it is important to note that we cannot strictly constrain the launching mechanism of the slow disk wind from our observations. For example, the characteristic onion-like velocity structure could be due to gas dragged by higher velocity flows, similar to the case of wind-driven shells. Depending on the gas density structure, wind-driven gas could be so extended that it does not show clear “shell” structures, which we use for categorizing wind-driven shells. So, while we use the name ``slow disk wind'' for this type of outflow, we do not exclude the possibility that it could consist of dragged material.

\begin{figure*}
    \centering
    \includegraphics[width=\linewidth]{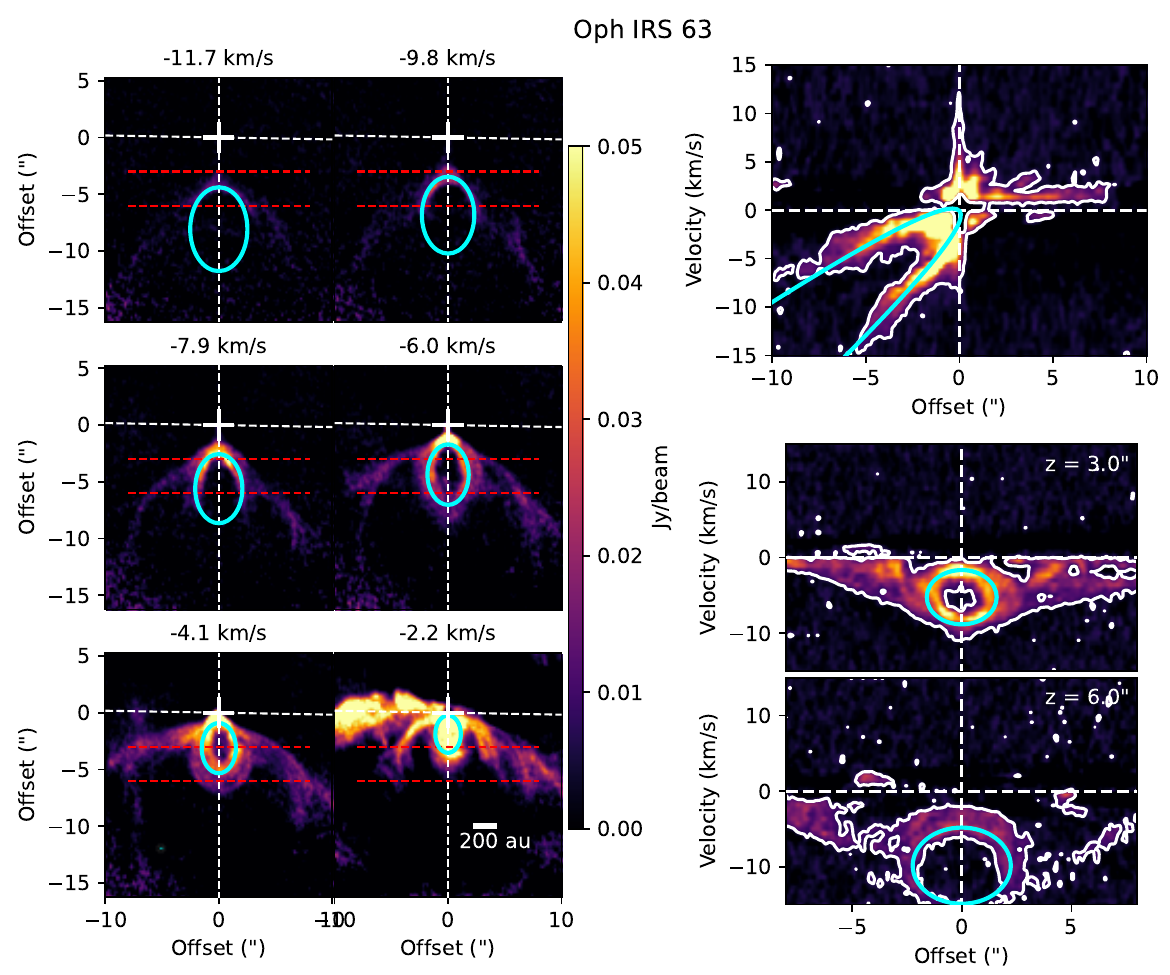}
    \caption{Maps of representative velocity channels (left), position-velocity diagram along the axis of the outflow (top right), and the P-V diagrams perpendicular to the outflow axis at 3$\arcsec$  (middle right) and 6$\arcsec$ from the protostar (bottom right) for Oph IRS 63. This is the best example of an outflow showing wind-driven shell emission in the eDisk sample. The white contours show the 5$\sigma$ level of the emission, where $\sigma$ is the root-mean-square noise level given in Table \ref{tab:image_properties}. 
    The channel map is rotated so that the outflow axis is in the vertical direction in the panel. The velocity is also shifted to the relative velocity to the systemic velocity. The position angle of the outflow axis and systemic velocity are listed in Table \ref{tab:PA_and_Vsys}. The white dashed lines in the channel map depict the outflow axis and PA$_{\rm cont}$, while the red dashed lines depict the offset positions for the P-V diagrams perpendicular to the outflow axis. Cyan lines depict the  model for shell B1 described in \S 5.1. The synthesized beam of the observation is shown by the cyan filled ellipse in the bottom left velocity channel map. A scale bar indicating a distance of 200 au is shown in the bottom right velocity map.}
    \label{fig:OphIRS63_combined_figure}
\end{figure*}

\section{Results}
\label{sec:results}
The molecular outflow emission was detected in 15 of the eDisk sources, and are divided into the
above mentioned 3 types based on the morphology in their channel maps and P-V diagrams.
The classifications are summarized in Table \ref{tab:sum}.  Columns 5 and 6 denote the emission type
for the blue- and red outflow lobes, respectively, where
``SDW'' stands for slow disk wind, ``BS'' for bow shock, and ``WS'' for wind-driven shell.
We found that the specific features of wind-driven shell and slow disk wind coexist in 4 sources, which are denoted as ``SDW/WS''. We also note that some sources show emission features quite different between blue and red components regarding dynamical time (see \S 5.1), opening angle, or coexistence of wind-driven shell and slow disk wind (e.g. TMC-1A). 

The channel maps of all the targets are shown in the Appendix A (Figures \ref{fig:appendix_BHR71IRS2} - \ref{fig:appendix_OphIRS63}). 
They are oriented so that the outflow axis is vertical in the panel, with the red-shifted outflow pointing up and blue-shifted outflow pointing down, and the velocity is shifted so that the velocity is relative to the systemic velocity.
Unless otherwise stated (see below), the axis of the outflow was determined by assuming that it is perpendicular to the major axis of the continuum emission, which are in turn determined by the Gaussian fits to the dust continuum as discussed in \citet{Ohashi2023}. 
We denote this position angle of the major axis of the dust continuum as PA$_{\rm cont}$ rather than PA$_{\rm disk}$, since Keplerian rotation is not confirmed  for some objects; i.e.  the continuum emission may be contaminated by envelope emission.
The systemic velocity, the position angle of the outflow axis, and PA$_{\rm cont}$ for each source are summarized in Table \ref{tab:PA_and_Vsys}.
References for the systematic velocity and the Gaussian fit of the continuum for each source are available in the first-look papers of the eDisk project, which are listed in Table \ref{tab:sum} with bold font.
PA$_{\rm cont}$ is indicated by
the dashed line which intersects the outflow axis (vertical dashed line) at the stellar position in Figures \ref{fig:appendix_BHR71IRS2} - \ref{fig:appendix_OphIRS63}.
For some sources we note that the outflow axis is obviously not perpendicular to PA$_{\rm cont}$, and/or the the blue-shifted and red-shifted components have different orientations. We set their outflow axis by finding the axis of symmetry of the cone- or parabola-shaped lobe by eye. The outflow axis for sources with wind-driven shell emission is determined by fitting the shells with a model (see \S \ref{sec:discussion} for details). While the error of the outflow axis determination is not easy to quantify, the symmetry of the lobe or the shell fitting clearly gets worse if we modify the axis by 5 degree. Considering this error, 5 sources show misalignment between the red- and blue-shifted outflows, and 7 sources show significant misalignment between the outflow axis (either or both of the red- and blue-shifted outflows) and the minor axis of
the dust continuum emission around the protostar (Table \ref{tab:PA_and_Vsys}).

In four objects, IRS 5N, Ced 110 IRS4, R CrA IRS 7B (IRS 7B hereinafter), and Oph IRS43 (hereinafter IRS43), we do not see any clear sign of outflow emission (\S \ref{sec:discussion}). For one source, IRAS04169+2702 (IRAS04169 hereinafter), we see peculiar features in the red-shifted lobe that cannot be categorized as either of the three types (see \S 4.5). 

We also note that the CO emission can be seen in ``molecular jets'', where they trace collimated, high-velocity material in the jet, particularly in heavily embedded sources \citep{Bally2016}. Among eDisk objects, molecular jets are observed in BHR71 IRS1, BHR71 IRS2, and IRAS04166+2760 (IRAS04166 hereinafter). BHR 71 IRS1 and IRS2, for example, show $^{12}$CO emission of high-velocity ($\gtrsim 40$ km s$^{-1}$) gas that is highly collimated compared to the lower velocity outflow \citep{Gavino2024}.
In the present work, however, we focus on molecular outflows with velocities of $\lesssim 30$ km/s, and do not discuss those molecular jets. For more details, see the eDisk first-look papers for these objects  \citep{Gavino2024,Phuong2025}

In this section, we present the channel maps and P-V diagrams of representative objects for emission from a wind-driven shell, a bow shock, a slow disk wind, and a combination of slow disk wind and wind-driven shell.

\begin{table*}
    \begin{center}
    \caption{Systemic velocity and PA of the outflow axis}
    \begin{tabular}{c c c c c c}
    \toprule
    Source & $v_{\rm sys}$ & PA\tablenotemark{a}  & PA method\tablenotemark{b} & PA$_{\mathrm{cont}}$\tablenotemark{c} & Deviation \tablenotemark{e}\\
           &   [km/s]   &   [\degr] &  & [\degr] & [\degr]\\
     \hline
    BHR71 IRS2 &  -4.45  & 152 & Sym & $67.6\pm3.7$ & $5.6\pm 6.2$\\
    B335 & 8.5   & 275 &  Sym & 163 & $22\pm 5$\\
    L1527 IRS  &  5.9  &  92 & Perp & 2 & - \\
    IRAS16253-2429 &   4.0 & $204.3$ & Perp & $114.3 \pm 0.6$ & - \\
    IRAS 16544-1604\tablenotemark{d} &  5  & $128 $ (B), $335/340$ (R) & Shell & $45$ & $7\pm 5$(B), $20\pm5$/$25\pm5$(R) \\
    GSS30 IRS3 &  4  &  $198$ & Shell & $109.36\pm0.30$ & $1.36\pm 5$\\
    IRAS15398-3359 &  5.24  & $233$ (B), $65$ (R) & Sym & $117.1 \pm 2.7$ & $25.9\pm 5.7$(B), $37.9\pm 5.7$(R)\\
    R CrA IRS 5N   &  6.5  &  $351.22 $ & Perp & $81.22 \pm 0.80$ & - \\
    IRAS04166+2706 &  7.0  &  $212$ & Shell & $121.5\pm0.5$ & $0.5\pm 5$\\
    R CrA IRAS 32  &  5.86  & $223$ & Shell & $135.3\pm0.4$ & $2.3\pm 5$\\
    BHR71 IRS1 &  -4.55  & 359 & Sym & $98.15\pm0.41$ & $9.15\pm 5$\\
    Ced 110 IRS 4  &  -4.67  &  $14$ & Perp & $104 \pm 1$ & -\\
    R CrA IRS7B    &  6.0  & 25 & Perp & 115 & -\\
    IRAS04302+2247 &  5.7 &  $264.7$ & Perp & $174.7 \pm 0.03$ & -\\
    IRAS04169+2702 &  6.8  & 230 (B), 70 (R) & Sym & 139 & $1\pm5$(B),$21\pm5$(R)\\
    TMC-1A         &  6.0  & $166$ & Shell & 76 & $0\pm5$\\
    Oph IRS43      &  4.0  & $43.5$ & Perp & $133.5\pm1.1$ & - \\
    L1489 IRS &  7.3  & 175 (B), 341 (R) & Sym & $67 \pm 0.3$ & $18\pm5$(B),$4\pm5$(R) \\
    Oph IRS63 &  2.5  &  $240$ (B) $90$ (R) & Shell & 149 & $1\pm5$(B), $31\pm5$(R) \\
    \hline
    \end{tabular}
    \label{tab:PA_and_Vsys}
    \end{center}
    \tablenotetext{a}{PA$=0$ is pointing north and PA increases counter-clockwise (i.e. east of north). Unless otherwise noted, the value given is the red-shifted PA with the blue-shifted PA being 180$\degr$ opposite. The error of PA is estimated to be $\lesssim 5\degr$ (see text).}
    \tablenotetext{b}{The method to determine the PA of the outflow axis. For ``Sym'', PA is determined by estimating the axis of symmetry of the outflow lobes by eye. Two PA values are listed when the axis is different between blue and red lobe. For ``Perp'', PA is taken to be perpendicular to PA$_{\rm cont}$. For objects without clear outflow emission (i.e. R CrA IRS5N, Ced 110 IRS 4, R CrA IRS7B, and Oph IRS43), we adopted the PA based on this method to draw the channel maps in Figures \ref{fig:appendix_IRS5N}, \ref{fig:appendix_Ced110IRS4}, \ref{fig:appendix_IRS7B}, and \ref{fig:appendix_OphIRS43} respectively in Appendix A. For ``Shell'', PA is determined by searching for the best parameters of the shell model for the velocity channel maps and P-V diagrams (see \S \ref{sec:discussion}). Two PA values are listed when the axis is different between blue and red lobe. }
    \tablenotetext{c}{The PA of the major axis of dust continuum fitted by a 2D Gaussian \citep{Ohashi2023}.  For some sources the error of PA$_{\rm cont}$ is not listed in the first-look papers. We estimate that the error is less than a few degree for those sources referring to the errors in other sources.}
    \tablenotetext{d}{Two PAs are listed for the red-shifted axis as a different PA value was derived for each of the two red-shifted shells (see \S \ref{sec:PA_Inclination}). The PA of the red-shifted shell closest to the source of the outflow (Shell R1) was used when creating Figure \ref{fig:IRAS16544_vel_map}.}
    \tablenotetext{e}{Deviation from orthogonality of PA and PA$_{\rm cont}$. The error is calculated by assuming  the error of 5\degr for PA.}
\end{table*}

\begin{figure*}
    \centering
    \includegraphics[width=\linewidth]{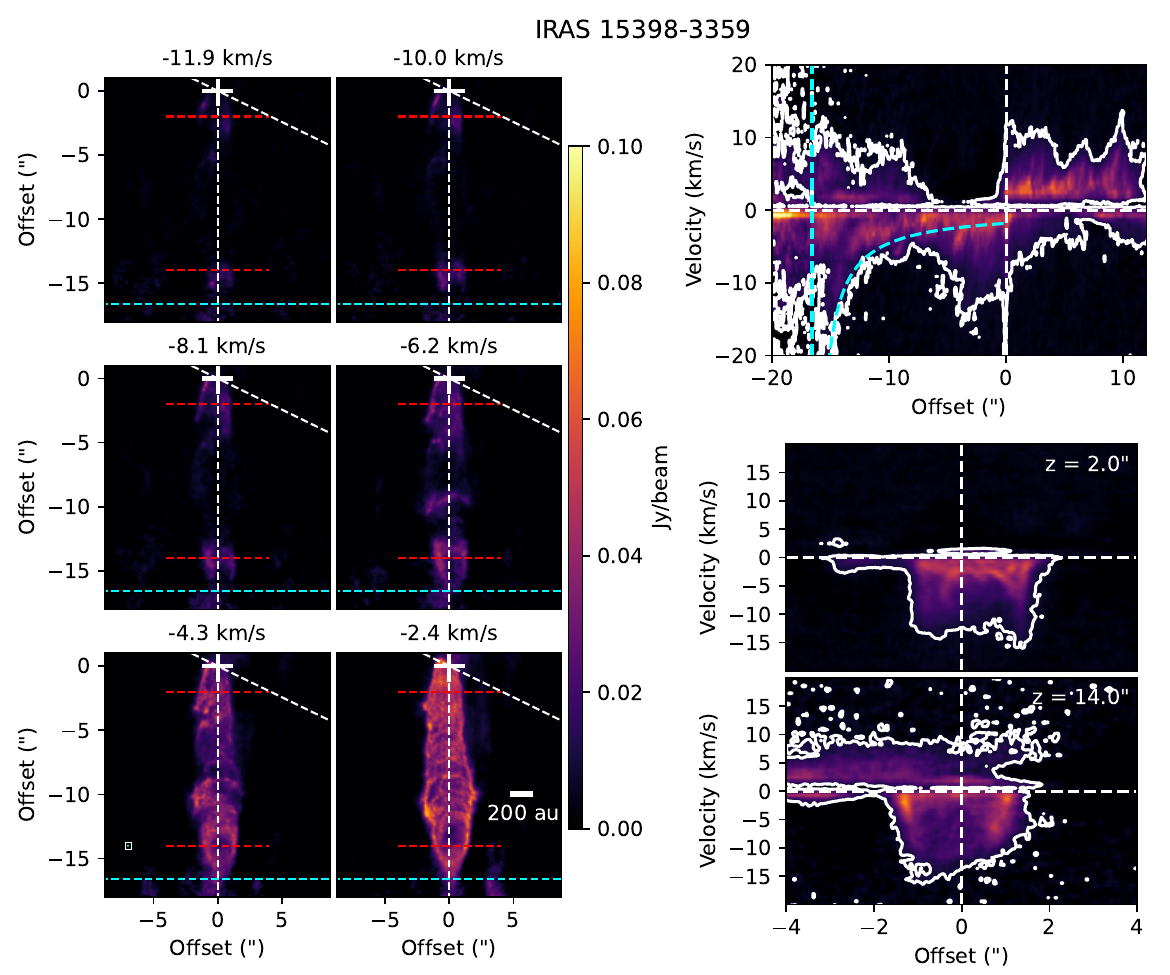}
    \caption{Similar to Figure \ref{fig:OphIRS63_combined_figure} but for IRAS15398, the only source in our sample to clearly show jet-driven bow shock emission. 
    The bottom-right panel shows the P-V diagrams perpendicular to the outflow axis at 2$\arcsec$  (middle right) and 14$\arcsec$ (bottom right) from the protostar.
    The position of the bow shock is indicated by the dotted cyan line at offset=16.6$\arcsec$, while the cyan dotted curve in the P-V diagram along the outflow axis (upper-right panel)
    shows a schematic emission shape as expected from the bow-shock model. 
    }
    \label{fig:IRAS15398_combined_figure}
\end{figure*}

\begin{figure*}
    \centering
    \includegraphics[height=7cm]{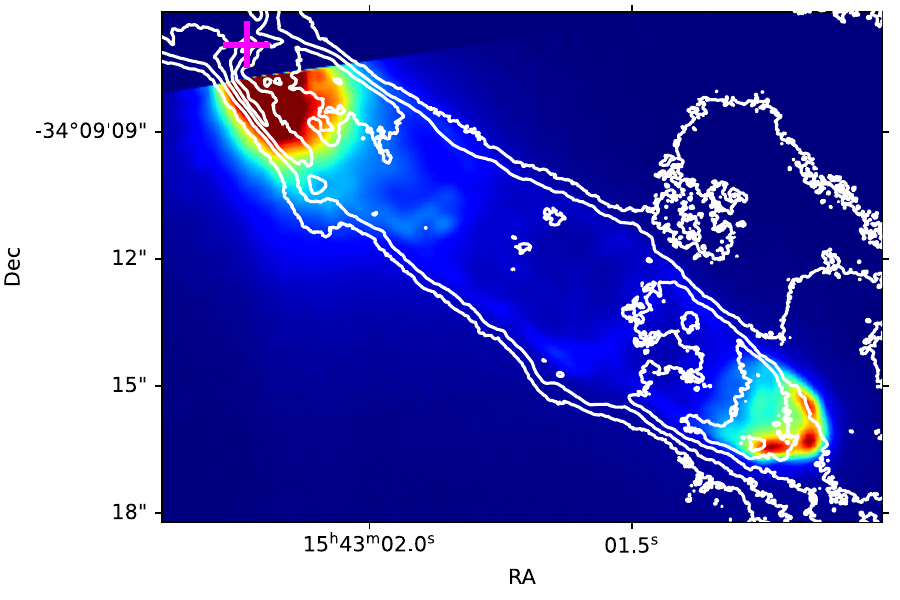}
    \caption{Integrated intensity map of $^{12}$CO of IRAS15398 (contours) overlaid on a JWST MIRI image taken with the F1000W filter (colorscale). The contours shown are $3\sigma_{\mathrm{rms}}$, $10\sigma_{\mathrm{rms}}$, $20\sigma_{\mathrm{rms}}$, $30\sigma_{\mathrm{rms}}$, and $40\sigma_{\mathrm{rms}}$, where $\sigma_{\mathrm{rms}} = 150\ \mathrm{mJy\ beam^{-1}\ km\ s^{-1}}$ is the root-mean-square noise of the image. The position of IRAS15398 is indicated by the magenta cross. The south lobe is the same lobe shown in the channel map in Figure \ref{fig:IRAS15398_combined_figure}.
    }
    \label{fig:IRAS15398_JWST}
\end{figure*}

\begin{figure*}
    \centering
    \includegraphics[width=\linewidth]{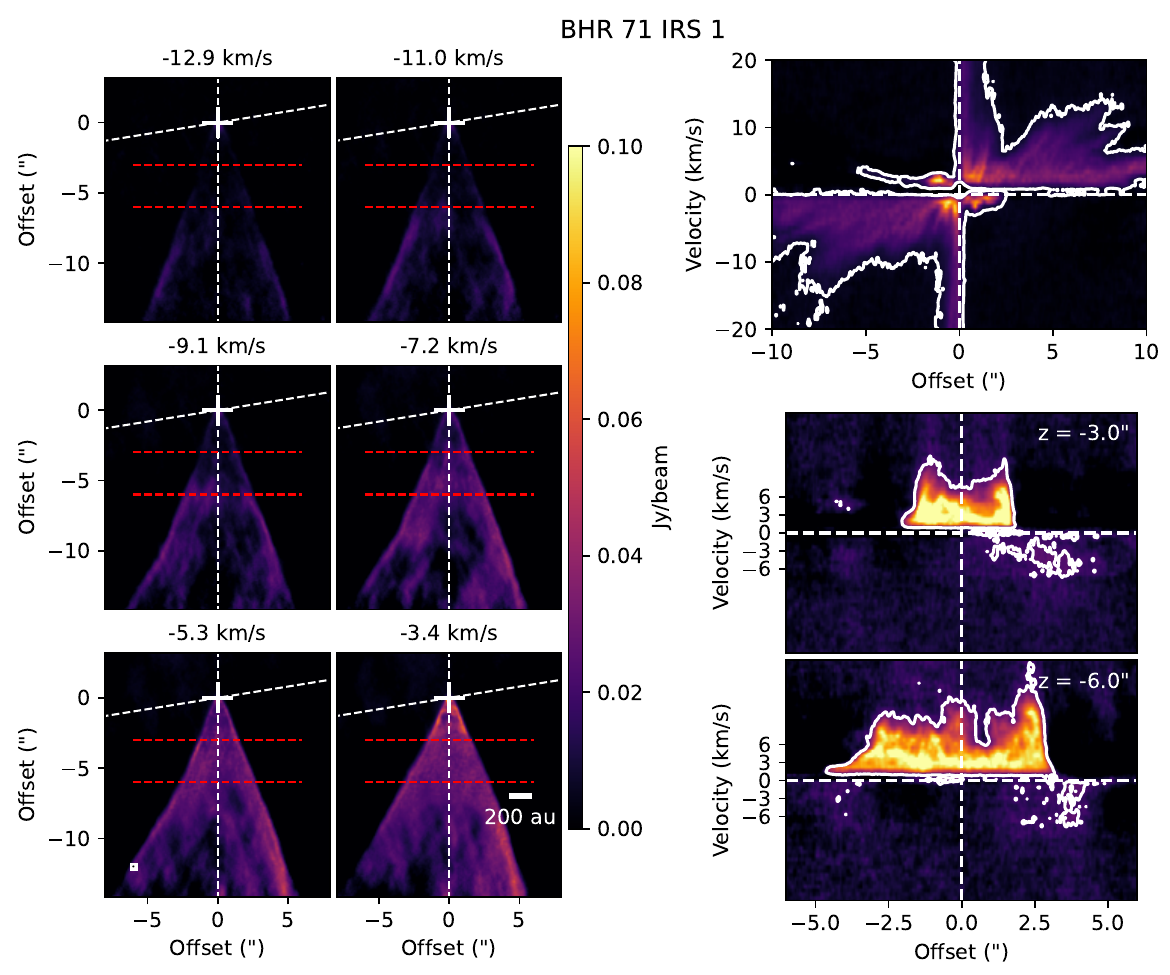}
    \caption{Similar to Figure \ref{fig:OphIRS63_combined_figure} but for BHR71 IRS1. This is an example of an outflow showing slow disk wind emission, where the outflow axis is inclined to the plane of the sky.
     The bottom-right panel shows the P-V diagrams perpendicular to the outflow axis at 3$\arcsec$  (middle right) and 6$\arcsec$ (bottom right) from the protostar.
    }
    \label{fig:BHR71_IRS1_combined_figure}
\end{figure*}

\begin{figure*}
    \centering
    \includegraphics[width=\linewidth]{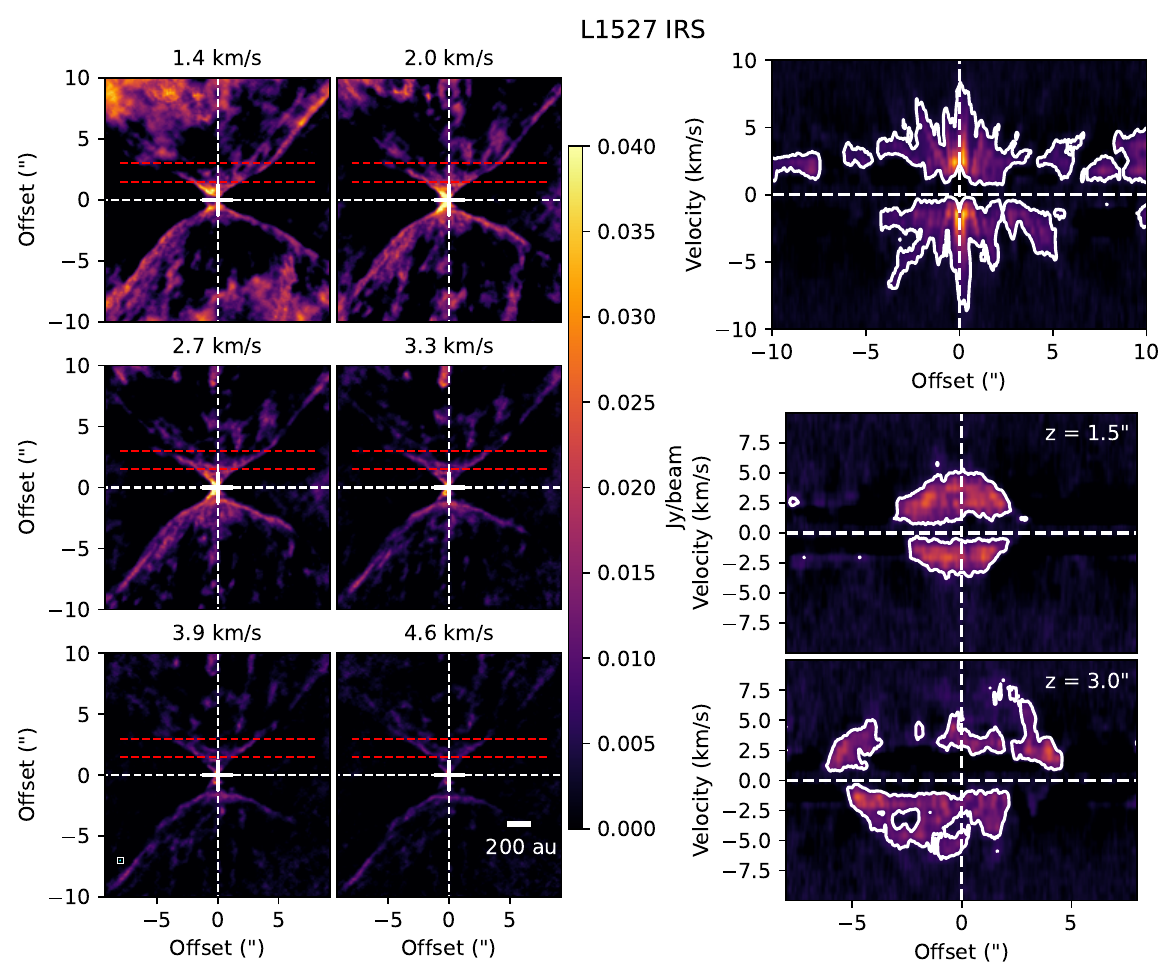}
    \caption{Similar to Figure \ref{fig:OphIRS63_combined_figure} but for L1527IRS. This is an example of an outflow showing slow disk wind emission, where the disk is edge-on, and the outflow is therefore in the plane of the sky.
     The bottom-right panel shows the P-V diagrams perpendicular to the outflow axis at 1.5$\arcsec$  (middle right) and 3$\arcsec$ (bottom right) from the protostar.
    }
    \label{fig:L1527IRS_combined_figure}
\end{figure*}

\begin{figure*}
    \centering
    \includegraphics[width=\linewidth]{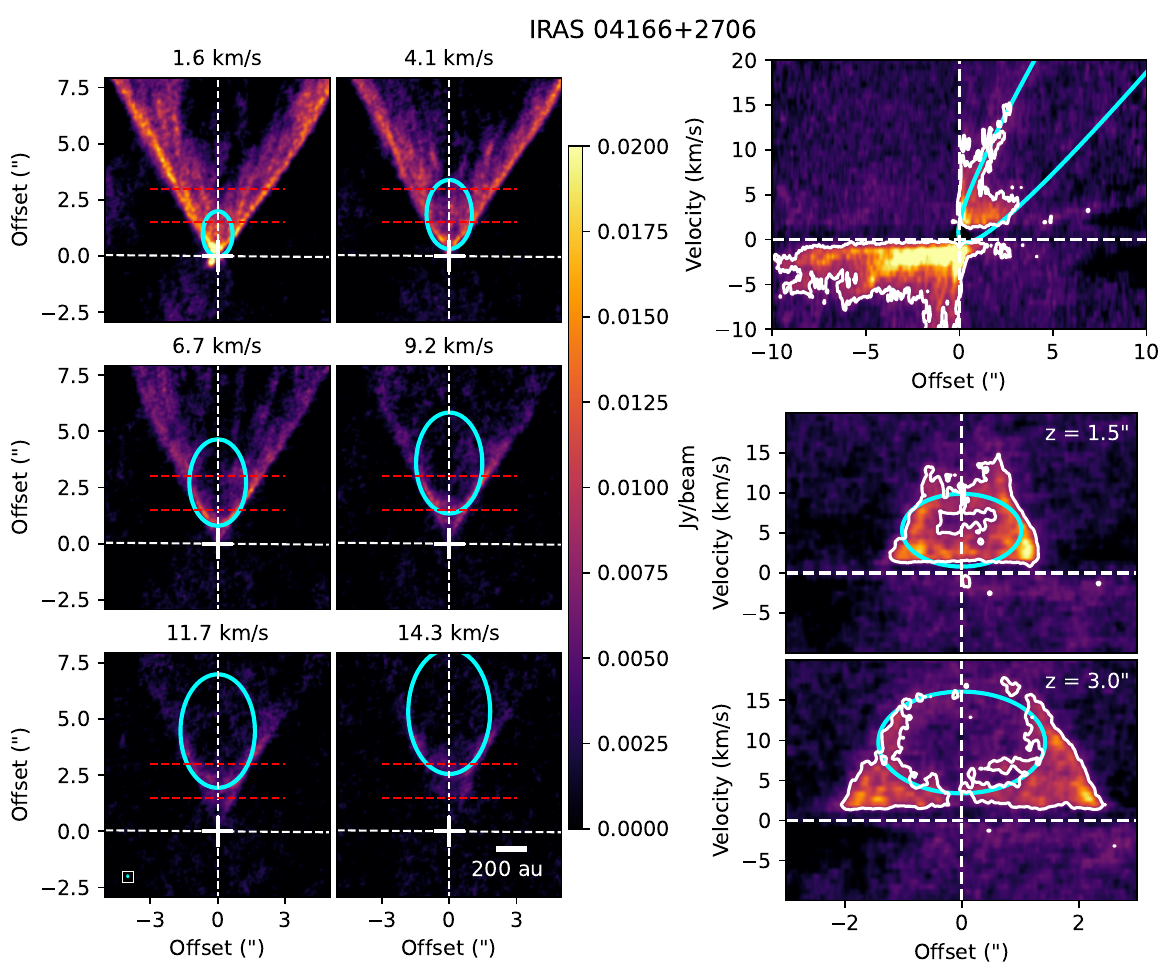}
    \caption{Similar to Figure \ref{fig:OphIRS63_combined_figure} but for IRAS04166. This is an example of an outflow showing a combination of slow disk wind and wind-driven shell emission.  The bottom-right panel shows the P-V diagrams perpendicular to the outflow axis at 1.5$\arcsec$  (middle right) and 3$\arcsec$ (bottom right) from the protostar.
    }
    \label{fig:IRAS04166_combined_figure}
\end{figure*}

\begin{figure*}
    \centering
    \includegraphics[width=\linewidth]{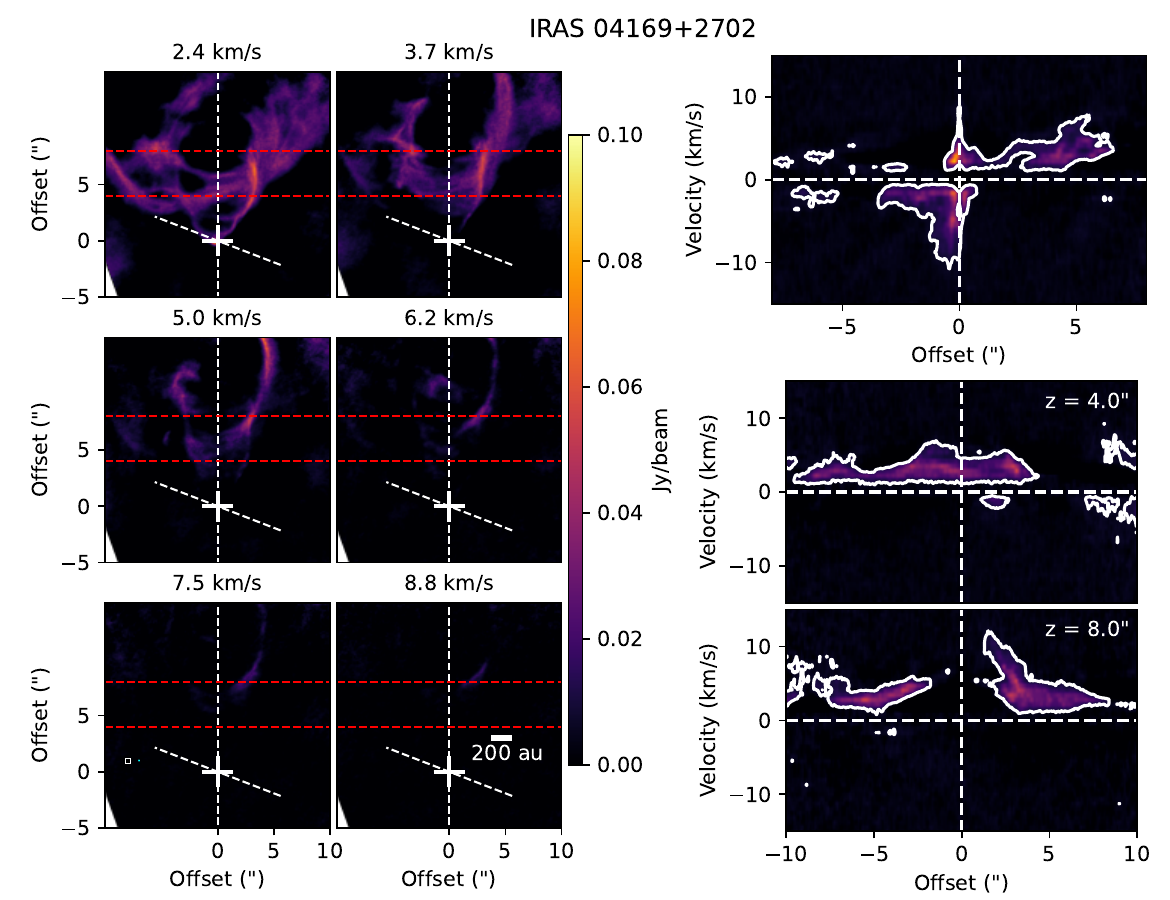}
    \caption{Similar to Fig. \ref{fig:OphIRS63_combined_figure} but for IRAS04169. It shows a complex outflow structure which is difficult to classify as one of the three outflow types considered.  The bottom-right panel shows the P-V diagrams perpendicular to the outflow axis at 4$\arcsec$  (middle right) and 8$\arcsec$ (bottom right) from the protostar.
}
    \label{fig:IRAS04169_combined_figure}
\end{figure*}

\subsection{Wind-driven shell emission}
The source that shows shell structure most clearly is Oph IRS 63. The left panel of Figure \ref{fig:OphIRS63_combined_figure} shows the maps of Oph IRS 63 at representative blue-shifted velocities. At least one shell-like structure can clearly be seen as an ellipse which shifts further away from the source and becomes larger in higher velocity channels, with emission visible at velocities up to $\approx 12\ \mathrm{km\ s^{-1}}$ relative to the systemic velocity (see also Figure \ref{fig:appendix_OphIRS63} in Appendix A). The bottom of the ellipse corresponds to the ``front side'' of the outflow, while the top of the ellipse corresponds to the ``back side'' of the outflow at a given velocity channel map. The P-V diagram along the outflow axis clearly shows a parabolic structure in the blue-shifted emission (the top right panel of Figure \ref{fig:OphIRS63_combined_figure}). The P-V diagrams perpendicular to the outflow axis (bottom right panels) show an elliptical feature, which becomes larger and shifts to higher velocity at larger offsets from the source. All of these properties are exactly what is expected of wind-driven shell emission described in \S \ref{sec:wind_driven_model} and Table \ref{tab:schematic}. 


A similar shell-like structure is also seen in the red-shifted outflow of Oph IRS 63. We note that the red-shifted outflow of Oph IRS 63 is much fainter and has a larger opening angle than the blue-shifted outflow \citep[Figure \ref{fig:appendix_OphIRS63} in Appendix A, see also][]{Flores2023}. Some other eDisk sources also show significant differences in morphologies between red and blue outflow components, which will be discussed in \S 4.5 and \S 5.1.

Other sources where the molecular outflow emission shows characteristics of the wind-driven shell include IRAS 16544-1604, GSS30 IRS 3, R CrA IRAS 32 (IRAS 32 hereinafter), IRAS04166, and TMC-1A. 
For IRAS04302, some $^{12}$CO emission is seen in the blue-shifted outflow on the outflow axis, but it is quite faint and has no distinct shape (Figure \ref{fig:appendix_IRAS04302} in Appendix A). Therefore, it is difficult to identify the type of emission. But the velocity channel map shows an emission which is shifted away from the protostar in the channel maps at higher velocities \citep[see also][]{Lin2023}. We thus possibly classify the outflow of IRAS04302 as a wind-driven shell.

\subsection{Jet-driven bow shock emission}
The molecular outflow emission observed in IRAS15398-3359 (IRAS15398 hereinafter) is more collimated than those of other objects (Figure \ref{fig:IRAS15398_combined_figure}). 
At the tip of the outflow located $\sim 15\arcsec$ from the source, a high-velocity emission
component up to $\sim 12\ \mathrm{km\ s^{-1}}$ from the systemic velocity is seen.
This relatively high-velocity emission at the tip of outflow observed in $^{12}$CO matches the position of a bow shock observed by JWST MIRI \citep[][]{Yang2022b, Okoda2025}(Figure \ref{fig:IRAS15398_JWST}). These features are consistent with the jet-driven bow shock, i.e. a U-shaped collimated outflow with its tip at the position of a bow shock in the jet. In addition, the P-V diagram along the outflow axis (the top right panel in Figure \ref{fig:IRAS15398_combined_figure}) shows a spur-like structure, where the velocity of the emission increases up to the position of the bow shock. This is also consistent with a jet-driven bow shock.

Jet-driven bow shock emission is also seen in the red-shifted outflow of IRAS15398, which also shows a collimated arc
at velocities of up to $\sim 10\ \mathrm{km\ s^{-1}}$ relative to the systemic velocity, at the tip of the outflow emission $5 - 10\arcsec$ from the source (Figure \ref{fig:appendix_IRAS15398} in Appendix A). This feature is particularly noticeable at the velocity channels of $5 - 7$ km s$^{-1}$. A spur-like structure can also be seen in the P-V diagram peaking at the tip of the emission. All these factors suggest that a bow shock is present in the jet at $\sim10\arcsec$.

With the possible exception of IRAS04169 (see \S 4.5), none of the other eDisk sources show evidence of jet-driven bow shock emission, suggesting perhaps that it is a relatively rare form of molecular outflow emission at radial distances of $\lesssim$ a few thousand au from the central protostar. 


\subsection{Slow disk wind emission}
Slow disk wind emission is the most common form of molecular outflow emission seen in the eDisk objects, with detections in at least 11 of the 19 objects. A typical example of slow disk wind emission is seen in the blue-shifted outflow of BHR71 IRS 1 (Fig. \ref{fig:BHR71_IRS1_combined_figure}). The velocity channel maps show a conical structure, with the opening angle of the cone decreasing at higher velocities (see also Figure \ref{fig:appendix_BHR71IRS1} in Appendix A). The P-V diagrams perpendicular to the outflow axis (the right bottom panels in Fig. \ref{fig:BHR71_IRS1_combined_figure}) show that the velocity of the emission decreases with distance from the outflow axis. These properties could reflect the onion-like velocity structure of the slow disk wind, where material closer to the outflow axis is moving at higher velocities than material further away from the axis \citep[e.g.][]{DeValon2020,Pascucci2023} (\S 3.3). 

As mentioned in \S \ref{sec:disk_wind_model}, the appearance of a slow disk wind depends strongly on the inclination of the source. While the disk of BHR71 IRS 1 is inclined relative to the plane of the sky ($i = 39\degr$), that of L1527 IRS is nearly edge-on \citep{VantHoff2023}. As a result, its appearance is somewhat different, as shown in Fig. \ref{fig:L1527IRS_combined_figure}. Similar to BHR71 IRS 1, the velocity channel maps show a conical structure with the opening angle decreasing at larger velocities (see also Figure \ref{fig:appendix_L1527IRS} in Appendix A). However, the velocity of the emission is lower than the inclined case, with the maximum velocity observed only $\sim 5\ \mathrm{km\ s^{-1}}$ relative to the systemic velocity. In the P-V diagrams perpendicular to the outflow axis (the right bottom panels in  Fig. \ref{fig:L1527IRS_combined_figure}), the emission traces an elliptical structure which is symmetric about the origin, with the ellipse being larger in P-V cuts at distances further from the source. We note that at this peculiar inclination, i.e. nearly edge-on angle, the P-V diagrams of L1527 IRS perpendicular to the outflow axis appear similar to those of wind-driven shells. We can discriminate the slow disk wind and wind-driven shell by taking into account both the channel maps and the P-V diagrams. The channel maps for a slow disk wind will show a conical structure, while for a wind-driven shell, they will show elliptical structures moving away from the source.

\subsection{Combination of slow disk wind and wind-driven shell emissions}
When categorizing the observed features, it is important to remember that multiple types of outflow emission could also be present in the same object. For example, wind-driven shells could be present alongside slow disk wind. This is seen in GSS30 IRS3, IRAS04166, IRAS 32, and TMC-1A.

The velocity channel maps and the P-V diagrams parallel to and perpendicular to the outflow axis for the red-shifted outflow of IRAS04166 are shown in Fig. \ref{fig:IRAS04166_combined_figure} as an example. The channel maps at low velocities show that the outflow has a conical shape that is typical of a slow disk wind. However, at the velocity of $|v-v_{\rm sys}|\sim 9$ km/s, the tip of the cone, which can be regarded as a shell-like structure, is shifted relative to the protostellar position. The shell-like structure is shifted farther away from the protostar at higher velocities
\footnote{The cyan lines in Figure \ref{fig:IRAS04166_combined_figure} show the shell model described in \S 5.1. We chose the parameters of the shell model referring to the high-velocity channels ($|v-v_{\rm sys}|\gtrsim 9$ km/s) and the P-V diagrams, while the cyan line is also plotted in the low-velocity channels.}. 
This is typical of a wind-driven shell. Evidence of disk wind emission and shell emission is also seen in the P-V diagrams. The P-V diagram parallel to the outflow axis (right top panel) shows a parabolic shape in the red-shifted outflow emission, which is evidence of shell emission, while the P-V diagrams perpendicular to the outflow axis (right bottom panels) show the velocity of the emission decreasing with distance from the outflow axis, characteristic of a disk wind, as well as a possible elliptical rim with an inner void in the emission, evidence of a wind-driven shell. 

Given that wind-driven shells are typically the result of a wide-angle wind blowing into an ambient medium, it is not surprising to see shell and disk wind emissions present in the same object. This could imply that the disk wind can interact with ambient material to drive shell structures. 
This statement then raises a question: why do we not see the disk wind feature in other objects showing wind-driven shell emission? In other words, what is the wide-opening angle flow driving the shell? Atomic gas flow would be a candidate. While we see only molecular gas in the millimeter observations, the centrifugal wind can consist of atomic or molecular gas depending on the physical conditions in the launching regions \citep{Delabrosse2024}. In this regard, the correlation between the features in mid-infrared emission and our $^{12}$CO data in IRAS 15398 is insightful \citep[see also][] {Harsono2023, Tychoniec2024}.

\subsection{Peculiar features in the outflow of IRAS04169}
When classifying the type of emission observed in the molecular outflow of each object, most of the sources can be classified as one or more of the three types of emission discussed in Section \ref{sec:models}. However, the type of outflow emission observed in IRAS04169 is more ambiguous.

While the blue-shifted outflow shows slow disk wind features with a parabolic shape in the velocity channel maps (see Figure \ref{fig:appendix_IRAS04169} in Appendix A), the emission in the red-shifted outflow exhibits a more complex structure. The red-shifted outflow shows some characteristics of a slow disk wind structure; the channel maps show a parabolic structure which becomes narrower at higher velocities (the left panel in Figure \ref{fig:IRAS04169_combined_figure}). The P-V diagram perpendicular to the outflow axis (the right bottom panel) shows a cone shape centered at offset$=0$. However, the velocity channel maps (the left panel) also show parabolic shell-like structures offset from the protostar and outflow axis, most clearly at velocities of $v-v_{\rm sys}\sim 2-5$ km/s.
In the channel maps at $v-v_{\rm sys}=6-9$ km/s , we also see a shell structure at $\sim 8 \arcsec$ offset from the protostar and $\sim 3 \arcsec$ offset from the outflow axis. 
These shells do not show a clear Hubble-law velocity structure, as their distances from the protostar do not significantly change between channels, probably indicating that they are not wind-driven shells.
They could be bow shocks, as emission in a shock is expected to exhibit a wide range of velocities. But the curvatures of the observed shells are inverse compared with that we expect for bow shocks driven by a fast flow from the central protostar.
Recently, \citet{Aizawa2025} found a bubble structure towards a Class II object WSB 52. They interpreted  the bubble as jet-driven expansion of gas, which is ejected during prior outflow events. The shells observed in IRAS 04169 could be relevant to such events.

\section{Discussion} \label{sec:discussion}
\subsection{Properties of wind-driven shells}
In this section, we discuss the morphological and kinematic properties of the outflows categorized as wind-driven shells using conventional approaches in the literature \citep[e.g.][]{Lee2000,Zhang2019}.

\subsubsection{Comparison with a parametric model}
For wind-driven shells there is a simple parametric model introduced by \citet{Lee2000}, which allows us to quantitatively discuss their physical properties such as the dynamical age of the shells \citep[e.g.][]{Zhang2019}:
\begin{equation}
    \left(\frac{z}{R_0}\right) = \left(\frac{R}{R_0}\right)^2, \hspace{1cm} v_Z = \frac{z}{t_0}, \hspace{1cm} v_R = \frac{R}{t_0}.
    \label{eqn:shell_model}
\end{equation}
The $z$-axis is along the outflow axis and the $R$-axis is perpendicular to the $z$-axis. The parameter $R_0$ is the characteristic radius which determines the width of the outflow and is the radius at which $z=R$. The parameter $t_0$ is the dynamic age and determines the velocity structure of the shell. For each shell we searched for the best values of the parameters $R_0$, $t_0$, and $i$ that can explain the observations by plotting the corresponding model on top of the velocity channel maps and P-V diagrams to compare by eye. Although the direction of the z-axis, i.e. PA, is determined by finding the axis of symmetry of the outflow lobe by eye (\S 4), we modify the PA if the shell model of equation (1) is misaligned with the observed shell until the shell model and observed shell align.

The best parameters are summarized for each source in Table \ref{tab:shell_parameters}, while the model shells overlaid on the observed channel maps and the P-V diagrams are shown in Appendix B (Figures \ref{fig:A2_GSS30IRS3_blue} -- \ref{fig:A2_TMC1A_blue}). We note that the actual shell structure depends on the density and velocity distributions of the driving wind and ambient gas \citep{Shang2006,Shang2020}. The agreement between our observations and this simple model thus varies among objects. For some objects, the model reasonably agrees with the P-V diagram, but a deviation is apparent in the channel map, and vice versa. When the shell structure is not clear in the channel map, but a Hubble flow or an ellipse is more clear in the P-V diagram, we prioritize the comparison of the model with the P-V diagram, rather than with the channel map. While it is difficult to quantify the error of the PA and inclination, since we find the best parameters by eye, the agreement between the model and the data clearly becomes worse if these values are changed by $\sim 5\degr$. Therefore, we adopt $5\degr$ as the approximate uncertainty for these parameters.

For Oph IRS 63, the representative object of wind-driven shell emission, the cyan lines in Figure \ref{fig:OphIRS63_combined_figure} depict the best model.
Assuming a distance to the source of $132\pm6\ \mathrm{pc}$ \citep{Zucker2020}, we find that the shell structure was best described by a model with $R_0 = 0.6\arcsec$, $t_0 \sim 380\ \mathrm{yrs}$, and $i \sim 47\degr$ (B1 in Table \ref{tab:shell_parameters}). This inclination is in agreement with the value of $i = 46.7\degr$ derived from the Gaussian fit to the continuum disk emission by \citet{Ohashi2023} and \citet{Flores2023}.

The channel map of Oph IRS 63 shows another shell, which is much wider than the one described above (see Figure \ref{fig:A2_OphIRS63_blue} in Appendix B). Its best parameters are $R_0 = 5\arcsec$ and dynamic age $t_0 = 1300$ yr (B2 in Table \ref{tab:shell_parameters}). However, the agreement is much worse compared with the narrower shell. 
While the model can explain the shell emission in the channel map at lower velocities ($|v-v_{\rm sys}|\lesssim 8$ km/s), at higher velocities the observed shell emission appears to be ``slower'' than the model shell. Such deviation is also found in larger, older shells in other objects in the eDisk survey. This may be explained by the deceleration of the flow due to interaction with ambient material, which is not considered in the simple model of equations (\ref{eqn:shell_model}). The deceleration would have a greater impact on higher-velocity material than the lower-velocity material, explaining the difference between the model and observed shells at high-velocity channels. It is also important to note that this deceleration indicates that the dynamic ages derived for the shells are upper limits on the age of the shell rather than their true age \citep{Zhang2019}.

\begin{figure}
    \includegraphics[width=\columnwidth]{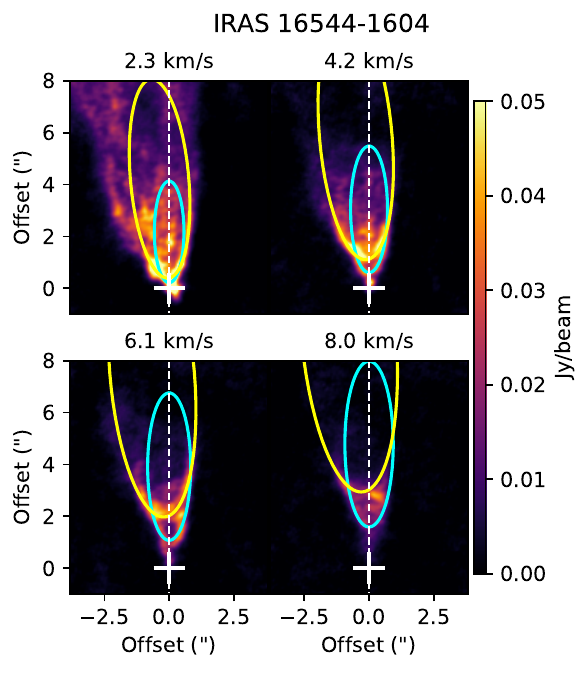}
    \caption{The velocity channel map of IRAS 16544-1604 (color scale) with the model shells R1 (cyan) and R2 (yellow) fitted to the emission shown by the ellipses.}
    \label{fig:IRAS16544_vel_map}
\end{figure}

\subsubsection{Dynamical timescales}

The dynamical timescales obtained above infer the time variability of the outflow.
The mass-loss rate in protostellar outflows is known to be highly variable. This is most notably seen in the form of a series of emission knots in the jets of YSOs, which are believed to trace bow shocks caused by variations in the velocity and direction of the flow \citep[e.g.][]{Zinnecker1998,Plunkett2015,Lee2018, Lee2024}. While evidence of variability has not been seen as frequently in the case of molecular outflows, \citet{Zhang2019} found multiple wide-angle shell structures in the CO emission of the HH46/47 molecular outflow. They argue that these shell structures are the result of entrainment of ambient material around the source by a series of outbursts from the central object. 
Our eDisk data shows such evidence of variability in the mass-loss rate of the outflows in the form of the wind-driven shell emission in 7 sources. The timescale of variability in these outflows can be inferred from the dynamical ages $t_0$ of the shells in each source.
In the case of Oph IRS63, for example, the observed shells in the blue-shifted outflow indicate that outbursts occurred in this object $\sim 380$ and $\sim 1300$ years ago. This indicates that the mass loss rate in the outflow varies on timescales of several hundred yrs. Similar timescales are seen in the other sources with wind-driven shell emission (as seen from the values of $t_0$ in Table \ref{tab:shell_parameters}).

\begin{table*}
    \caption{Shell Model fits for the eDisk sources with detected wind-driven shell emission.}
    \begin{center}
    \begin{tabular}{c c c c c c c c}
    \toprule
    Source\tablenotemark{a} & $i_{\mathrm{cont}}$\tablenotemark{b} & PA$_{\mathrm{cont}}$ & Shell & $t_0$ & $R_0$ & $i$\tablenotemark{c} & PA\tablenotemark{c} \\
     &  $\degr$ & $\degr$ & & yr & arcsec & $\degr$ & $\degr$ \\
     \hline
    IRAS 16544-1604 & 73 & 45 & B1 & 2500 & 4 & 73 & 128\\
     & & & R1 & 100 & 0.2 & 73 & 335\\
     & & & R2 & 190 & 0.4 & 73 & 340\\
    \hline
    GSS30 IRS3 & 64 & 109.36$\pm 0.30$ & B1 & 1400 & 18 & 82 & 18 \\
     & & & R1 & 40 & 0.7 & 85 & 198 \\
     & & & R2 & 400 & 4 & 85 & 198 \\
    \hline
    IRAS04166+2706 & 47 & 121.5$\pm 0.5$ & R1 & 250 & 0.5 & 49 & 212\\
    \hline
    R CrA IRAS 32 & 69 & 135.3$\pm 0.4$ & B1 & 1200 & 4 & 69 & 43 \\
     & & & R1 & 850 & 3.5 & 69 & 223 \\
    \hline
    IRAS04302+2247 & 84 & 174.7$\pm 0.03$ & B1 & $>200$ &$>0.3$ & 84 & 85 \\
    \hline
    TMC-1A & 52 & 76 & B1 & 400 & 0.5 &  52 & 346 \\
     & & & B2 & 800 & 1.0 & 52 & 346 \\
    \hline
    Oph IRS63 & 47 & 149 & B1 & 380 & 0.6 & 47 & 240\\
     & & & B2 & 1300 & 5 & 47 & 240\\
     & & & R1 & 1600 & 6 & 47 & 90\\
    \hline
    \end{tabular}
    \end{center}
    \tablenotetext{a}{The sources are listed in order of increasing bolometric temperature.}
    \tablenotetext{b}{Disk inclination angles adopted from \citet{Ohashi2023}. They are consistent with the values derived in the first-look papers within an error of $\le 1$ degree.}
    \tablenotetext{c}{The error of $i$ and PA is estimated to be $\lesssim$ 5\degr (see text).}
    \label{tab:shell_parameters}
\end{table*}

Future studies of shell structures within outflows could benefit from comparing the dynamic ages of shells to the dynamic ages of emission knots observed in collimated jets of the same source. This could tell us if the wide-angle wind that drives the shell is related to the jet. The dynamical ages could also be compared with the dates of accretion outbursts detected in the source, e.g. FUor or EXor outbursts, if any, given that the accretion and mass loss processes in star formation are closely linked.
\citet{Kim2024}, for example, recently observed B335 and found that the dynamical age of the high-velocity $^{12}$CO outflow is similar to that of the accretion burst probed by the mid-infrared brightness.


As noted in \S \ref{sec:results}, the properties of blue-shifted and red-shifted outflows differ in some objects. Specifically, Table \ref{tab:shell_parameters} indicates that the dynamical ages of the shells are different between the red- and blue-shifted outflows.
In the case of IRAS 16544-1604, for example,  the blue-shifted outflow has one shell of dynamical age of 2500 yrs, while the red-shifted component has two shells of dynamical age of 100 yr and 190 yr (R1 and R2 in Figure \ref{fig:IRAS16544_vel_map}). These differences may be due to different properties of the ambient medium on either side of the outflow. The swept-up shells could be too faint to be detected if the density of the ambient material is low. 
Such differences between outflow lobes have been observed previously in other objects. An extreme example is HH30/31, which shows no molecular outflow in its southern lobe because it is at the edge of its parental cloud \citep{Louvet2018}.
Density distribution of the ambient material also affects the circumstellar magnetic fields, which plays an essential role in driving winds. Another possibility is that in weakly magnetized cloud cores, unipolar outflows can be formed due to turbulent accretion \citep{Takaishi2024}.


\subsubsection{Position angle and inclination}
\label{sec:PA_Inclination}
Other parameters of the shell model (Eq. \ref{eqn:shell_model}) can also yield interesting information on the geometry of the outflows, i.e. their position angles (PA) to be compared with that of the dust continuum, i.e. PA$_{\rm cont}$.
Among our targets, IRAS 16544-1604 and Oph IRS 63 are categorized as wind-driven shells without any indications of coexistence with a slow disk wind. It is interesting that the PA of their outflows (i.e. shells) are not perpendicular to that of the dust continuum.
For example, the PAs of the red-shifted shells in IRAS 16544-1604 are 335 \degr and 340 \degr, while the PA of its dust continuum is 45 \degr \citep[Table \ref{tab:PA_and_Vsys}, see also][]{Kido2023}; the latter value should be 65-70 \degr if the disk is perpendicular to the outflow. For this source, the dust continuum is thought to be tracing a disk. \citet{Kido2023} analyzed C$^{18}$O (2-1) emission to find that the gas rotation in the continuum emitting region is close to Keplerian rotation. Similarly, Keplerian rotation is confirmed around Oph IRS 63 \citep{Flores2023}.

Interestingly, the PAs of the two red-shifted shells of IRAS 16544-1604 also differ by $\sim 5\degr$, suggesting that the outflow axis changed between the two outbursts that created the shells, while the dynamical ages of the two shells differ only by 90 yrs (Figure \ref{fig:IRAS16544_vel_map}). 
We note that temporal changes in the outflow axis have been found and discussed in another eDisk target IRAS 15398; in addition to the outflow shown in Figure \ref{fig:IRAS15398_combined_figure}, there are two sets of old outflow candidates (with dynamical timescales of $10^3-10^4$ yr) misaligned by $20-90 \degr$ \citep{Okoda2021, Thieme2023,Sai2024}.
Such temporal variation of the outflow axis could be due to the accretion of gas with various angular momentum vectors, which could change the angular momentum axis of the star-disk system.
The impact should be more significant for lower mass protostars, in which the ratio of newly added momentum to the total momentum should be relatively large. Indeed, the protostellar masses of IRAS 16544-1604 and IRAS 15398 are estimated to be relatively small ($\sim 0.1 M_{\odot}$) among eDisk objects \citep{Okoda2021,Thieme2023,Kido2023}.


From the comparison with the simple shell model, the inclination angle $i$ of an outflow is also estimated, and it can be compared with the disk inclination ($i_{\rm cont}$ in Table \ref{tab:sum} and \ref{tab:shell_parameters}) that is estimated from the major and minor axis of the dust continuum emission.  While these angles are similar to each other for most of the sources, for GSS30 IRS 3 we obtained $i \sim 82$ and $\sim 85 \degr$ for the blue and red-shifted shells respectively, compared to $i_{\rm cont} = 64.3\pm1.5\degr$ \citep{Santamaria-Miranda2024}. However, given that the disk may be geometrically thick, $i_{\rm cont}$ is only a lower limit of the disk inclination. The inclination angles of the disk and the outflow thus could be consistent, and the disk could be nearly edge-on.

For IRAS04302, it is difficult to derive a shell model due to the faint nature of the emission and lack of distinct elliptical shape in the channel maps. The value of $R_0$ is not very well constrained, and a wide range of values of $R_0$, $t_0$ and $i$ can be consistent with the data.
Assuming that the inclination angle of the disk and outflow is the same, we obtain a lower limit for $R_0$ and $t_0$ based on the smallest shell that can fit the data.

\subsection{Objects with no outflow detected}
While most of the sources show a certain type of molecular outflow emission, four sources, Ced 110 IRS 4, IRS 5N, IRS 7B, and IRS43, show no clear $^{12}$CO outflow in our data. An important question to address is whether this is because the emission has been resolved out or because there is indeed no outflow present. 

$^{12}$CO molecular outflows have previously been detected at larger scales \citep{Yildiz2015,VanKempen2009,Bontemps1996} for Ced 110 IRS 4 and IRS 7B.
So, it is reasonable to assume that the $^{12}$CO emission is simply resolved out in the eDisk observations as the spatial scale of these previous studies is larger than our maximum resolvable scale of $2 \arcsec - 3 \arcsec$. Observations obtained with ALMA in short-baseline configurations (Plunkett et al., in prep.) will help confirm this.

As for IRS43, \citet{Bontemps1996} observed a velocity gradient of the $^{12}$CO (2-1) emission in the east-to-west direction using IRAM 30 m and interpreted it as an outflow. This velocity gradient, however, is consistent with the velocity of an infalling rotating envelope found in \citet{Narayanan2023} using $^{12}$CO, $^{13}$CO, and C$^{18}$O (2-1) data of eDisk. The V-shape structures in the channel in Figure \ref{fig:appendix_OphIRS43} correspond to the envelope.
\citet{Narayanan2023} also showed that the axis of the outflow cavity runs from north to south, even though outflow emission is not clearly detected in this direction. This example shows having both large-scale and high-resolution observations is crucial for identifying outflows.

For IRS5N, extended $^{12}$CO emission is observed surrounding the source. However, it does not appear to trace an outflow or jet, as discussed by \citet{Sharma2023}. Previous studies \citep[e.g][]{Bontemps1996,Yang2018} have also failed to obtain any clear molecular outflow detection, and there has been no clear detection of other jet/outflow tracers for this source. While cm-wavelength radio emission was detected by \citet{Miettinen2008}, which could at least partially be thermal free-free emission from an ionized base of the jet, the negative spectral index and rapid variability of the emission suggest it is mainly gyrosynchrotron from the stellar magnetosphere. IRS 5N was also suggested as the possible source of a H$_2$ shock detected by \citet{Kumar2011}. The high density of sources in the R CrA region, however, makes it difficult to be certain whether it is associated with IRS 5N. In addition, if this shock were associated with IRS 5N, the PA of the outflow would be almost parallel to the major axis of the disk. Therefore, it seems unlikely that it traces an outflow originating from IRS 5N. Based on the lack of detection in this and other studies, it is tempting to speculate that IRS 5N lacks an outflow or has a very weak outflow, although the complex spatial and kinematic structures of the R CrA region due to the high density of sources could hamper the clear detection. In theoretical studies, it has been proposed, for example, that a misalignment between the cloud rotation axis and the magnetic field could suppress outflow activities \citep{Hirano2020,Takaishi2024}.

It is also worth noting that Ced 110 IRS 4, IRS 7B, and IRS 43 are three of the four close-binary systems with an apparent separation of $\lesssim 250$ au on the plane of the sky in the eDisk targets. It indicates that the existence of a close binary may have an effect on the structure of the molecular outflow on small scales. Another close-binary system in eDisk targets, IRAS 32, however, shows clear outflow emission categorized as a combination of a slow disk wind and wind-driven shells, which we assume are launched from the primary (IRAS 32 A).

\begin{figure}
    \includegraphics[width=\columnwidth]{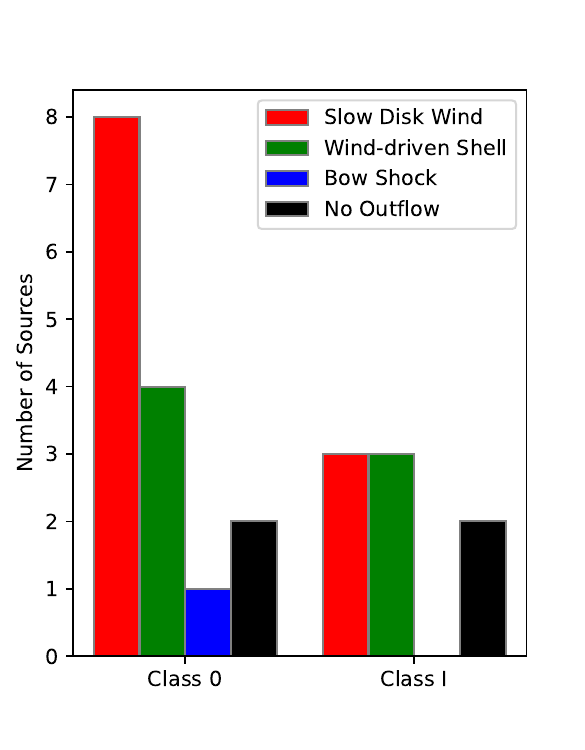}
    \caption{A bar chart showing the number of sources with slow disk wind (SDW), wind-driven shell (WS), and bow shock (BS) emission in Class 0 and I. The black bars labeled "No" depict the number of sources without clear outflow emission.}
    \label{fig:class_0_vs_class_I}
\end{figure}

\subsection{Is there a correlation between the class of the source and the outflow features?}

An important question is whether there is a correlation between the type of outflow emission observed and the evolutionary state of the central protostar. Naively, one might expect the wind-driven shell to be more common in Class 0 objects as there should be more ambient material present than in Class I. Similarly, for wind-driven shells one might expect the older Class I objects to possess shells with an older dynamical age compared to the younger Class 0 objects. Our eDisk data, however, does not immediately support this hypothesis: Figure \ref{fig:class_0_vs_class_I} summarizes the number of sources with slow disk wind, wind-driven shell, bow shock, and no clear outflow emission for each class. Differences between Class 0 and I are not clear. While 4 out of the 12 Class 0 sources ($33\%$) show wind-driven shells, 3 out of the 7 Class I sources ($43\%)$ also show such shells. 
Slow disk wind emission is common in both Class 0 ($67\%$) and  Class I sources ($43\%$).
No relation between the class of the source and the dynamical age of the shells is apparent either, with Class 0 objects possessing shells with a range of ages from 40 to 2500 yrs and Class I objects possessing shells with a range of ages from 200 to 1600 yrs. Data of larger samples and with a larger maximum recoverable scale (e.g. combination of 12m-array with ACA) would be desirable for further studies.




\clearpage

\section{Conclusions}

We have analyzed outflow features observed with $^{12}$CO (2-1) towards 19 protostellar sources covered in eDisk. Various emission features are detected towards 15 objects. 
We categorized the outflow emission of the 15 objects into three types, wind-driven shell, bow shock, and slow disk wind, based on the specific features found in the velocity channel maps and the P-V diagrams along and perpendicular to the outflow axis. 
Seven objects are categorized as wind-driven shell, 1 object as bow shock, and 11 objects as slow disk wind, among which 4 objects show both slow disk wind and wind-driven shell. For the bow shock emission of IRAS15398, the shell structures found in our $^{12}$CO data correlate well with a bow shock feature observed with JWST. We do not find any clear correlation between outflow features with the evolutionary stage, i.e. Class 0 and I. 

Some objects show different features in their blue- and red-shifted outflows. For example, GSS 30 IRS3, IRAS04166, and TMC-1A show mixed features of disk wind and wind-driven shell only in blue or red components. The blue-shifted outflow of IRAS04169 is categorized as disk wind, while its red-shifted outflow shows peculiar features that are difficult to categorize. The blue-shifted and red-shifted outflows are not aligned with each other in IRAS 16544-1604, IRAS 15398, IRAS 04169, L1489, and Oph IRS63. Misalignment between the outflow axis and the disk minor axis, which is determined by 2D Gaussian fitting of the dust continuum emission, is also found for 7 objects (e.g. L1489 IRS and IRAS 16544-1604).

We adopted the model of \citet{Lee2000} and compared it to the emission categorized as wind-driven shell to derive e.g. dynamical ages of the shells. The derived dynamical ages are different between the shells in the blue- and red-shifted outflows, which is possibly caused by either differences in the amount of ambient material with which the wind interacts, or by the red-shifted and blue-shifted shells being launched at different times. At high velocity channels, old shells show signatures of deceleration (e.g. $|v-v_{\rm sys}| \gtrsim 8$ km s$^{-1}$ in the blue-shifted wide shell of Oph IRS 63). In IRAS 16544-1604, two shells with dynamical ages of 100 yr and 190 yr are misaligned, indicating variation of the outflow axis on a short timescale.


In our eDisk samples, four objects, Ced110 IRS4, R CrA IRS 7B, R CrA IRS 5N and Oph IRS43, do not show outflow emission. Among them, large-scale outflows that could be resolved out in eDisk data are previously observed towards Ced110 IRS4 and R CrA IRS 7B. As for Oph IRS43, the large scale CO emission previously interpreted as an outflow \citep{Bontemps1996}
turned out to be a rotating envelope component described 
in the eDisk first-look paper \citep{Narayanan2023}.
These results imply that it is critical to sample both large- and fine-scale structures
for unambiguous identification of molecular outflows.

While we adopted a simple categorization of outflows and a simple parametric model of wind-driven shell for comparison with the observation, we do not aim to exclude any models or physical mechanisms not explored in the present work. For example, the outflows categorized as slow disk wind could be explained by gas dragged by a higher velocity flow.
Our motivation is to report a variety of structures and features, some of which are previously observed individually but now with a group of sources with a uniform data quality. 
Further observational and theoretical studies are necessary to understand this variety.
For example, it is important to compare our results with observations of the eDisk sample at shorter baselines. This will allow the detection of large-scale molecular emission in the outflows that may have been resolved out in the data presented here, particularly for those objects where no molecular outflow emission was detected. Such observations have already been carried out and will be presented in a future publication.



\appendix

\section{Channel Maps}
\begin{figure}
    \centering
    \includegraphics[width=\linewidth]{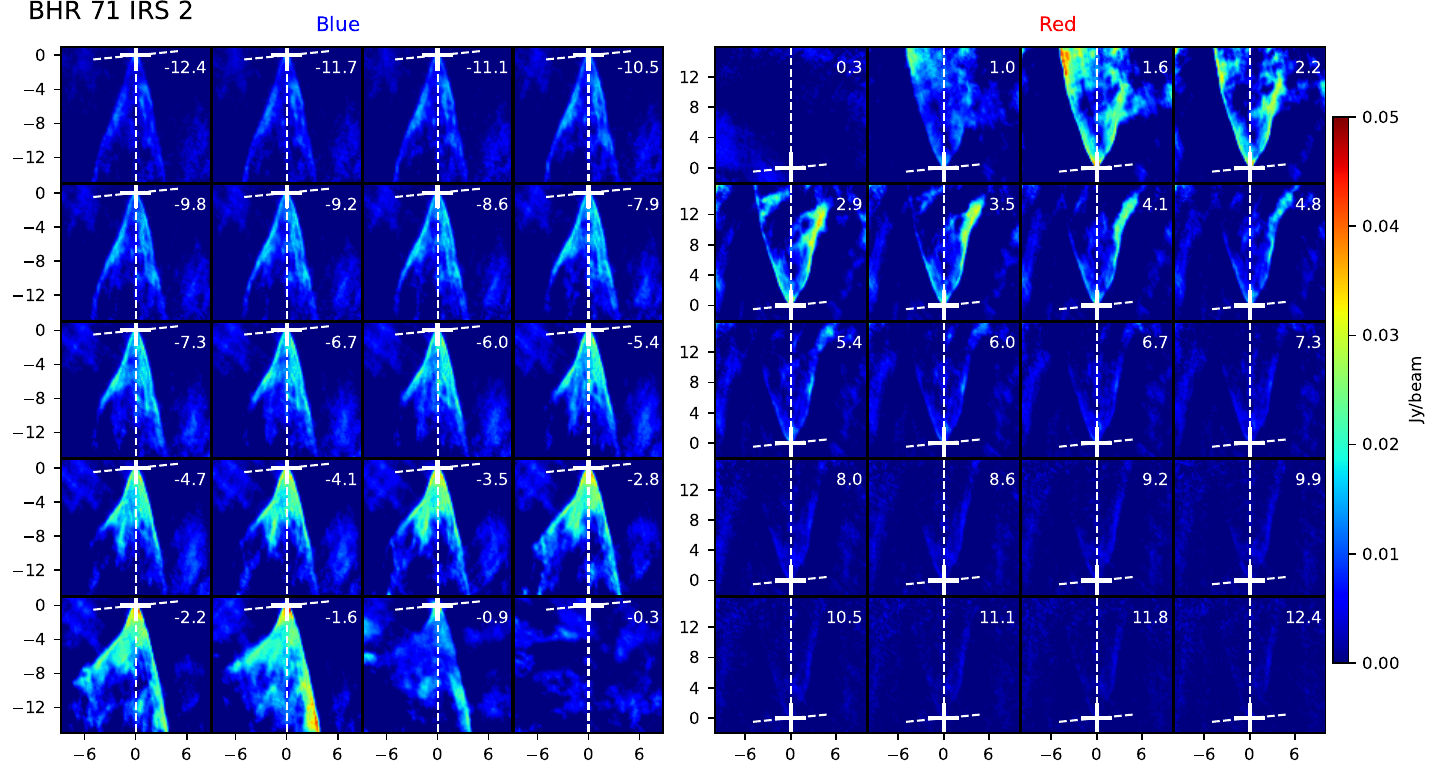}
    \caption{Channel map of BHR71 IRS 2. The stellar position is indicated by the cross at co-ordinated (0, 0). The axis of the jet is indicated by the vertical dashed line. The position angle of the major axis of the dust continuum emission is indicated by the dashed line which intersects the vertical dashed line at the stellar position. Higher velocity channels, i.e. molecular jet, are described in \citet{Gavino2024}.}
    \label{fig:appendix_BHR71IRS2}
\end{figure}

\begin{figure}
    \centering
    \includegraphics[width=\linewidth]{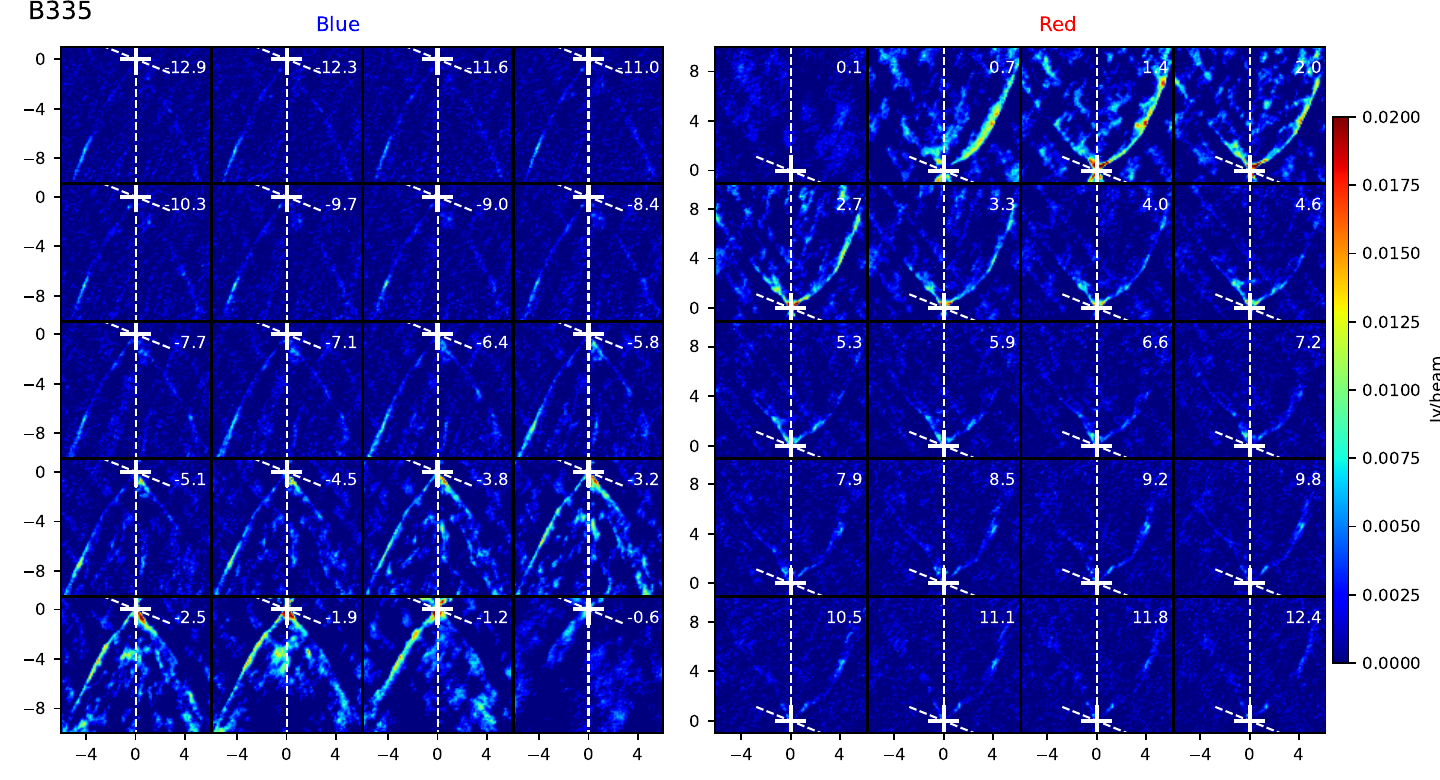}
    \caption{Same as Figure \ref{fig:appendix_BHR71IRS2} but for B335.}
    \label{fig:appendix_B335}
\end{figure}

\begin{figure}
    \centering
    \includegraphics[width=\linewidth]{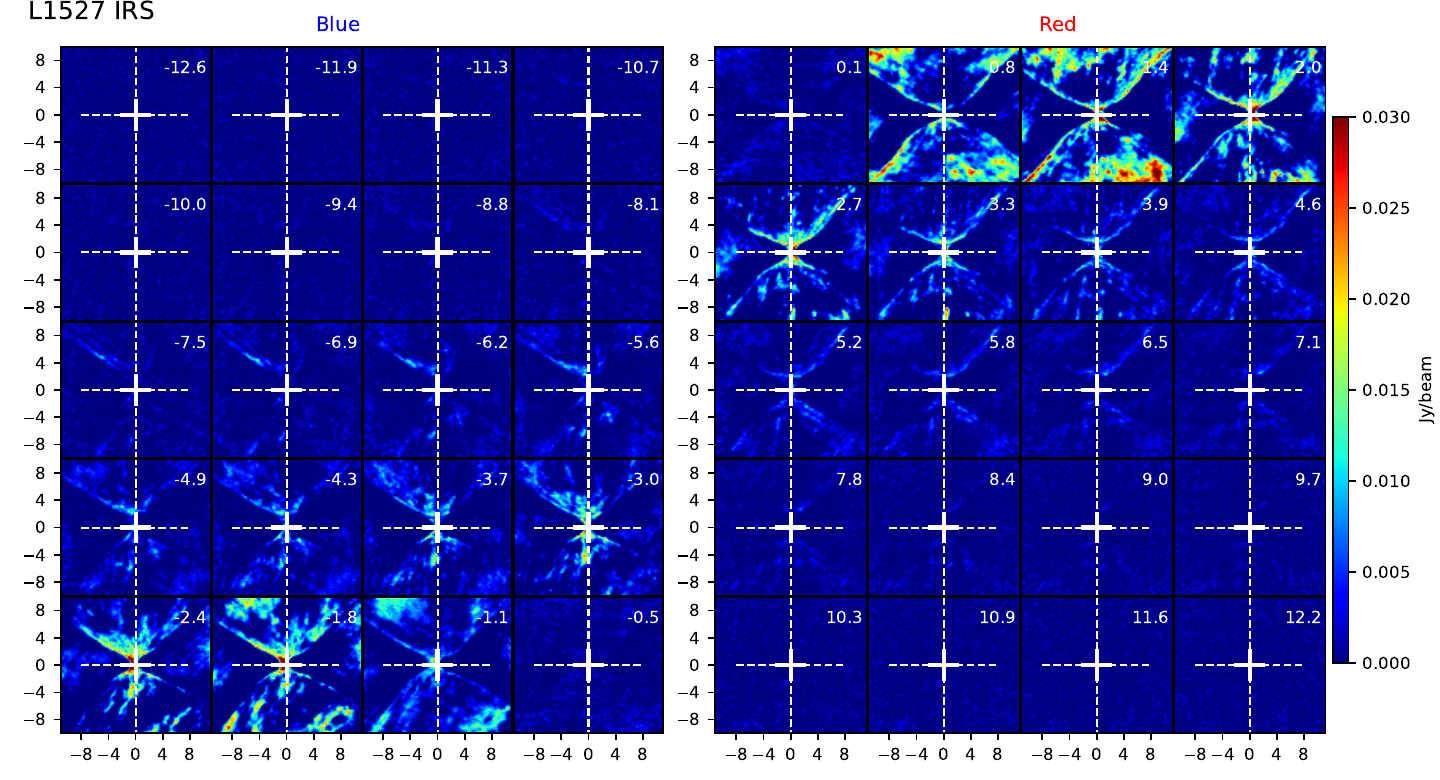}
    \caption{Same as Figure \ref{fig:appendix_BHR71IRS2} but for L1527IRS.}
    \label{fig:appendix_L1527IRS}
\end{figure}

\begin{figure}
    \centering
    \includegraphics[width=\linewidth]{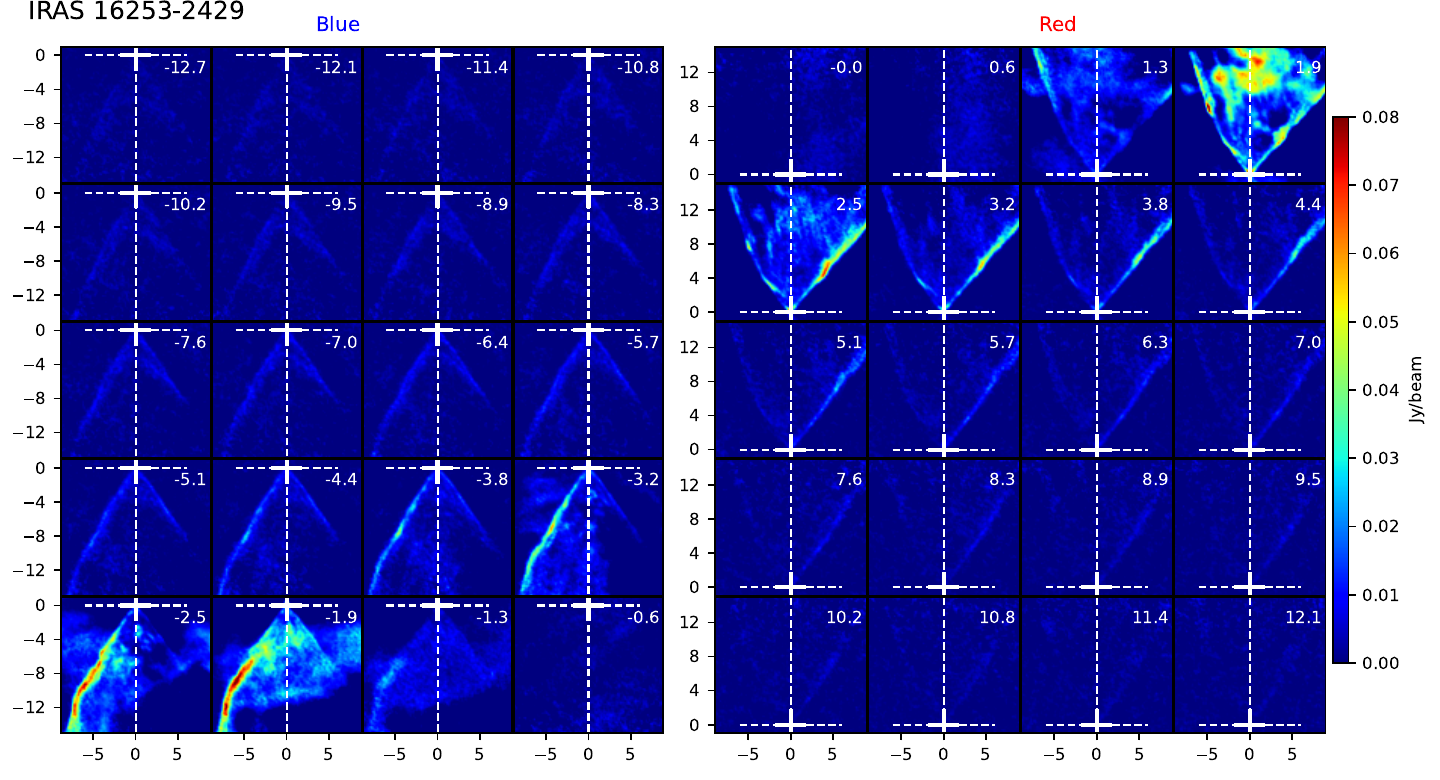}
    \caption{Same as Figure \ref{fig:appendix_BHR71IRS2} but for IRAS16253.}
    \label{fig:appendix_IRAS16253}
\end{figure}

\begin{figure}
    \centering
    \includegraphics[width=\linewidth]{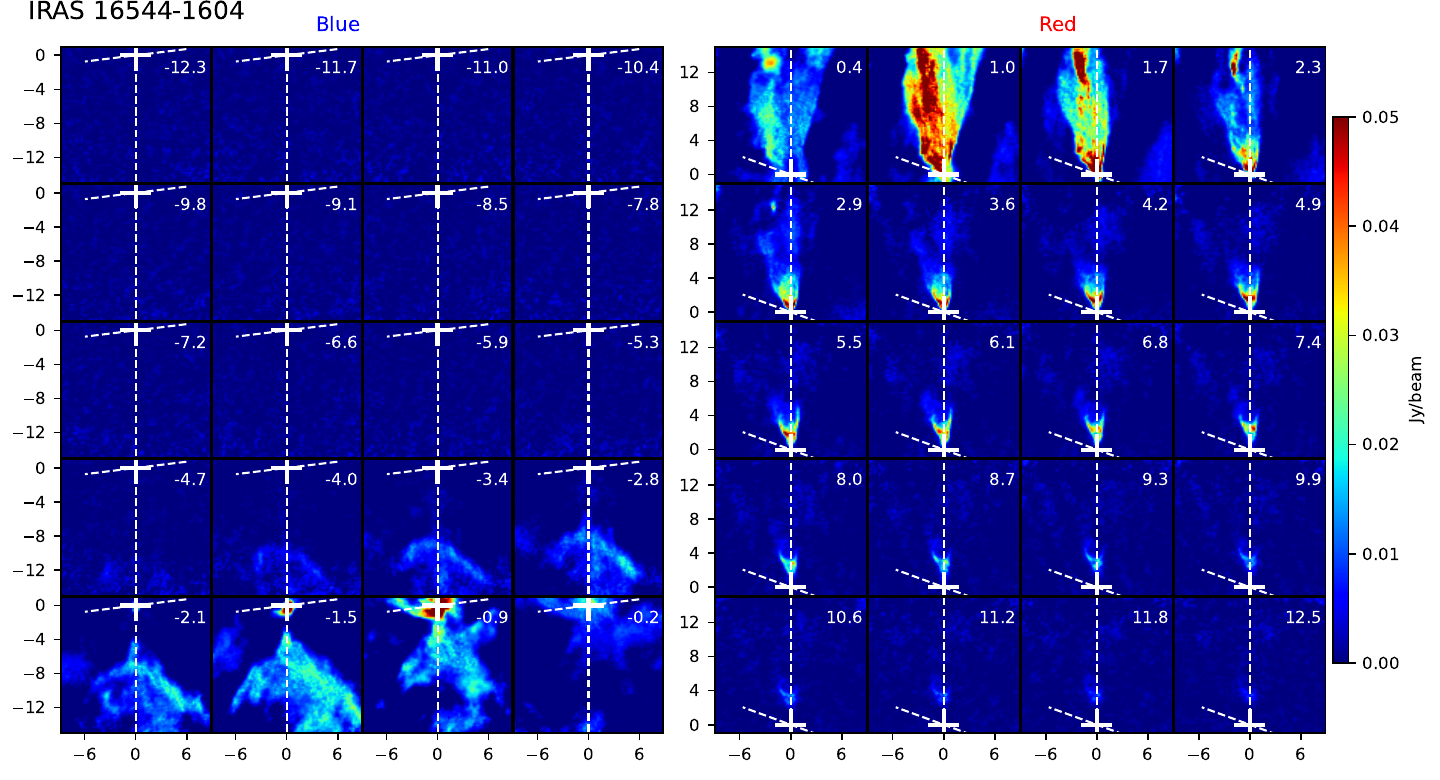}
    \caption{Same as Figure \ref{fig:appendix_BHR71IRS2} but for IRAS16544.}
    \label{fig:appendix_IRAS16544}
\end{figure}

\begin{figure}
    \centering
    \includegraphics[width=\linewidth]{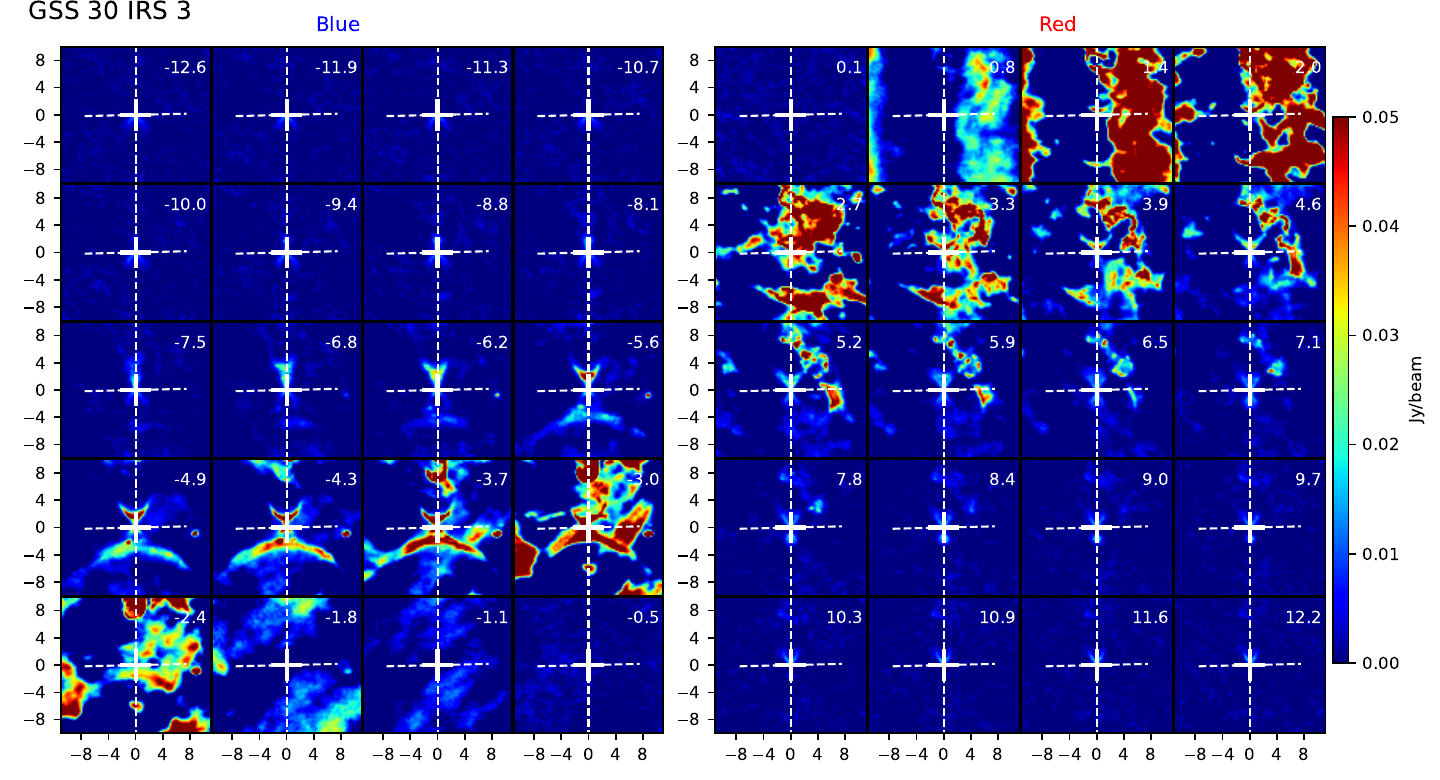}
    \caption{Same as Figure \ref{fig:appendix_BHR71IRS2} but for GSS30 IRS3.}
    \label{fig:appendix_GSS30IRS3}
\end{figure}

\begin{figure}
    \centering
    \includegraphics[width=\linewidth]{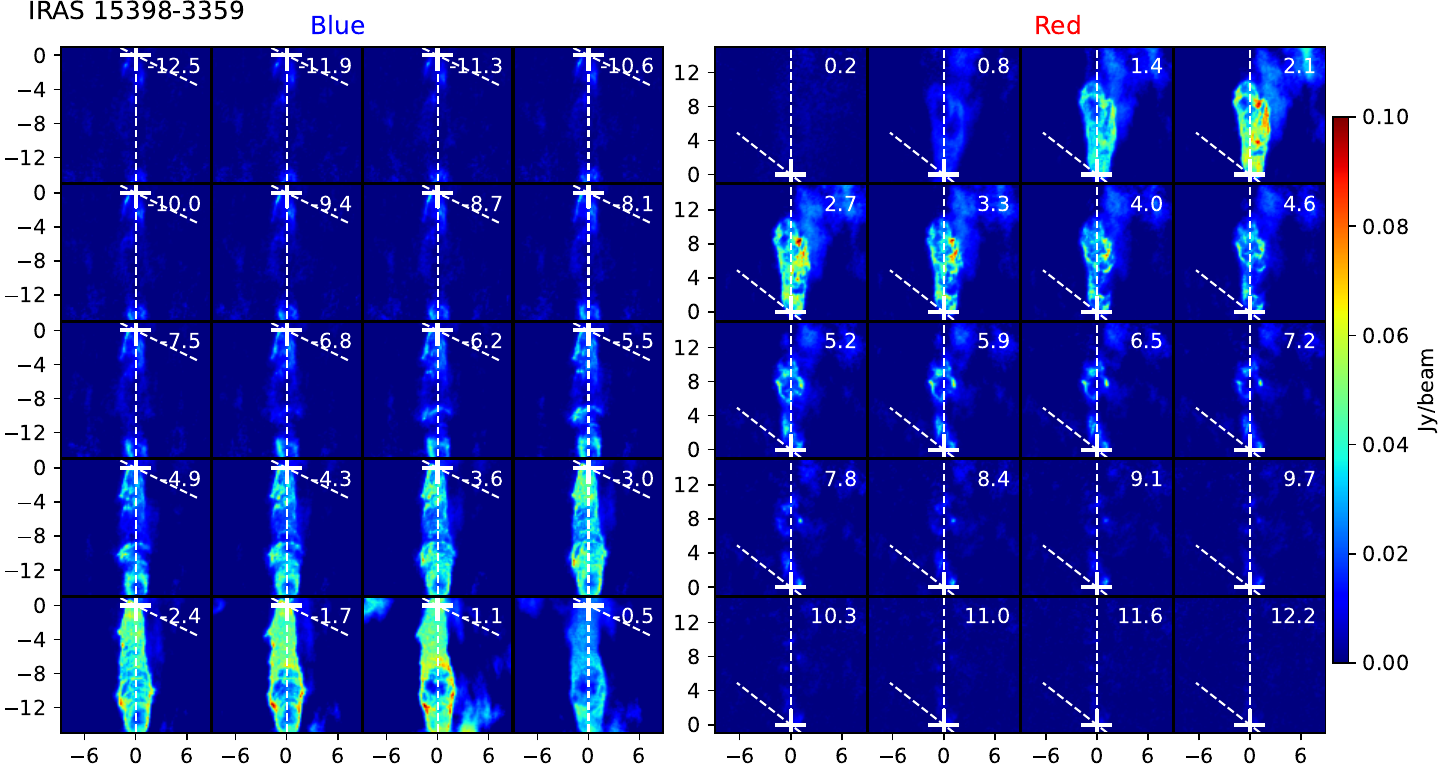}
    \caption{Same as Figure \ref{fig:appendix_BHR71IRS2} but for IRAS15398.}
    \label{fig:appendix_IRAS15398}
\end{figure}

\begin{figure}
    \centering
    \includegraphics[width=\linewidth]{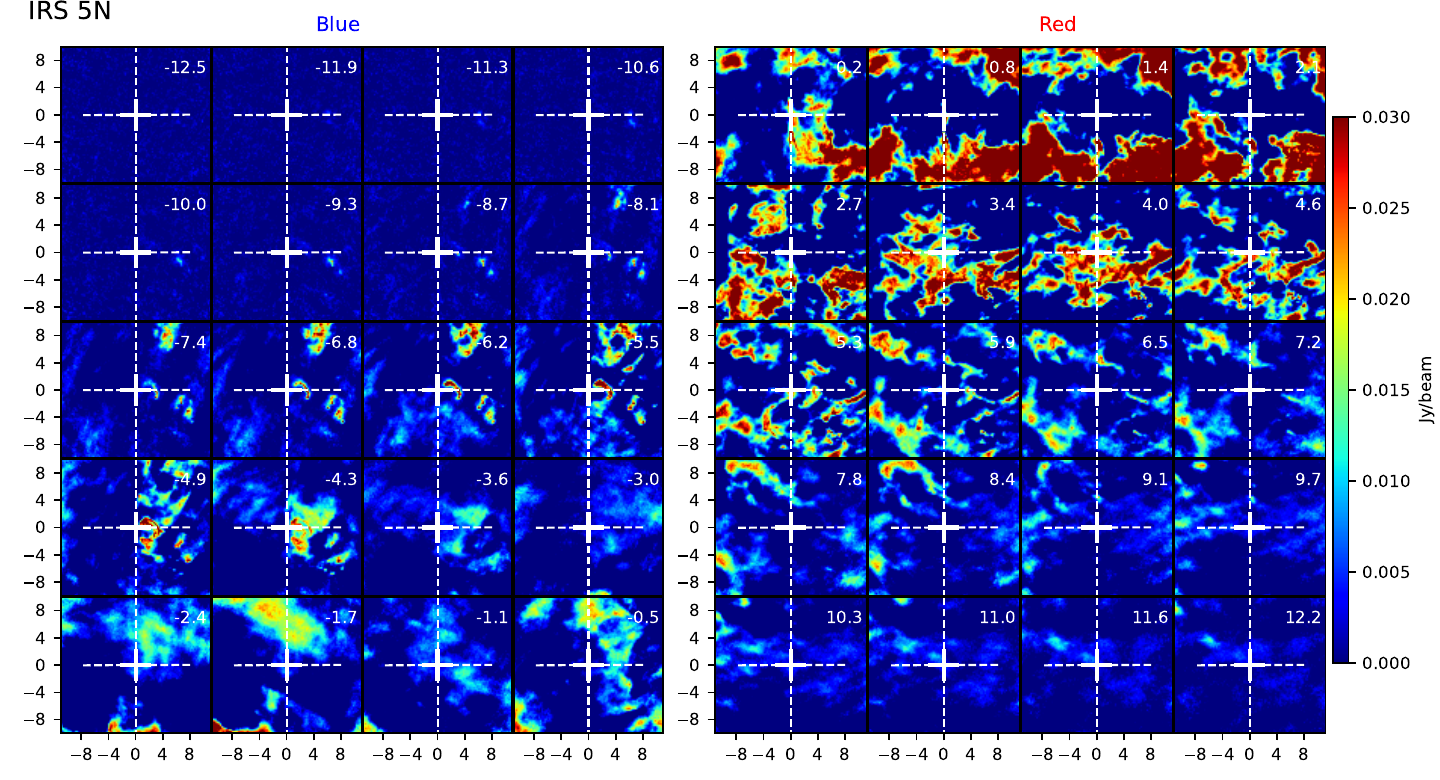}
    \caption{Same as Figure \ref{fig:appendix_BHR71IRS2} but for R CrA IRS 5N.}
    \label{fig:appendix_IRS5N}
\end{figure}

\begin{figure}
    \centering
    \includegraphics[width=\linewidth]{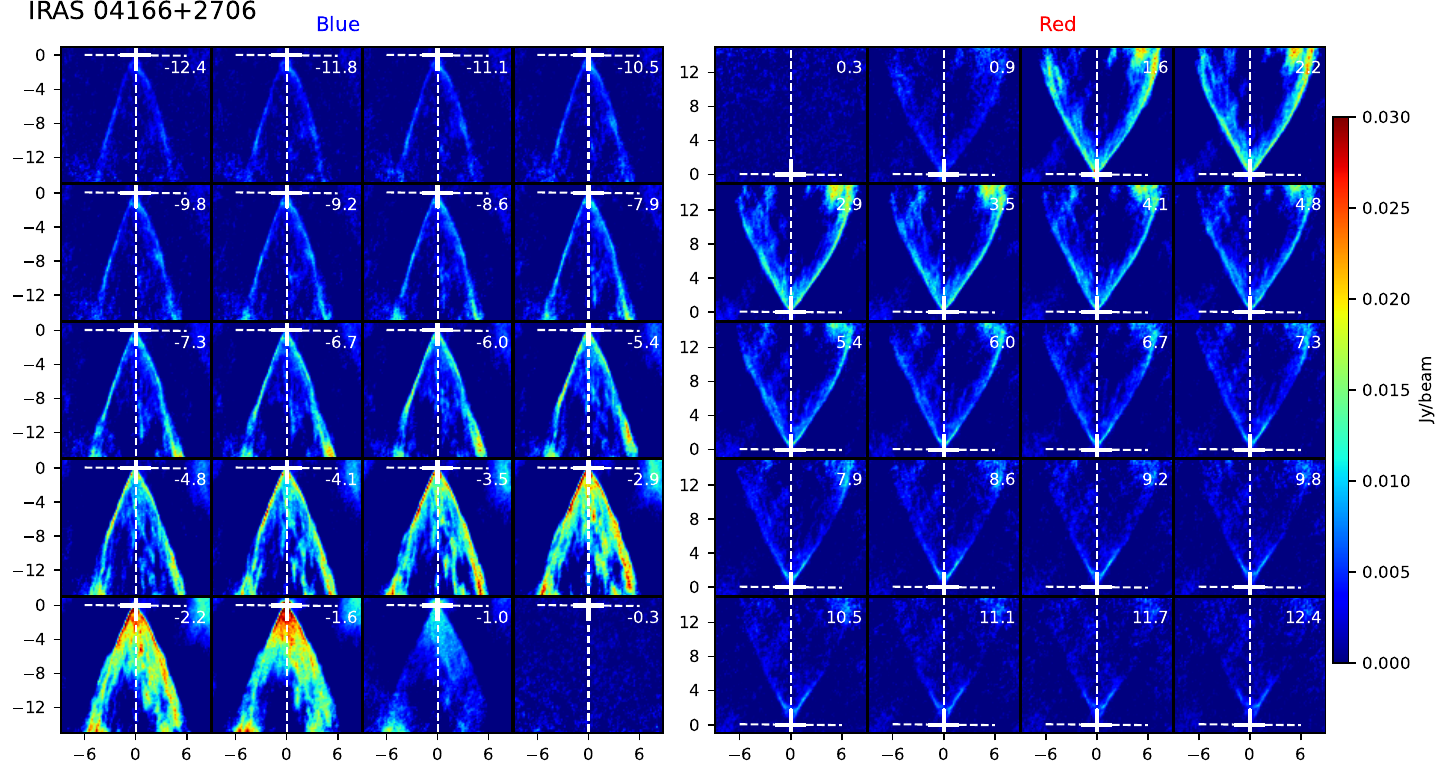}
    \caption{Same as Figure \ref{fig:appendix_BHR71IRS2} but for IRAS04166.}
    \label{fig:appendix_IRAS04166}
\end{figure}

\begin{figure}
    \centering
    \includegraphics[width=\linewidth]{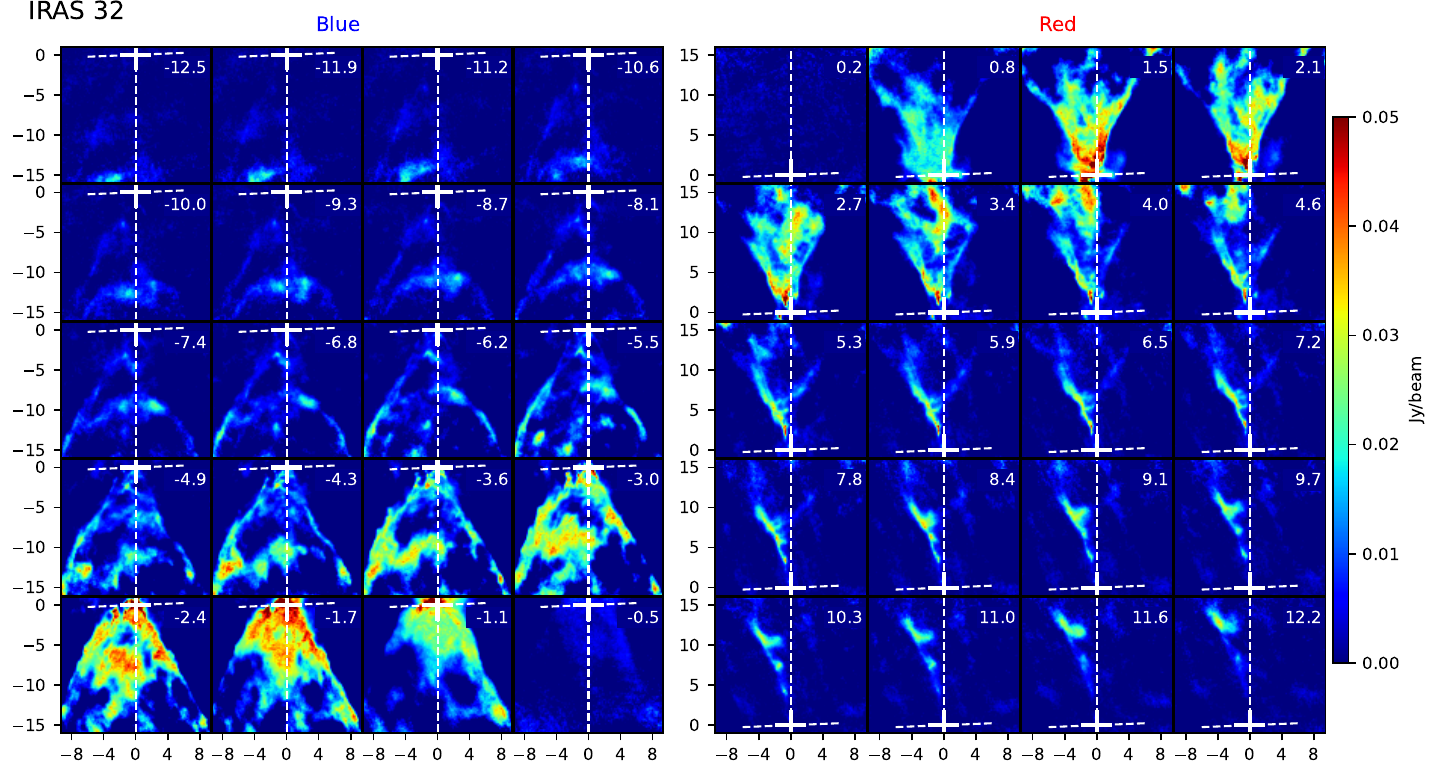}
    \caption{Same as Figure \ref{fig:appendix_BHR71IRS2} but for R CrA IRAS 32.}
    \label{fig:appendix_IRAS32}
\end{figure}

\begin{figure}
    \centering
    \includegraphics[width=\linewidth]{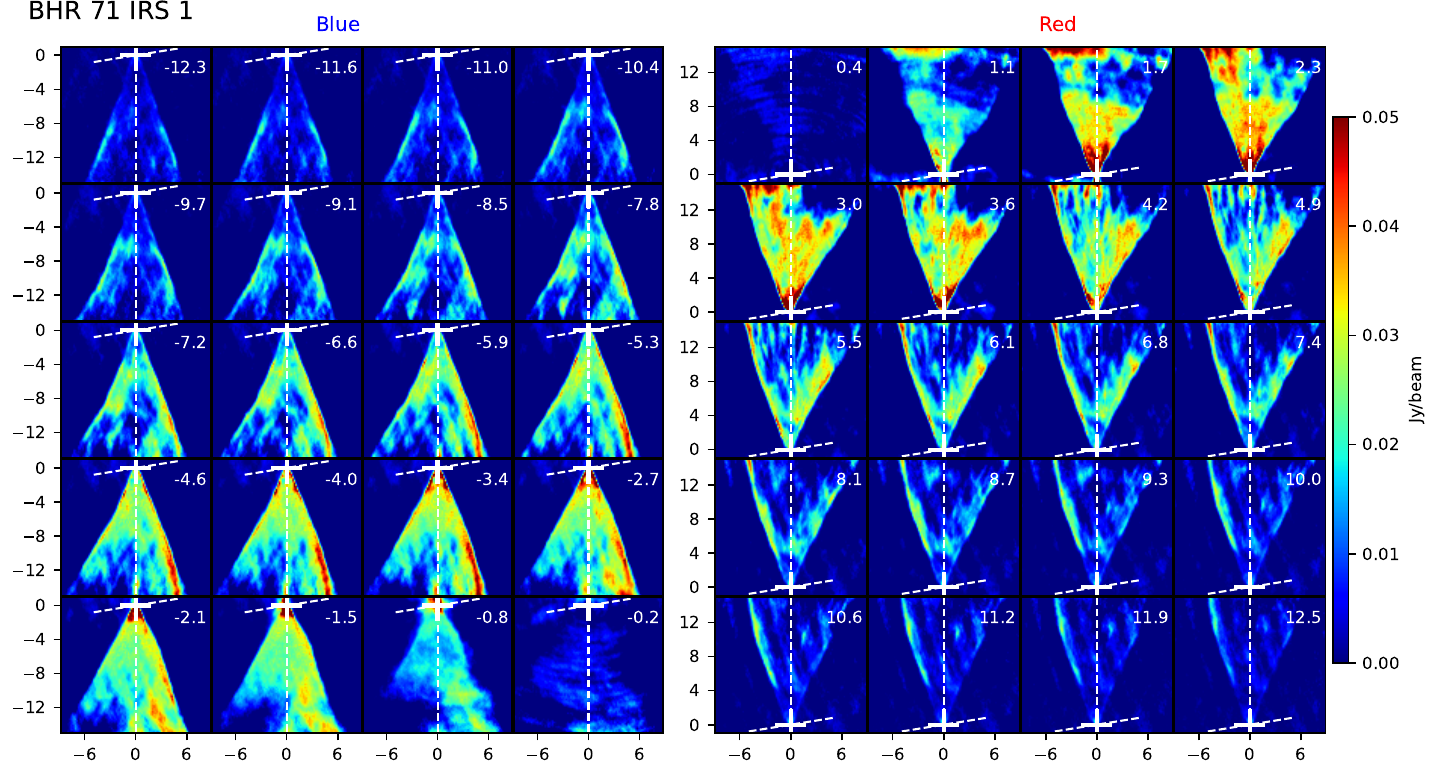}
    \caption{Same as Figure \ref{fig:appendix_BHR71IRS2} but for BHR71 IRS 1.  Higher velocity channels, i.e. molecular jet, are described in \citet{Gavino2024}.}
    \label{fig:appendix_BHR71IRS1}
\end{figure}

\begin{figure}
    \centering
    \includegraphics[width=\linewidth]{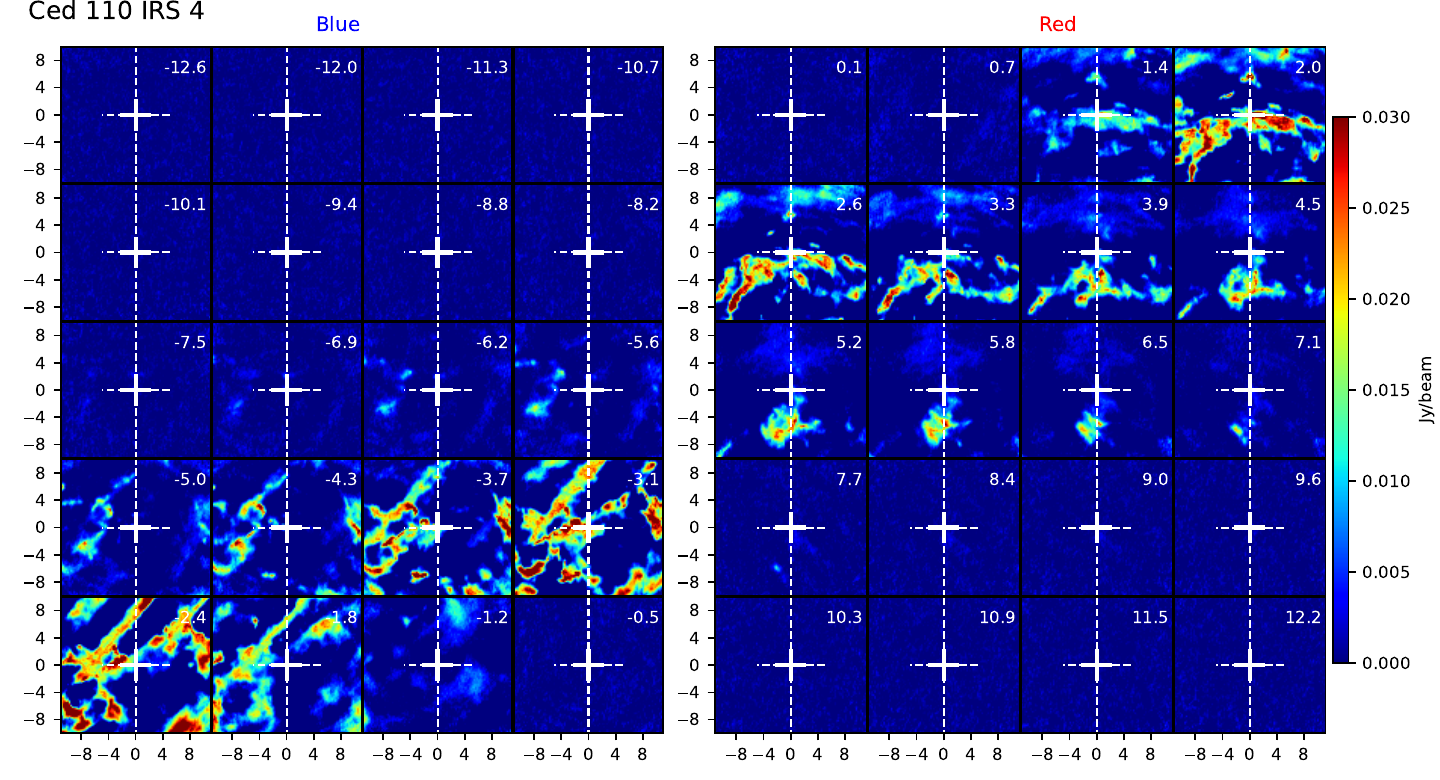}
    \caption{Same as Figure \ref{fig:appendix_BHR71IRS2} but for Ced110 IRS4.}
    \label{fig:appendix_Ced110IRS4}
\end{figure}

\begin{figure}
    \centering
    \includegraphics[width=\linewidth]{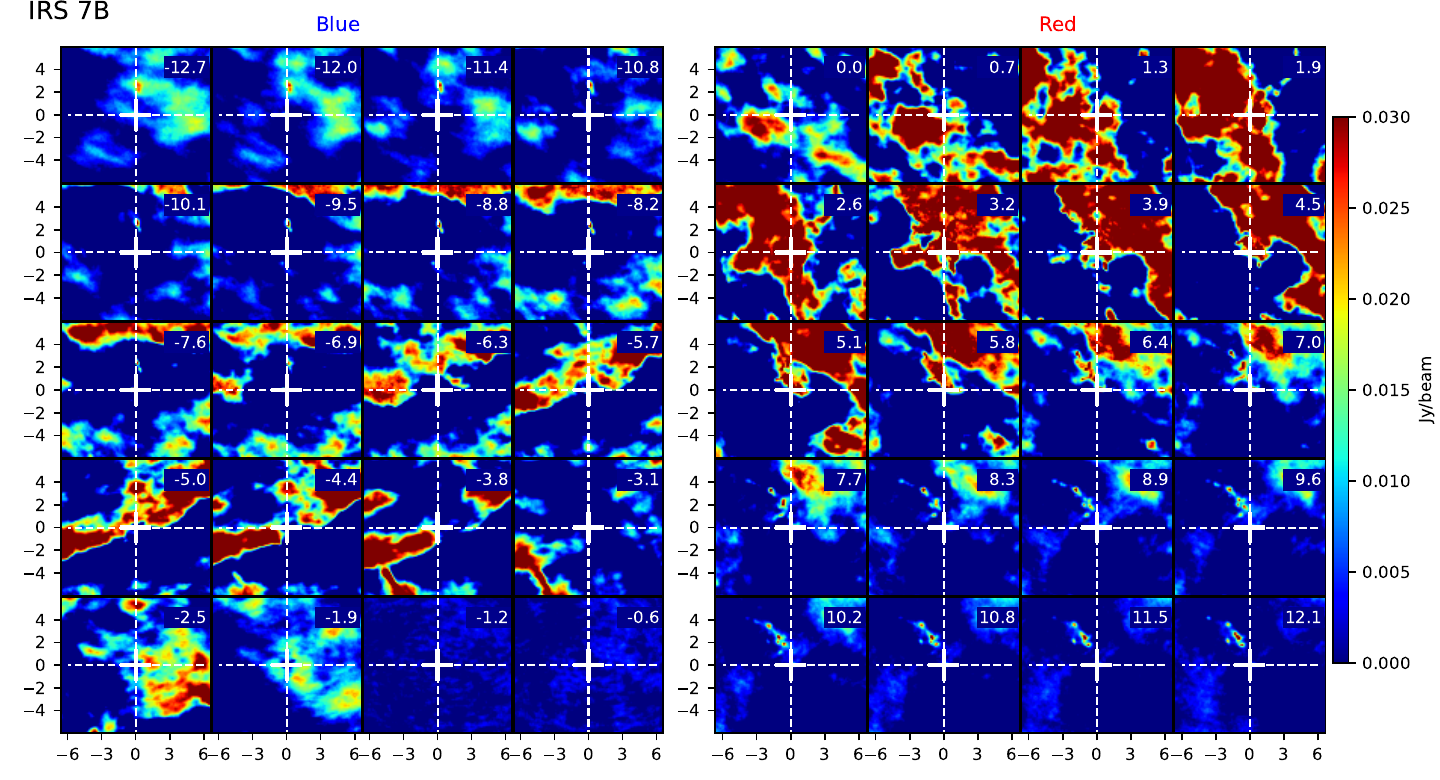}
    \caption{Same as Figure \ref{fig:appendix_BHR71IRS2} but for R CrA IRS 7B.}
    \label{fig:appendix_IRS7B}
\end{figure}

\begin{figure}
    \centering
    \includegraphics[width=\linewidth]{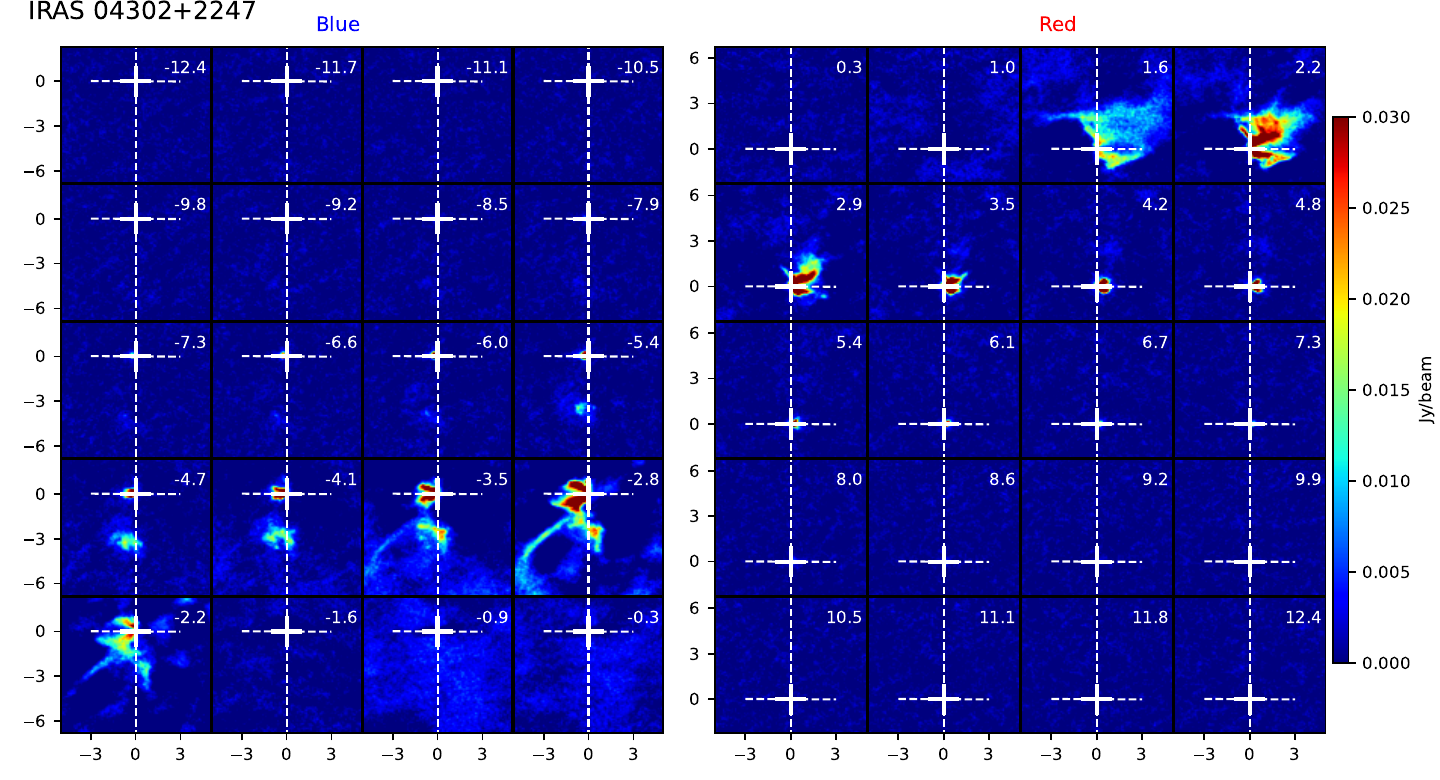}
    \caption{Same as Figure \ref{fig:appendix_BHR71IRS2} but for IRAS04302.}
    \label{fig:appendix_IRAS04302}
\end{figure}

\begin{figure}
    \centering
    \includegraphics[width=\linewidth]{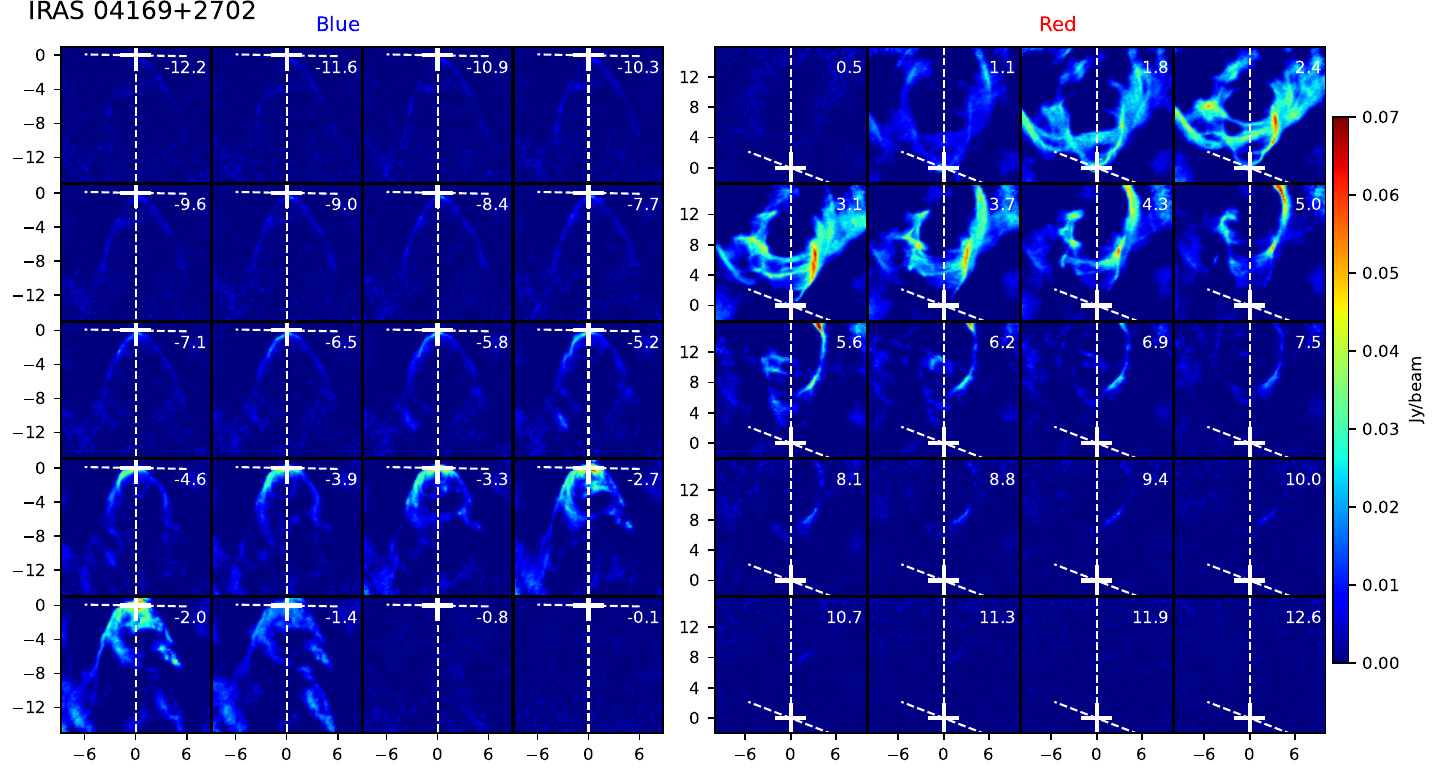}
    \caption{Same as Figure \ref{fig:appendix_BHR71IRS2} but for IRAS04169.}
    \label{fig:appendix_IRAS04169}
\end{figure}

\begin{figure}
    \centering
    \includegraphics[width=\linewidth]{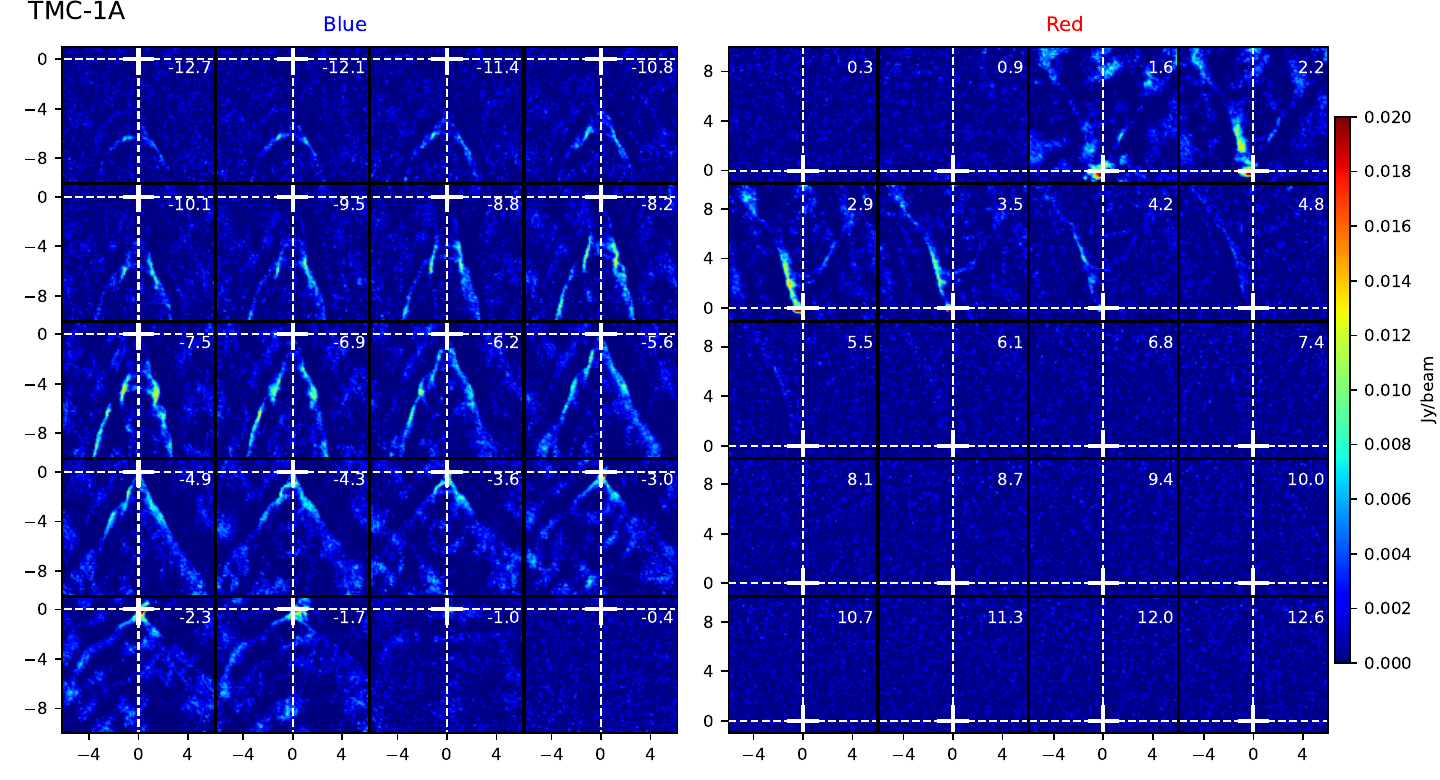}
    \caption{Same as Figure \ref{fig:appendix_BHR71IRS2} but for TMC-1A.}
    \label{fig:appendix_TMC1A}
\end{figure}

\begin{figure}
    \centering
    \includegraphics[width=\linewidth]{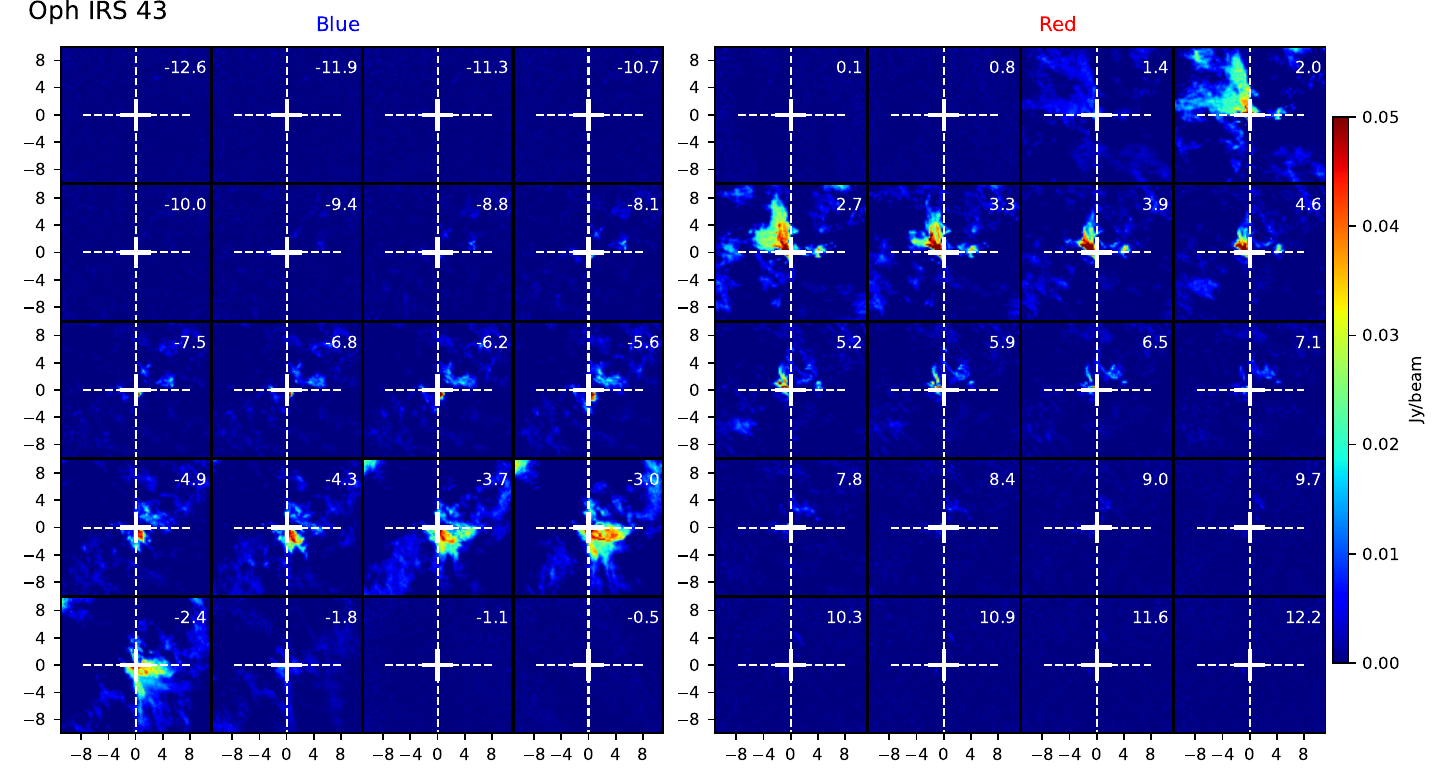}
    \caption{Same as Figure \ref{fig:appendix_BHR71IRS2} but for Oph IRS 43.}
    \label{fig:appendix_OphIRS43}
\end{figure}

\begin{figure}
    \centering
    \includegraphics[width=\linewidth]{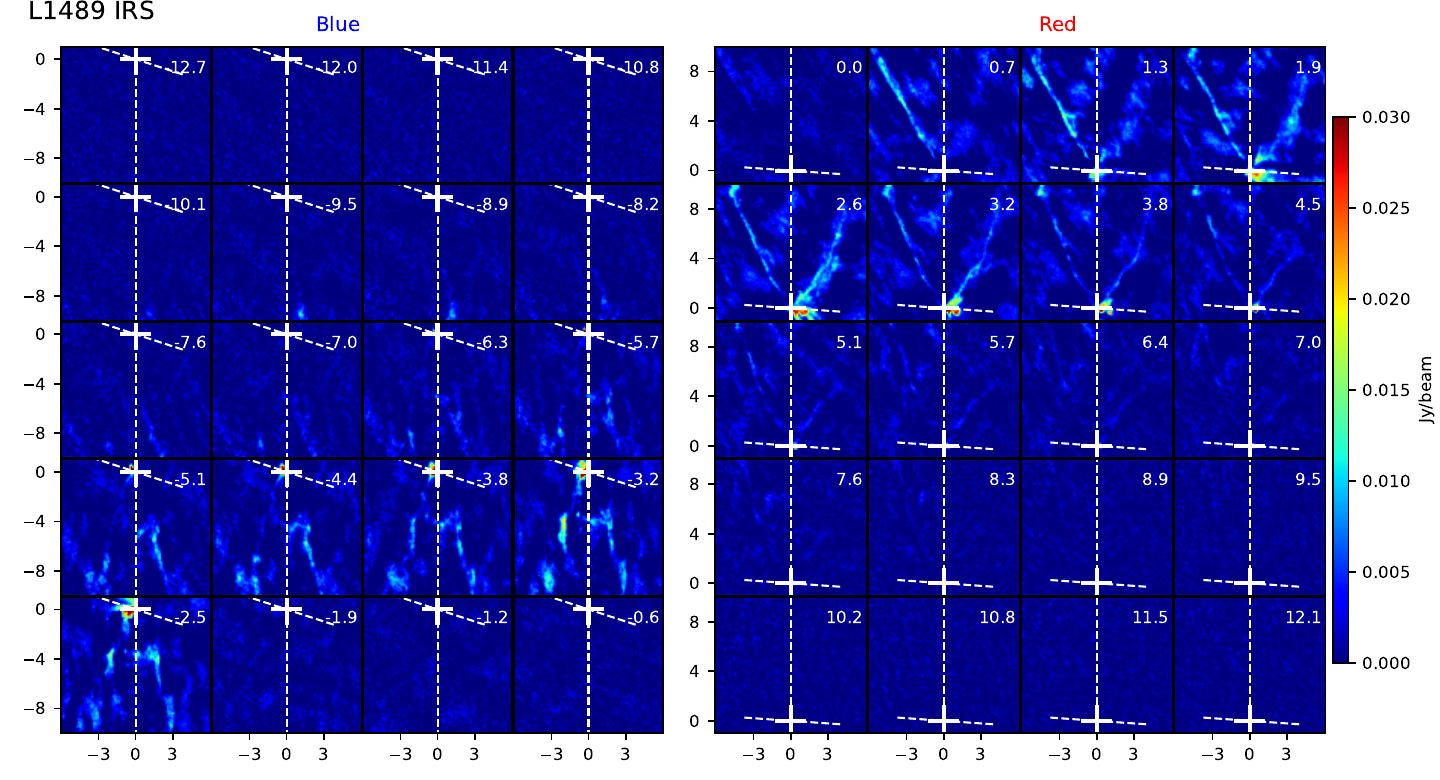}
    \caption{Same as Figure \ref{fig:appendix_BHR71IRS2} but for L1489IRS.}
    \label{fig:appendix_L1489IRS}
\end{figure}

\begin{figure}
    \centering
    \includegraphics[width=\linewidth]{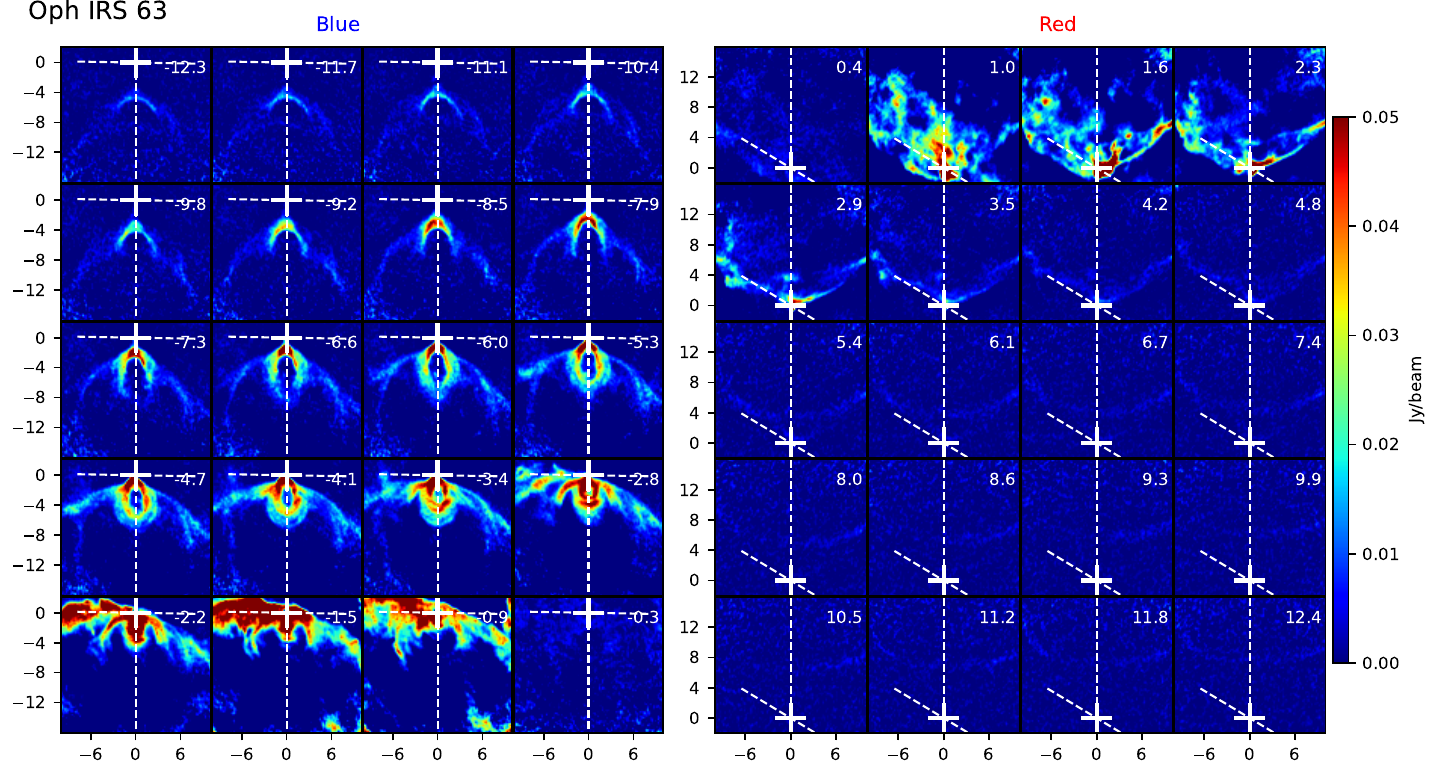}
    \caption{Same as Figure \ref{fig:appendix_BHR71IRS2} but for Oph IRS 63.}
    \label{fig:appendix_OphIRS63}
\end{figure}

Figures \ref{fig:appendix_BHR71IRS2} -- \ref{fig:appendix_OphIRS63} show the velocity channel maps for each of the sources. The maps are oriented so that the axis of the blue-shifted outflow points upwards, with the outflow axis of each source indicated by a vertical dashed line. The outflow axis is expected to be perpendicular to the major axis of the dust continuum emission. For several objects, however, the outflow is significantly misaligned to that direction. For those objects the outflow axis is determined by finding the axis of symmetry of the outflow by eye, or by fitting the data with a simple shell model in the cases of objects with wind-driven shells (Table \ref{tab:PA_and_Vsys}). The position angle of the the major axis of the dust continuum emission is also indicated by a dashed line to show whether the outflow axis is aligned perpendicular to the disk.
The velocities are relative to the systemic velocity. Table \ref{tab:PA_and_Vsys} summarizes the systemic velocity and position angle adopted in drawing the channel maps.

\section{Model Fitting of Wind-driven shells}

\begin{figure}
    \centering
    \includegraphics[width=\linewidth]{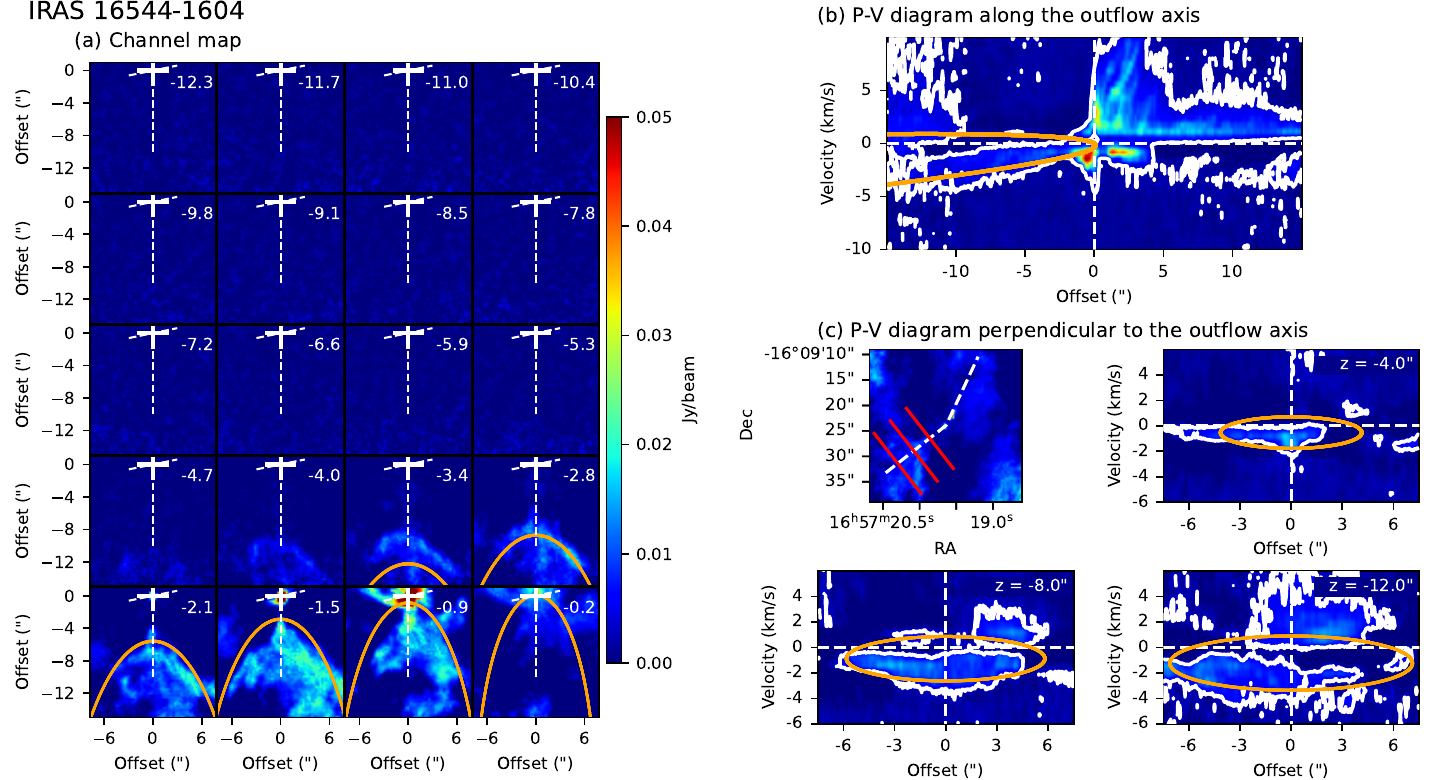}
    \caption{(a) the velocity channel map, (b) the P-V diagram along the outflow axis, and (c) the P-V diagrams perpendicular to the outflow axis of the blue-shifted outflow of IRAS 16544-1604. The position of the PV cuts are indicated in the channel map in the top right of (c). The distance in arcseconds of the PV cut from the source along the jet axis is indicated in the top right of each PV diagram. The white contours in the P-V diagrams depict 4-sigma and 5-sigma. The feature categorized as a wind-driven shell is fitted with a simple model described by Eq. \ref{eqn:shell_model}, which is indicated by the orange lines.}
    \label{fig:A2_IRAS16544_blue}
\end{figure}

\begin{figure}
    \centering
    \includegraphics[width=\linewidth]{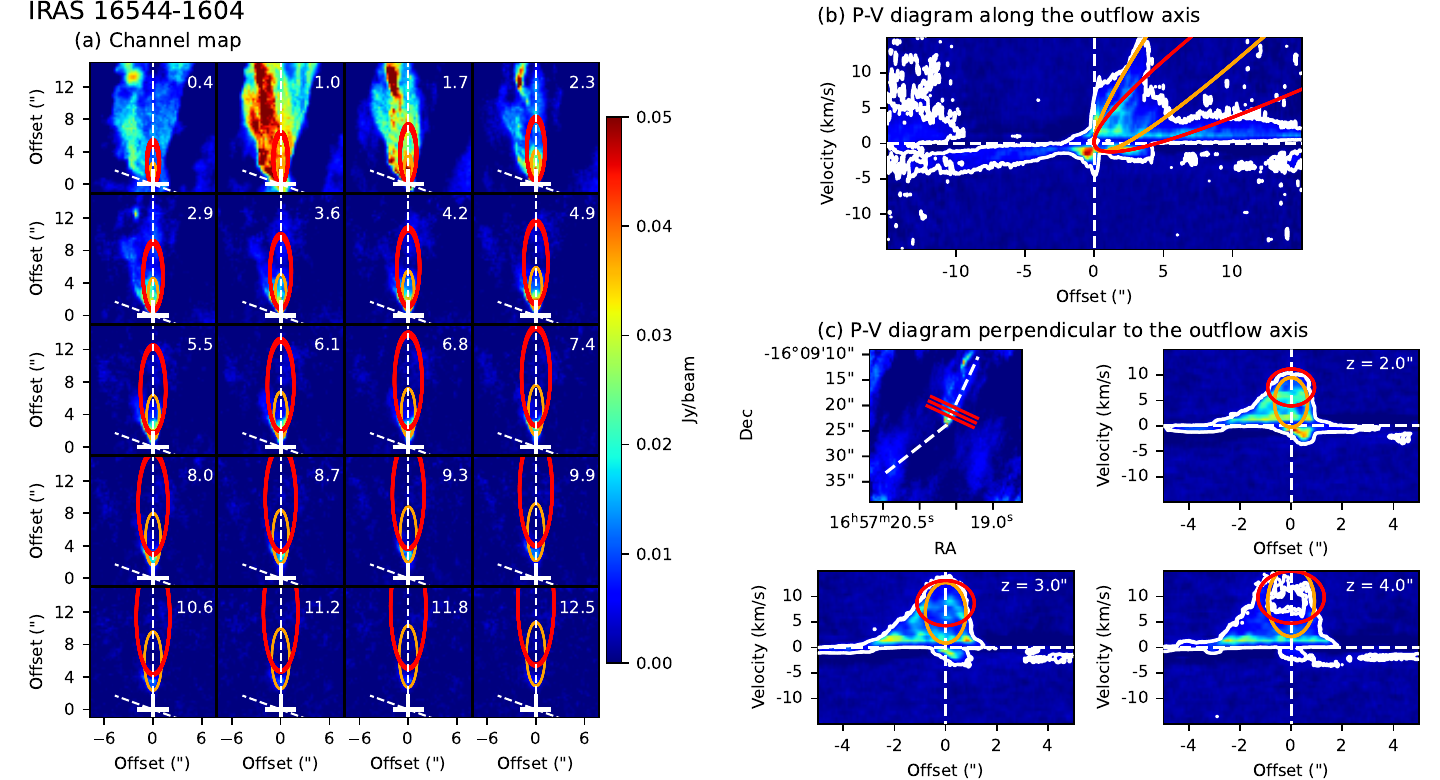}
    \caption{Same as Figure \ref{fig:A2_IRAS16544_blue} but for red-shifted outflow of IRAS16544. Shell R1 is indicated by the red lines, whild shell R2 is indicated by the orange lines.}
    \label{fig:A2_IRAS16544_red}
\end{figure}

\begin{figure}
    \centering
    \includegraphics[width=\linewidth]{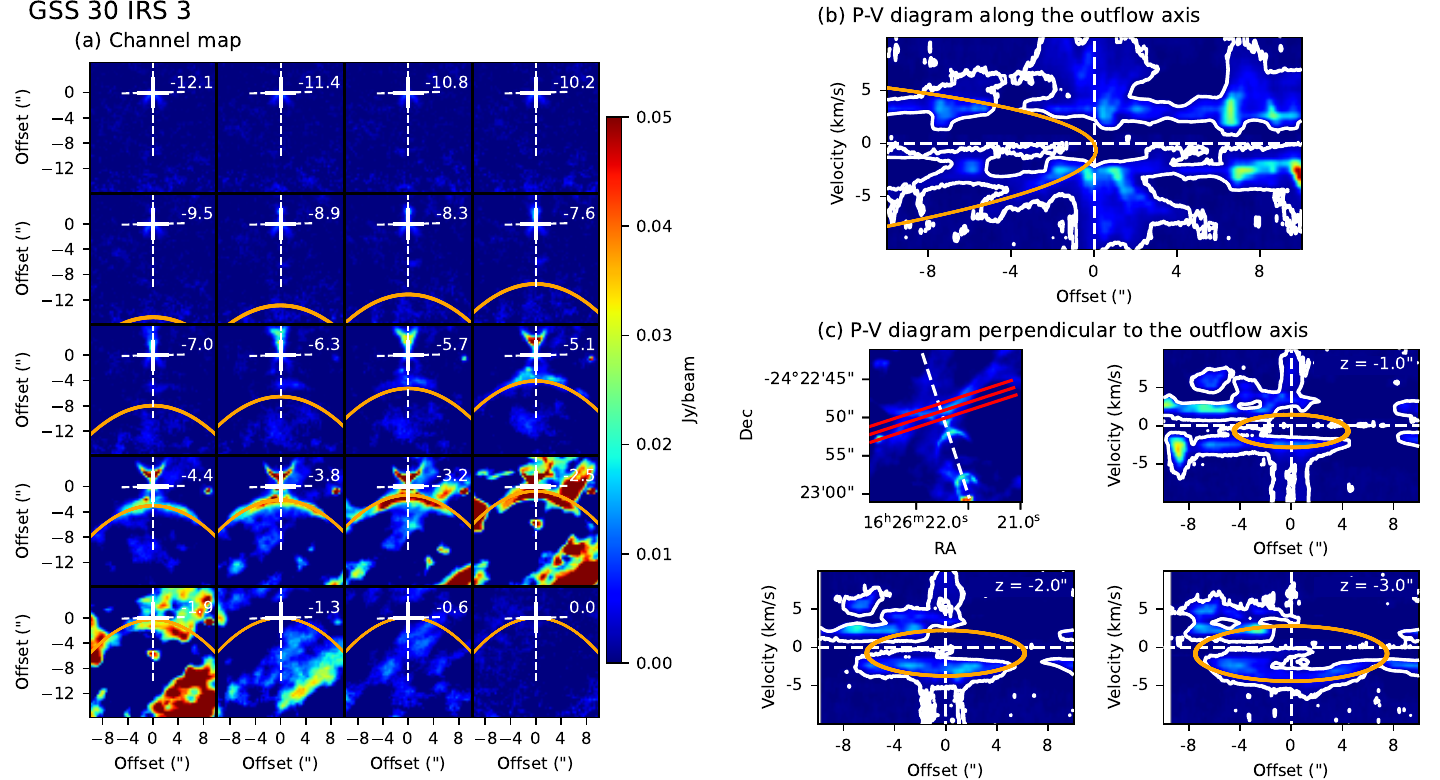}
    \caption{Same as Figure \ref{fig:A2_IRAS16544_blue} but for blue-shifted outflow of GSS30 IRS3.}
    \label{fig:A2_GSS30IRS3_blue}
\end{figure}

\begin{figure}
    \centering
    \includegraphics[width=\linewidth]{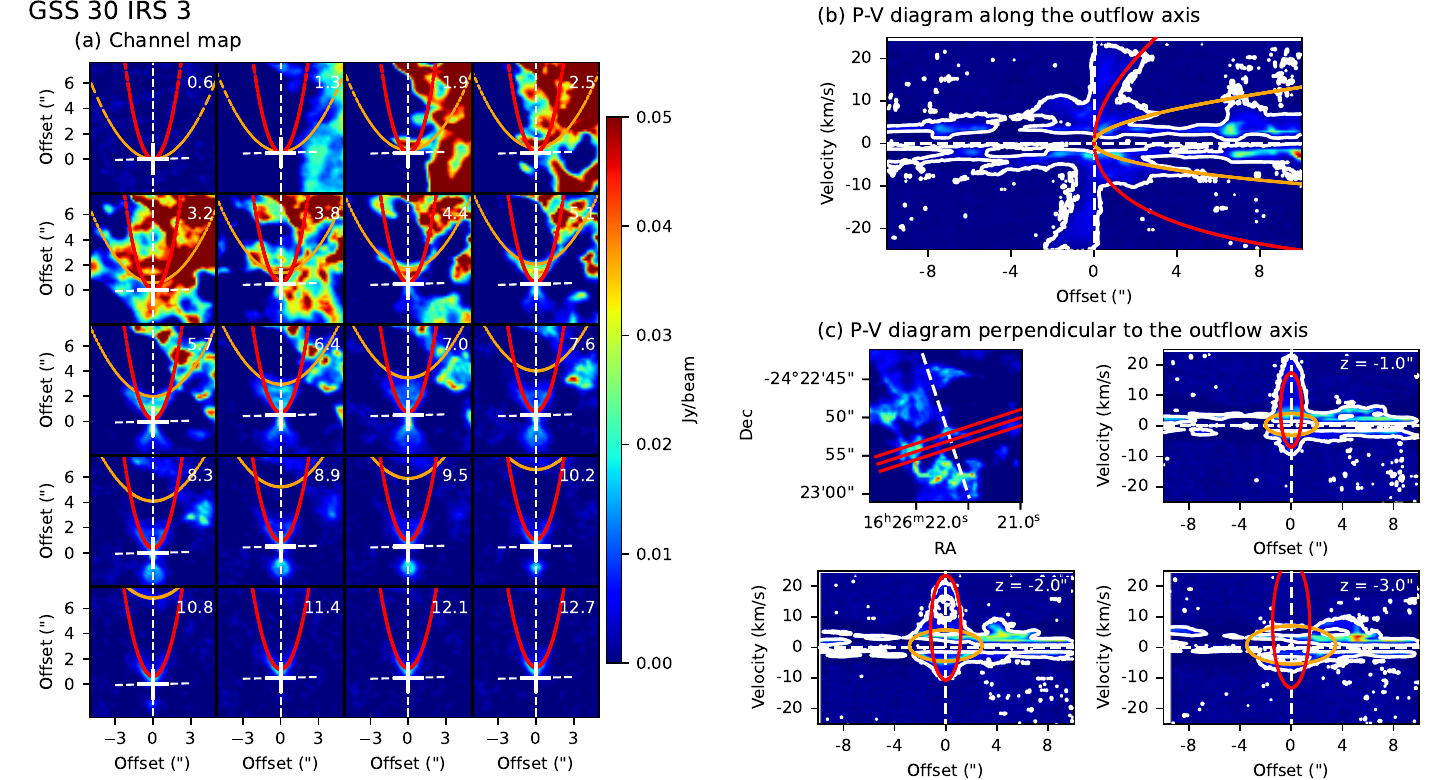}
    \caption{Same as Figure \ref{fig:A2_IRAS16544_blue} but for red-shifted outflow of GSS30 IRS3. Shell R1 is indicated by the red lines, while shell R2 is indicated by the orange lines.}
    \label{fig:A2_GSS30IRS3_red}
\end{figure}

\begin{figure}
    \centering
    \includegraphics[width=\linewidth]{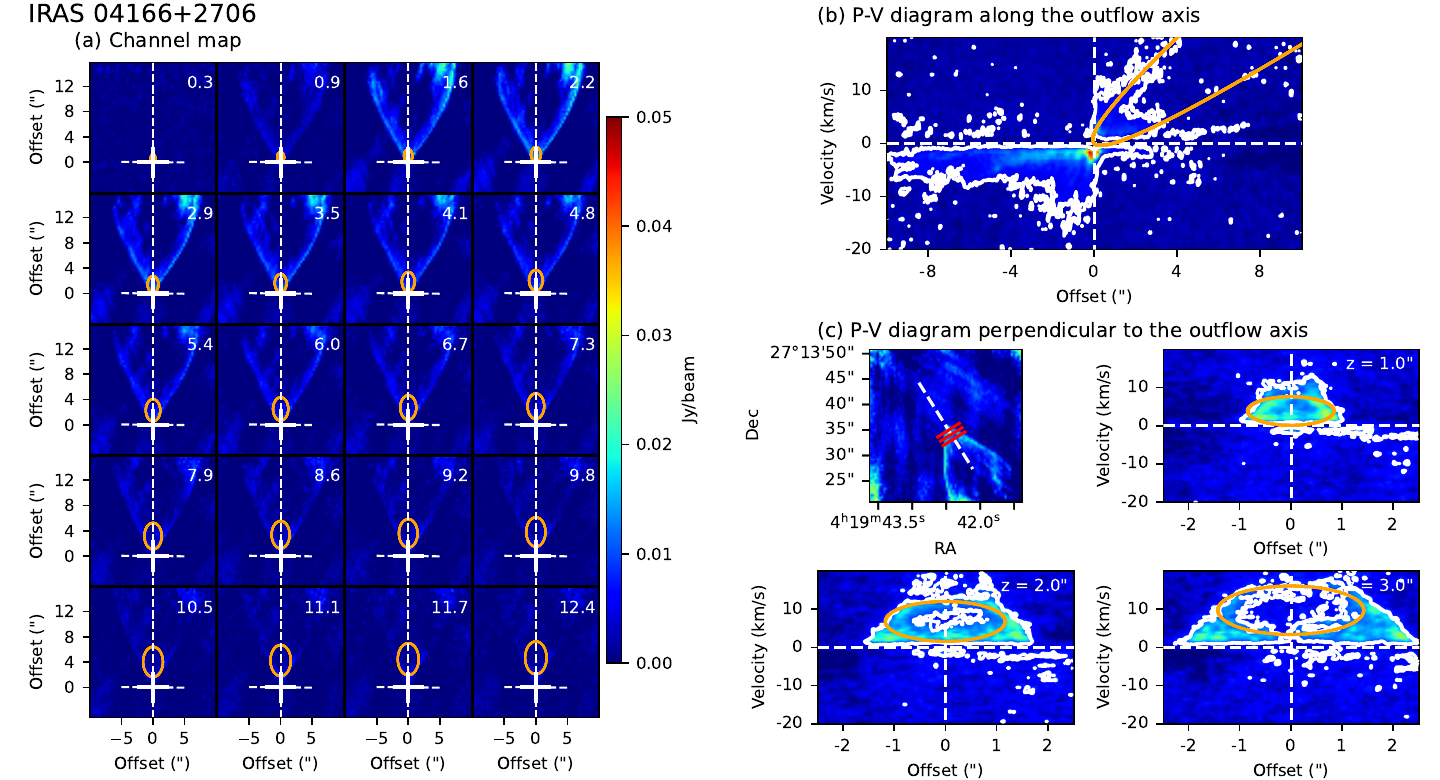}
    \caption{Same as Figure \ref{fig:A2_IRAS16544_blue} but for red-shifted outflow of IRAS04166.}
    \label{fig:A2_IRAS04166_red}
\end{figure}

\begin{figure}
    \centering
    \includegraphics[width=\linewidth]{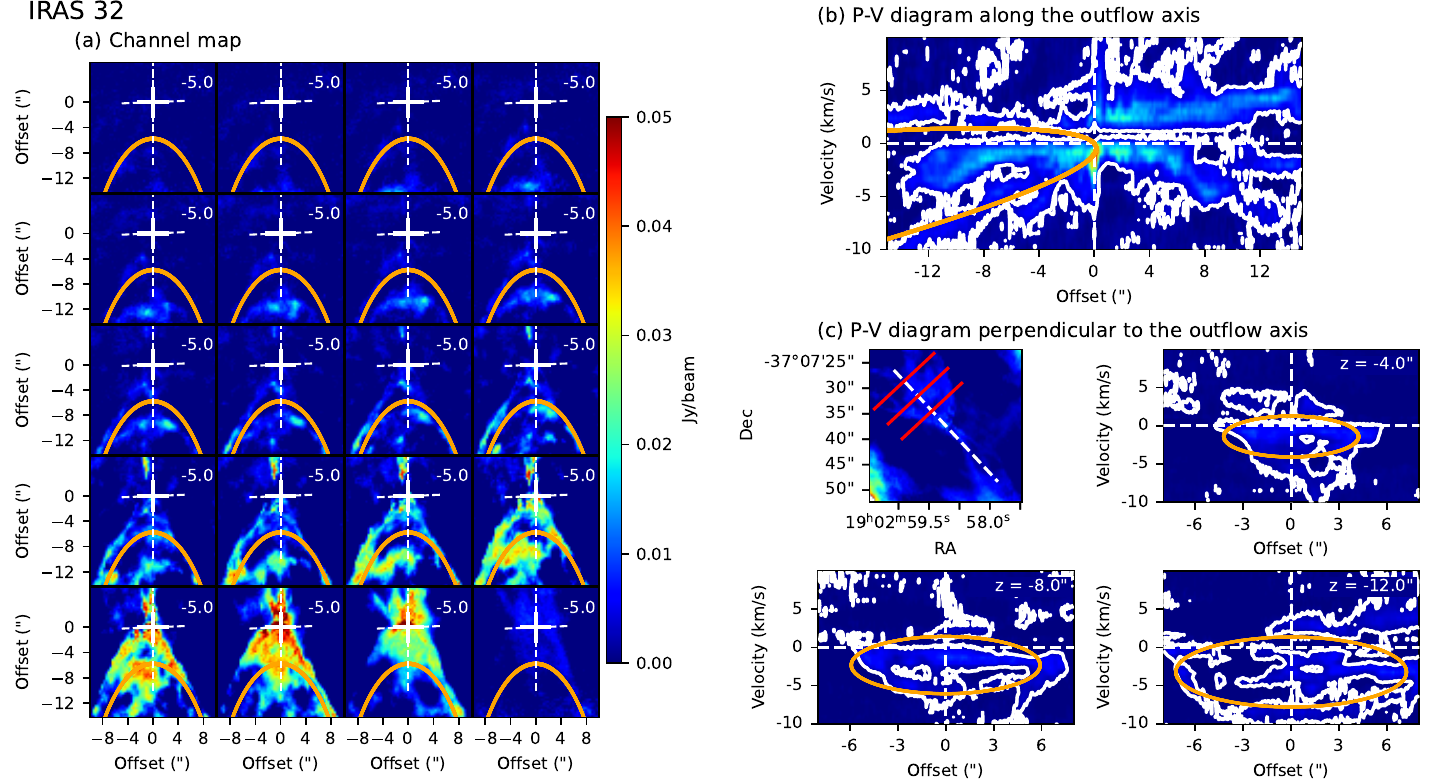}
    \caption{Same as Figure \ref{fig:A2_IRAS16544_blue} but for blue-shifted outflow of IRAS32.}
    \label{fig:A2_IRAS32_blue}
\end{figure}

\begin{figure}
    \centering
    \includegraphics[width=\linewidth]{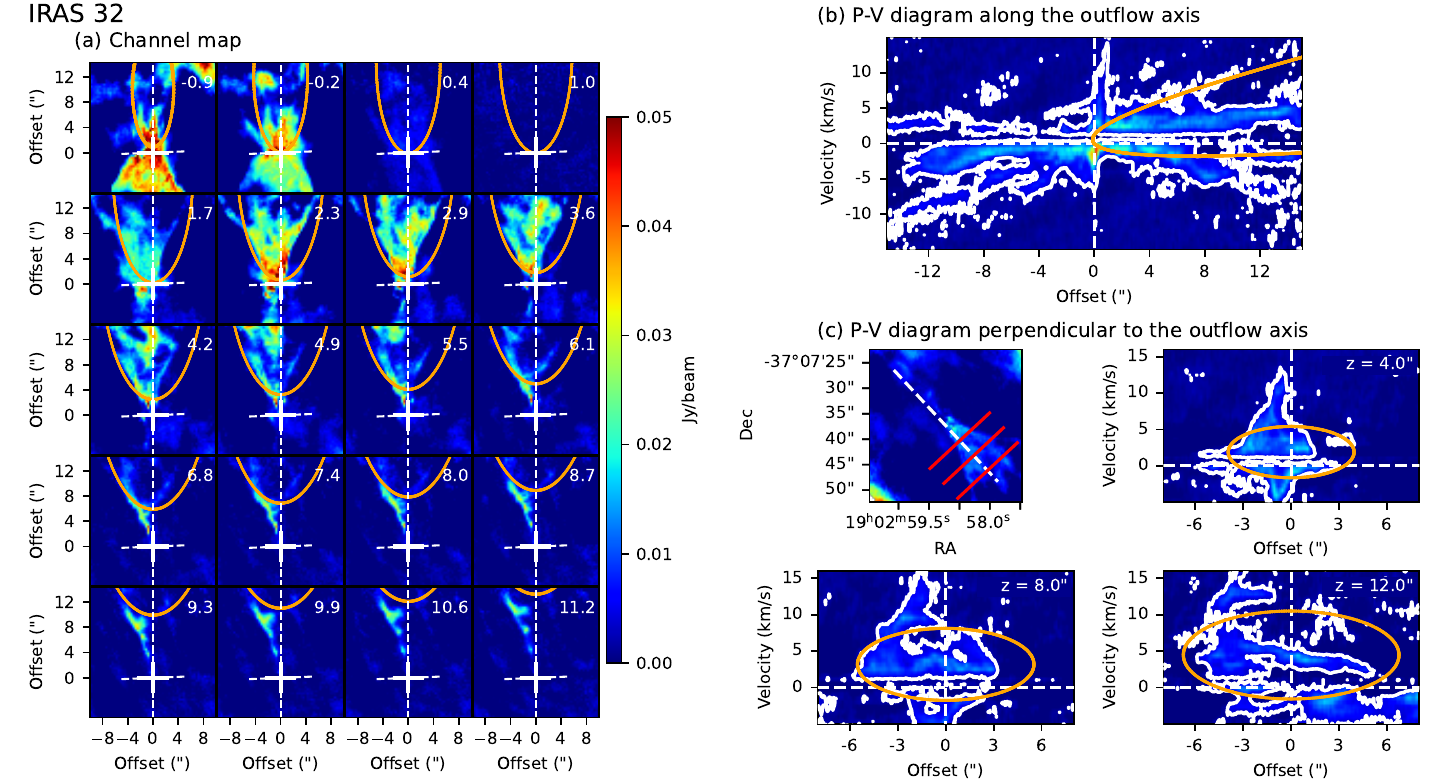}
    \caption{Same as Figure \ref{fig:A2_IRAS16544_blue} but for red-shifted outflow of IRAS32.}
    \label{fig:A2_IRAS32_red}
\end{figure}

\begin{figure}
    \centering
    \includegraphics[width=\linewidth]{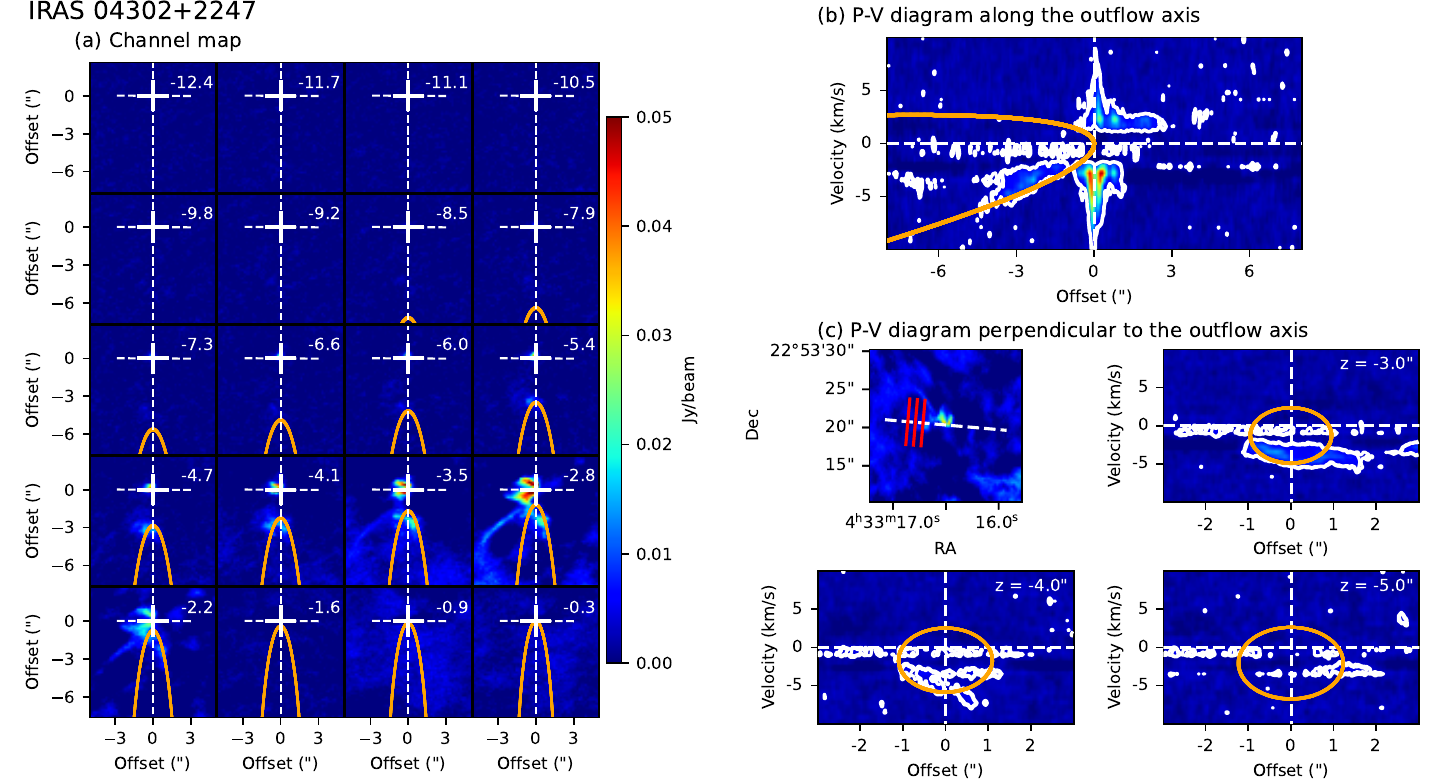}
    \caption{Same as Figure \ref{fig:A2_IRAS16544_blue} but for blue-shifted outflow of IRAS04302.}
    \label{fig:A2_IRAS04302_blue}
\end{figure}

\begin{figure}
    \centering
    \includegraphics[width=\linewidth]{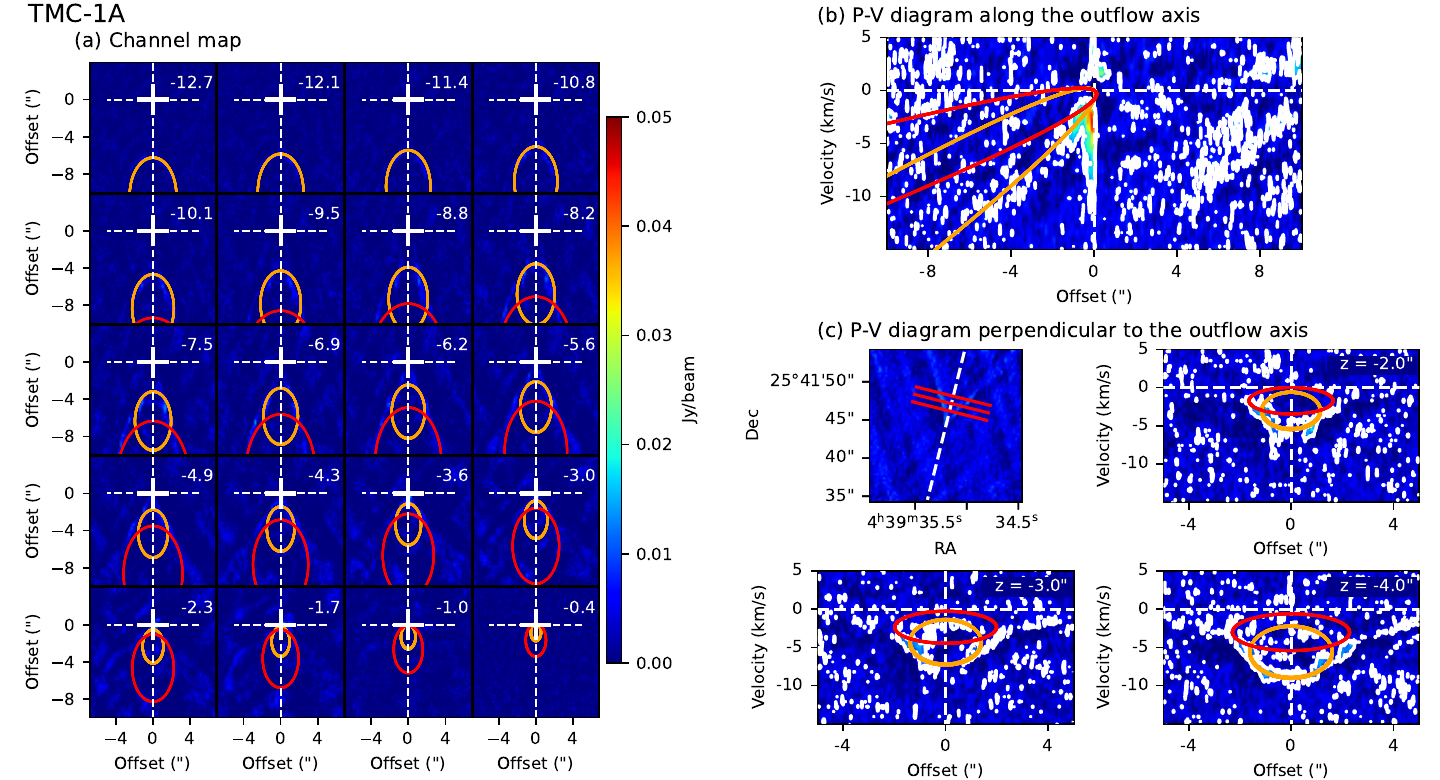}
    \caption{Same as Figure \ref{fig:A2_IRAS16544_blue} but for blue-shifted outflow of TMC-1A. Shell B1 is indicated by the orange lines, whild shell R2 is indicated by the red lines.}
    \label{fig:A2_TMC1A_blue}
\end{figure}

\begin{figure}
    \centering
    \includegraphics[width=\linewidth]{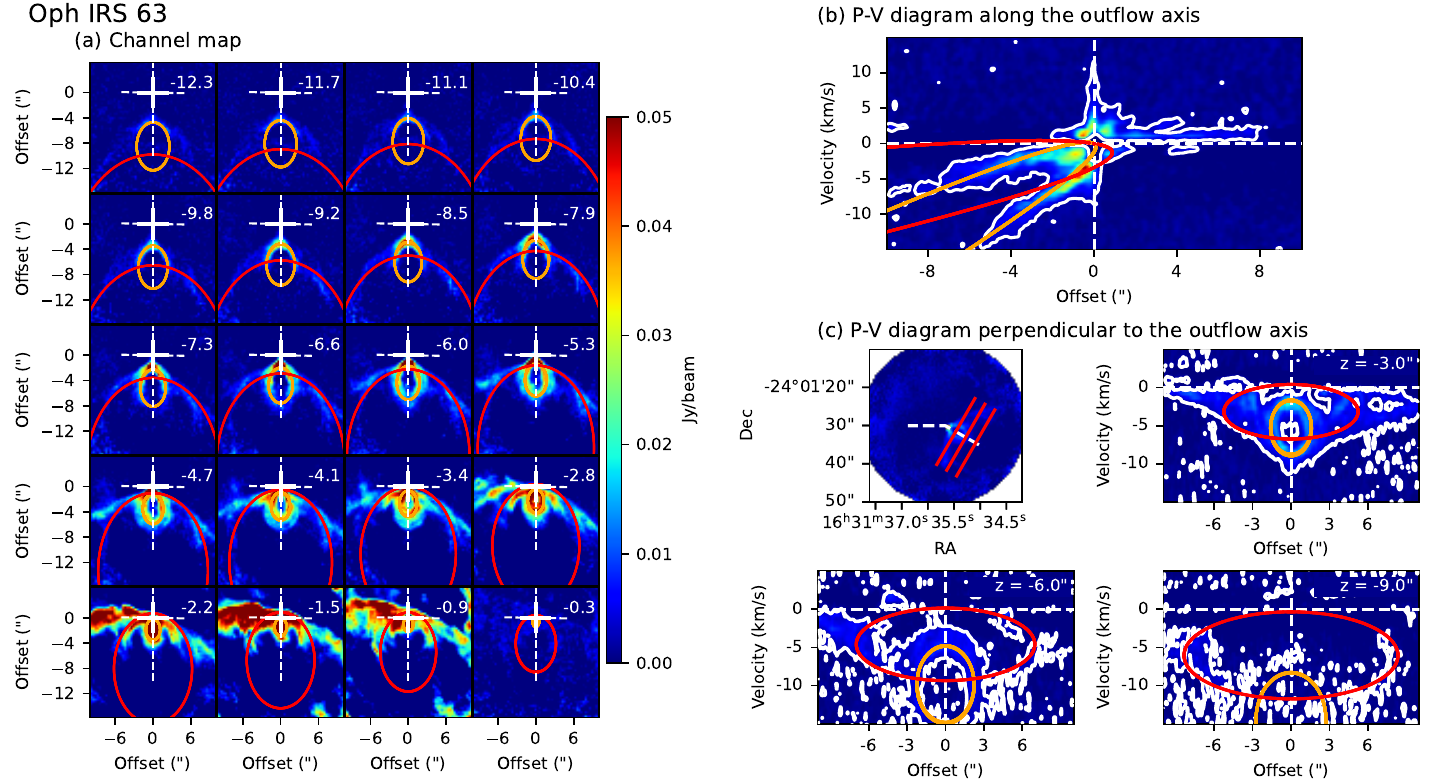}
    \caption{Same as Figure \ref{fig:A2_IRAS16544_blue} but for blue-shifted outflow of Oph IRS63. Shell B1 is indicated by the orange lines, whild shell B2 is indicated by the red lines.}
    \label{fig:A2_OphIRS63_blue}
\end{figure}

\begin{figure}
    \centering
    \includegraphics[width=\linewidth]{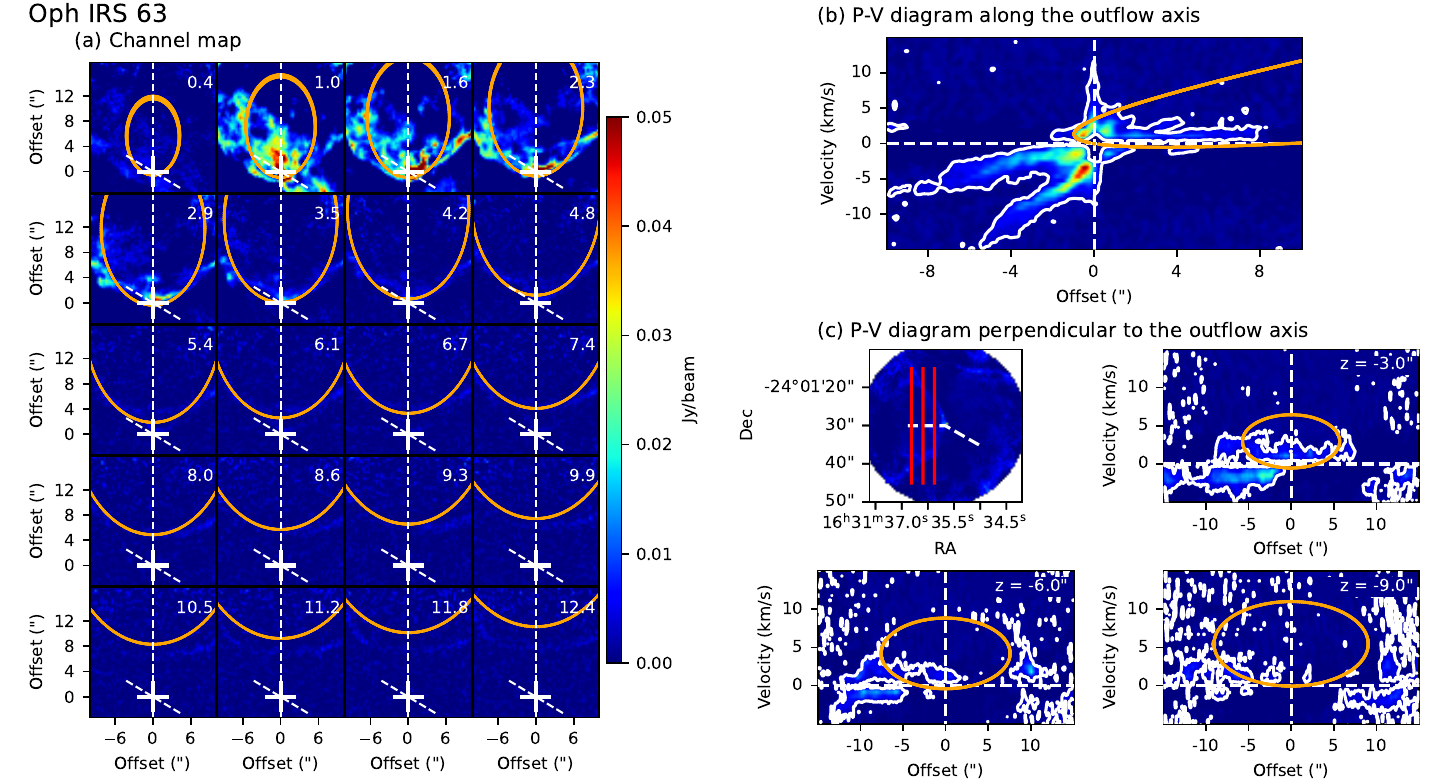}
    \caption{Same as Figure \ref{fig:A2_IRAS16544_blue} but for red-shifted outflow of Oph IRS63.}
    \label{fig:A2_OphIRS63_red}
\end{figure}

Figures \ref{fig:A2_IRAS16544_blue} -- \ref{fig:A2_OphIRS63_red} show (a) the velocity channel map, (b) the P-V diagram along the outflow axis, and (c) the P-V diagram perpendicular to the outflow axis for each of the outflows showing emission categorized as wind-driven shell emission. The features categorized as wind-driven shells are fitted with a simple model of Eq. \ref{eqn:shell_model} in \S 5.1.

\begin{acknowledgments}

A.F.-J. was supported by the NAOJ ALMA Scientific Research grant code
2019-13B.
Y.A. acknowledges support by Grant-in-Aid for Transformative Research Areas (A) grant Nos. 20H05844 and 20H05847 and JSPS KAKENHI grant No. 24K00674.
S.T. acknowledge the support by JSPS KAKENHI grant Nos. JP21H00048 and JP21H04495 and by NAOJ ALMA Scientific Research grant No. 2022-20A.
N.O. and M.N. acknowledge acknowledge support from the National Science and
Technology Council (NSTC) in Taiwan through the grant NSTC 113-2112-M-001-037 and the
Academia Sinica Investigator Project Grant (AS-IV-114-M02).
LWL acknowledges support from NSF AST-2108794.
This paper makes use of the following ALMA data: ADS/JAO.ALMA\#2019.1.00261.L, and \#2019.A.00034.S. ALMA is a partnership of ESO (representing its member states), NSF (USA) and NINS (Japan), together with NRC (Canada), NSTC and ASIAA (Taiwan), and KASI (Republic of Korea), in cooperation with the Republic of Chile. The Joint ALMA Observatory is operated by ESO, AUI/NRAO and NAOJ

\end{acknowledgments}

%

\vspace{5mm}
\facilities{ALMA}


\software{astropy \citep{astropy} 
          }




\bibliography{edisk_outflow}{}
\bibliographystyle{aasjournal}



\end{document}